\documentclass[12pt]{article}
\RequirePackage{amsthm,amsmath,natbib}
\usepackage{amsmath, amsthm, amssymb}
\usepackage{verbatim,color,amssymb,epsfig}
\usepackage[usenames,dvipsnames]{xcolor}
\usepackage{multirow,booktabs}

\usepackage{xr}
\RequirePackage{amsthm,amsmath,natbib}
\usepackage{amsmath, amsthm, amssymb}
\usepackage[auth-sc, affil-sl]{authblk}
\usepackage{epsfig,natbib}

\def\boxit#1{\vbox{\hrule\hbox{\vrule\kern6pt
          \vbox{\kern6pt#1\kern6pt}\kern6pt\vrule}\hrule}}

\def\trans{^{\rm T}}

\numberwithin{equation}{section}
\theoremstyle{plain}

\newtheorem{Theorem}{\underline{\bf Theorem}}[section]

\newtheorem{Lemma}{\underline{\bf Lemma}}

\newtheorem{Condition}{\underline{\bf Condition}}

\newtheorem{Algorithm}{\underline{\bf Algorithm}}[section]
\begin{document}

\thispagestyle{empty}
\baselineskip=28pt
\begin{center}
{\LARGE{\bf Signal extraction approach for sparse multivariate response regression}}
\end{center}
\vskip 10mm

\baselineskip=15pt
\vskip 2mm
\begin{center}
 Ruiyan Luo\\
\vskip 2mm
Division of Epidemiology and Biostatistics, Georgia State University School of Public Health, One Park Place, Atlanta, GA 30303\\
rluo@gsu.edu\\
\hskip 5mm\\
Xin Qi\\
\vskip 2mm
Department of Mathematics and Statistics, Georgia State University, 30 Pryor Street, Atlanta, GA 30303-3083\\
xqi3@gsu.edu \\
\end{center}

\begin{abstract}
\baselineskip=15pt
In this paper, we consider multivariate response regression models with high dimensional predictor variables. One way to model the correlation among the response variables is through the low rank decomposition of the coefficient matrix, which has been considered by several papers for the high dimensional predictors. However, all these papers focus on the singular value decomposition of the coefficient matrix. Our target is the decomposition of the coefficient matrix which leads to the best lower rank  approximation to the regression function, the signal part in the response. Given any rank, this decomposition has nearly the  smallest expected prediction error among all approximations to the the coefficient matrix with the same rank. To estimate the decomposition, we formulate a penalized generalized eigenvalue problem to obtain the first  matrix in the decomposition and then obtain the second one by a least squares method.  In the high-dimensional setting, we establish the oracle inequalities for the estimates. Compared to the existing theoretical results, we have less restrictions on the distribution of the noise vector in each observation and allow correlations among its coordinates. Our theoretical results do not depend on the dimension of the multivariate response. Therefore, the dimension is arbitrary and can be larger than the sample size and the dimension of the predictor.  Simulation studies  and  application to real data show that the proposed method has good prediction performance and is efficient in dimension reduction for various reduced rank models.
\end{abstract}
\vskip 5mm

\noindent{\bf\underline{Keyword: } }
{multivariate regression; high dimensional predictors; signal extraction; dimension reduction; best lower rank  approximation; oracle inequalities.}

\pagenumbering{arabic}
\newlength{\gnat}
\setlength{\gnat}{22pt}
\baselineskip=\gnat

\section{Introduction}
 
In this paper, we consider multivariate response regression models with high dimensional predictor variables. Several methods have been proposed to estimate the coefficient matrix and select a common subset of explanatory variables.
The sparse partial least square method \citep{chun} finds sparse linear combinations of the original predictors to maximize their covariance with response variables. 
\cite{turlach2005simultaneous}, \cite{simila2007input}, \cite{peng2010regularized},   \cite{chen2012sparse}, \cite{chen2012reduced}, and \cite{bunea2012joint} estimate the coefficient matrix by minimizing the penalized (joint) residual sum of squares with different penalties. 
\cite{simila2007input} and \cite{turlach2005simultaneous} assume row sparsity of the coefficient matrix and  use  group-Lasso type penalties with $l_2$ or $l_{\infty}$ norm that treat each row of the regression coefficient matrix as a group. 
 \cite{peng2010regularized} imposes both row-wise and element-wise sparsity on the coefficient matrix.  In addition to the row-wise sparsity assumption, \citet{chen2012sparse}, \cite{chen2012reduced} and \cite{bunea2012joint}  make the reduced rank assumption on the regression coefficient matrix \citep{izenman1975reduced, reinsel1998multivariate}. Under this assumption, \citet{chen2012sparse} and  \cite{chen2012reduced} aim to estimate the singular value decomposition (SVD) of the coefficient matrix.
\citet{chen2012sparse}  specify the rank by the cross-validation method,   and use a group-Lasso type penalty on the first  matrix in the SVD of the coefficient matrix to achieve sparsity. 
  \cite{chen2012reduced} pre-specifies the rank by existing methods as given in  \cite{anderson2002specification}, \cite{camba2003tests} or \cite{bunea2011optimal}, and  impose  an adaptive-lasso type penalty. \cite{bunea2012joint} penalizes on the rank sparsity and the variable sparsity simultaneously, and provides theoretical results in the high-dimensional settings.

We also  assume the reduced-rank structure and the row-wise sparsity on the coefficient matrix. Instead of the SVD of the coefficient matrix, our target is the decomposition of the coefficient matrix which leads to the best lower rank  approximation to the regression function, the product of the design matrix and the coefficient matrix.    Given any rank, this decomposition has the smallest approximation error to the regression function and nearly the smallest expected prediction error among all approximations  to the coefficient matrix with the same rank.  Therefore, our proposed method  is
expected to have good prediction performance and be efficient in dimension reduction. To estimate this decomposition,  we first propose a penalized generalized eigenvalue problem   to obtain the first lower rank matrix in this decomposition, and then  obtain  the second by a least squares method.  In the high-dimensional setting, we establish the oracle inequalities for the estimators of the lower rank matrices, the coefficient matrix and the estimated regression function, respectively. \cite{bunea2012joint} provides the convergence rate of the estimate of the regression function under the assumption that the coordinates of the noise vector are identically independently distributed normal randoms variables. The convergence rate depends on the dimension of the multivariate response variable. In order that the convergence rate goes to zero, the increase of the dimension of the multivariate response variable has to be slower than the sample size.  We make weaker assumption on the distribution of the noise vector  and allow correlations among its coordinates. Our theoretical results do not depend on the dimension of the multivariate response. Therefore, the dimension of the multivariate response can be arbitrary and even larger than the sample size and the dimension of the predictor.  Through simulation studies on the reduced rank models with various settings, we demonstrate that our method has competitive predictive ability and is efficient in dimension reduction.
		
		The paper is organized as follows. In Section 2, we formulate the best lower rank approximation problem and establish its equivalence with a generalized eigenvalue problem. In Section 3, in the high dimensional settings,  we propose sparse estimates of the decomposition and the coefficient matrix. We  establish the oracle inequalities for the estimates. In Section 4, we discuss the choice of the number of components and tuning parameters. We conduct simulation studies and a case study in Sections 5 and 6, respectively, and summarize the paper with discussion in Section 7.  All the proofs can be found in the supplementary material.

\section{Signal extraction approach to multivariate regression (SiER)}\label{section.2}
  We consider the following  linear regression model with responses taking values in $\mathbf{R}^q$, where $q \ge 1$. Suppose that the $i$-th observation satisfies 
	$\mathbf{y}_i=x_{i1}\boldsymbol{\beta}_1+x_{i2}\boldsymbol{\beta}_2+\cdots+x_{ip}\boldsymbol{\beta}_p+\boldsymbol{\varepsilon}_i$,  $1\le i\le n$. 
	Here  $x_{ij} \in \mathbb{R}$ is the $i$-th observed value of the $j$-th predictor variable, 
 $\mathbf{y}_i=({y}_{i1},{y}_{i2},\cdots,{y}_{iq})\trans \in\mathbb{R}^q$ and $\boldsymbol{\varepsilon}_i=({\varepsilon}_{i1},{\varepsilon}_{i2},\cdots,{\varepsilon}_{iq})\trans\in\mathbb{R}^q$ are the $i$-th observed response vector and the $i$-th noise vector, respectively, 
and each coefficient $\boldsymbol{\beta}_j=({\beta}_{j1},{\beta}_{j2},\cdots,{\beta}_{jq})\trans$, $j=1, \ldots, p$, is a $q$-dimensional vector.  Let $\mathbf{x}_i=({x}_{i1},{x}_{i2},\cdots,{x}_{ip})\trans$ denote the $i$-th observation of the $p$-dimensional predictor vector, $\mathbf{Y}=[\mathbf{y}_1,\cdots,\mathbf{y}_n]\trans$   the  $n \times q$ response matrix, $\mathbf{X}=[\mathbf{x}_1,\cdots,\mathbf{x}_n]\trans$ the $n\times p$  design matrix, $\mathcal{B}=[\boldsymbol{\beta}_1, \cdots, \boldsymbol{\beta}_p]\trans$ the $p \times q$ coefficient matrix,  and $\boldsymbol{\varepsilon}=[\boldsymbol{\varepsilon}_1,\cdots, \boldsymbol{\varepsilon}_n]\trans$ the $n \times q$ random noise matrix, respectively.   We assume that $\boldsymbol{\varepsilon}_1,\cdots, \boldsymbol{\varepsilon}_n$, are i.i.d. random vectors with $E[\boldsymbol{\varepsilon}_i]=\mathbf{0}$. Correlations are allowed among coordinates of $\boldsymbol{\varepsilon}_i$, $1\le i \le n$. Throughout this paper, we will assume that $\mathbf{X}$ is nonrandom and  the column means of $\mathbf{X}$ are all zeros, the same setting as in \citet{bickel2009simultaneous}. Then the model can be written as 
\begin{align}
\mathbf{Y} = \mathbf{X}\mathcal{B}+\boldsymbol{\varepsilon}\;.\label{100003}
\end{align}

 In this paper, we make the reduced-rank assumption, that is, the rank of $\mathcal{B}$, denoted by $K$, is small compared to $n$, $p$ and $q$. As the dimension of the subspace spanned by $\{\boldsymbol{\beta}_1,\cdots,\boldsymbol{\beta}_p\}$ in $\mathbb{R}^q$ is equal to $K$, given any $K$ linearly independent vectors $\mathbf{w}_1,\cdots,\mathbf{w}_K$ in this subspace, each coefficient vector $\boldsymbol{\beta}_j$ can be expressed as a linear combination of $\mathbf{w}_1,\cdots,\mathbf{w}_K$. So we have the decomposition
\begin{align}
 \mathcal{B}=\mathbf{A}\mathbf{W}\trans=\boldsymbol{\alpha}_1\mathbf{w}_1\trans+\cdots+\boldsymbol{\alpha}_K\mathbf{w}_K\trans,\label{70001}
\end{align}
where $\mathbf{A}=[\boldsymbol{\alpha}_1,\cdots, \boldsymbol{\alpha}_K]$ and $\mathbf{W}=[\mathbf{w}_1,\cdots,\mathbf{w}_K]$  are $p\times K$ and $q\times K$ matrices, respectively. There are infinitely many choices of $\mathbf{w}_1,\cdots,\mathbf{w}_K$, hence the decomposition $\eqref{70001}$ is not unique. \citet{chen2012sparse} and  \cite{chen2012reduced} consider the SVD, $\mathcal{B}=\mathbf{U}\mathbf{D}\mathbf{V}\trans$, where $\mathbf{U}$ and $\mathbf{V}$ are $p\times K$ and $q\times K$ matrices with orthonormal columns, respectively, and $\mathbf{D}$ is a $K\times K$ nonnegative diagonal matrix. This SVD is a special case of $\eqref{70001}$ with 
\begin{equation}
\mathbf{A}=\mathbf{U}\mathbf{D} \text{  and  } \mathbf{W}=\mathbf{V}.
\label{svd.eq}
\end{equation}

 We will consider a different decomposition which leads to the best lower rank approximation to  $\mathbf{X}\mathcal{B}$, the signal in $\mathbf{Y}$. Specifically, we want to find $\mathbf{A}$ and $\mathbf{W}$ such that for any $1\le k\le K$, we have 
\begin{align}
 \|\mathbf{X}\mathcal{B}-\sum_{j=1}^k\mathbf{X}\boldsymbol{\alpha}_j\mathbf{w}_j\trans\|_{F}^2=\min_{\substack{\mathbf{r}_j\in\mathbb{R}^n, \mathbf{v}_j\in \mathbb{R}^q,\\ 1\le j\le k}} \|\mathbf{X}\mathcal{B}-\sum_{j=1}^k\mathbf{r}_j\mathbf{v}_j\trans \|_F^2,\label{70018}
\end{align}
where $\|\cdot\|_F$ is the Frobenius norm and is define as $\|\mathbf{M}\|_F=\sqrt{\sum_{i=1}^s\sum_{j=1}^tM_{ij}^2}$ for any $s\times t$ matrix $\mathbf{M}$. 
Therefore, $\eqref{70018}$ implies that $\sum_{j=1}^k\mathbf{X}\boldsymbol{\alpha}_j\mathbf{w}_j\trans$ is the best rank $k$ approximation to $\mathbf{X}\mathcal{B}$ for any $1\le k\le K$. Note that when we change $\boldsymbol{\alpha}_j$ to $\boldsymbol{\alpha}_j/c$ and $\mathbf{w}_j$ to $c\mathbf{w}_j$, where $c$ is any nonzero scalar, $\boldsymbol{\alpha}_j\mathbf{w}_j\trans$ is unchanged. Hence,   we  restrict that $\boldsymbol{\alpha}_j\trans \mathbf{S}\boldsymbol{\alpha}_j=1$ for any $1\le j\le K$, where $\mathbf{S}=\mathbf{X}\trans\mathbf{X}/n$. To find $\mathbf{A}$ and $\mathbf{W}$, we consider the SVD of $\mathbf{X}\mathcal{B}$,
\begin{align}
 \mathbf{X}\mathcal{B}=\sigma_1\boldsymbol{\gamma}_1\mathbf{u}_1\trans+\sigma_2\boldsymbol{\gamma}_2\mathbf{u}_2\trans+\cdots+\sigma_K\boldsymbol{\gamma}_K\mathbf{u}_K\trans,\label{70006}
\end{align}
where $\sigma_1\ge \sigma_2\ge \cdots\ge \sigma_K\ge 0$ are singular values of $\mathbf{X}\mathcal{B}$, $\boldsymbol{\gamma}_k\in\mathbb{R}^n$ and $\mathbf{u}_k\in\mathbb{R}^q$ are the left-singular and right-singular vectors corresponding to $\sigma_k$, respectively, with $\|\mathbf{u}_k\|_2=1$, $\|\boldsymbol{\gamma}_k\|_2=1$.  By the Eckart-Young Theorem, $\sum_{j=1}^k\sigma_j\boldsymbol{\gamma}_j\mathbf{u}_j\trans$ is the best rank $k$ approximation to $\mathbf{X}\mathcal{B}$. We  define the columns of  $\mathbf{W}$ and $\mathbf{A}$ as
\begin{align}
\mathbf{w}_k=\frac{\sigma_k}{\sqrt{n}} \mathbf{u}_k,\quad   \boldsymbol{\alpha}_k=\frac{n}{\sigma_k^2}\mathcal{B}\mathbf{w}_k,\quad   1\le k\le K, \label{70002}
\end{align}
respectively.  As $\mathbf{u}_k$, $1\le k\le K$, are orthonormal, by $\eqref{70006}$ and $\eqref{70002}$,
\begin{align}
   \mathbf{X}\boldsymbol{\alpha}_k=\frac{n}{\sigma_k^2}\mathbf{X}\mathcal{B}\mathbf{w}_k=\frac{\sqrt{n}}{\sigma_k}\mathbf{X}\mathcal{B}\mathbf{u}_k=\sqrt{n}\boldsymbol{\gamma}_k,\quad  1\le k\le K.  \label{77000}
\end{align}
Therefore, $\sum_{j=1}^k\mathbf{X}\boldsymbol{\alpha}_j\mathbf{w}_j\trans=\sum_{j=1}^k\sigma_j\boldsymbol{\gamma}_j\mathbf{u}_j\trans$ is the best rank $k$ approximation to $\mathbf{X}\mathcal{B}$ and $\boldsymbol{\alpha}_k\trans \mathbf{S}\boldsymbol{\alpha}_k=1$, for any $1\le k\le K$. 

\begin{figure}[h]
\begin{center}
\includegraphics[height=3in,width=7.5in, bb=-0.5in 0in 15in 7.5in]{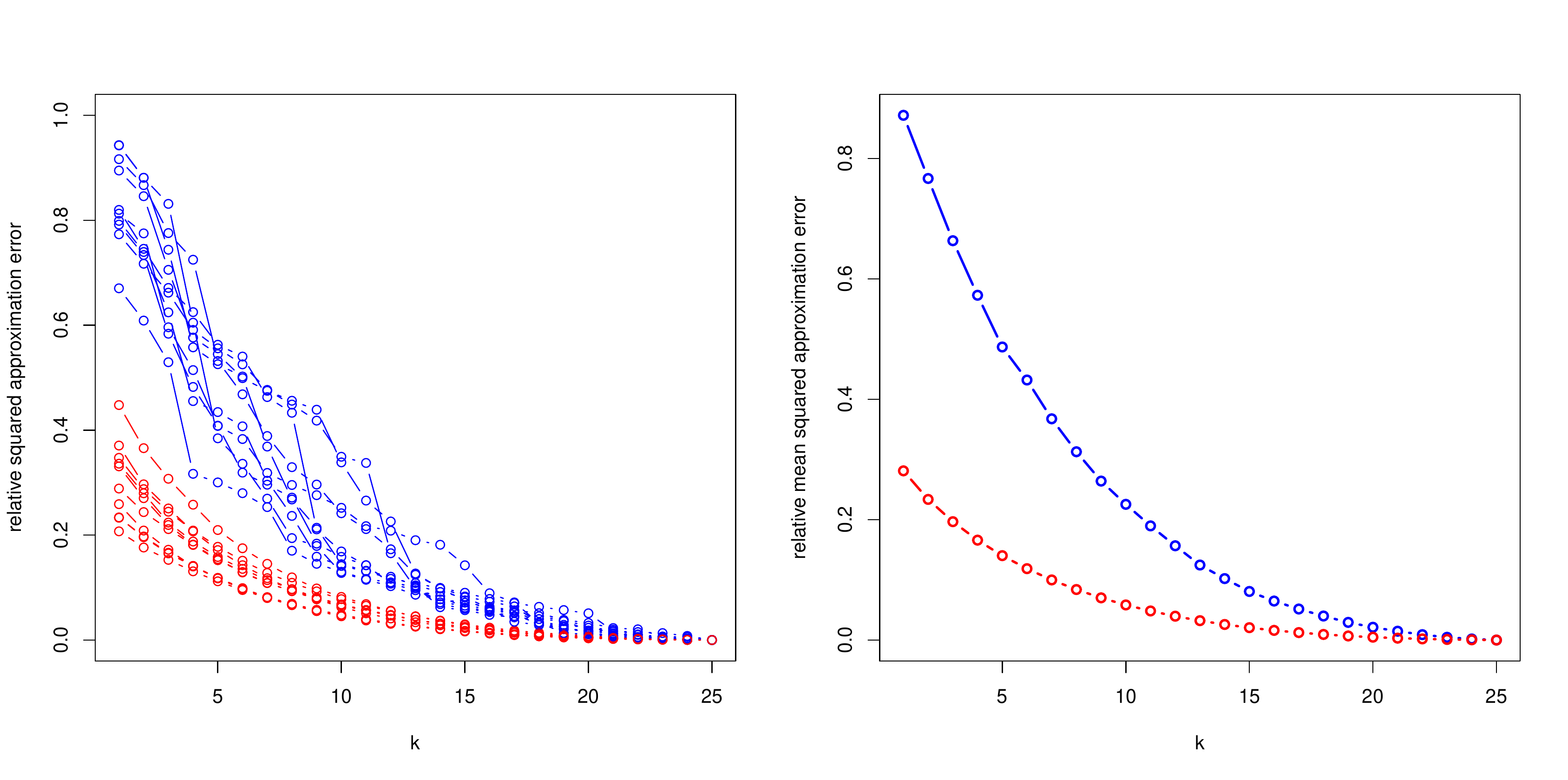}
\end{center}
    \caption{ \baselineskip=10pt  Relative squared low rank approximation error to $\mathbf{X}\mathcal{B}$ based on our decomposition (red) and the SVD (blue) of $\mathcal{B}$. The x-axis is the rank of the approximation matrix. }
\label{approx.fig}
\end{figure}

 We visually illustrate the difference in approximating $\mathbf{X}\mathcal{B}$ using the two decompositions of the coefficient matrix: SVD of $\mathcal{B}$  given by $\eqref{svd.eq}$ and our decomposition given by $\eqref{70002}$ based on the SVD of  $\mathbf{X}\mathcal{B}$, in Figure \ref{approx.fig} through an example. We take $n=100$, $p=1000$ and $q=100$. Each row of $\mathbf{X}$ is generated from a $p$-dimensional multivariate normal distribution with mean zero. Its covariance matrix has all the diagonal elements equal to 1 and all the off-diagonal elements equal to   0.7.  We generate $\mathcal{B}$ using $\mathcal{B}=\mathbf{CD}$, where $\mathbf{C}$ is $p \times 25$ and $\mathbf{D}$ is $25 \times q$. Elements in the first 40 rows of $\mathbf{C}$ are independently generated from $N(0,1)$, and the other $p-40$ rows of $\mathbf{C}$ are zeros.  Elements in $\mathbf{D}$ are independently generated from the uniform distribution between $-1$ and $1$. Therefore, in this example, the rank of $\mathcal{B}$ is $K=25$. We perform the simulation 100 times. In each repeat, for each $1\le k\le K$,  we calculate the relative squared  approximation error to  $\mathbf{X}\mathcal{B}$ defined by $\|\mathbf{X}\mathcal{B}-\mathbf{X}\mathcal{B}_k\|_F^2/\|\mathbf{X}\mathcal{B}\|_F^2$, where $\mathcal{B}_k$ is the rank $k$ approximation matrix to $\mathcal{B}$ using our decomposition (red in Figure \ref{approx.fig}) or the SVD (blue). 
In   Figure \ref{approx.fig}, we plot the relative error for 10 repeats in the left panel, and plot the mean relative error over 100 repeats in the right panel, when $k$ changes from 1 to $K=25$. 

In the following, for any $1\le k\le K$, let $\mathcal{B}_k=\boldsymbol{\alpha}_1\mathbf{w}_1\trans+\cdots+\boldsymbol{\alpha}_k\mathbf{w}_k\trans$, the sum of the first $k$ terms of our decomposition. We will show in Section \ref{sect.4} that the property that $\mathbf{X}\mathcal{B}_k$ is the best rank $k$ approximation to $\mathbf{X}\mathcal{B}$  leads to the property that $\mathcal{B}_k$ has  nearly the  smallest expected prediction errors among all possible rank $k$ approximation to $\mathcal{B}$ under the row-wise sparsity assumption on $\mathcal{B}$.  Our decomposition of $\mathcal{B}$ leads to the following model transformation,  
\begin{align}
\mathbf{Y} = \mathbf{X}\mathcal{B}+\boldsymbol{\varepsilon}= \mathbf{T}\mathbf{W}\trans+\boldsymbol{\varepsilon}=\mathbf{t}_1\mathbf{w}_1\trans+\cdots+\mathbf{t}_K\mathbf{w}_K\trans+\boldsymbol{\varepsilon},
\label{modeltw.eq}
\end{align}
where 
\begin{align}
\mathbf{T}=[\mathbf{t}_1, \cdots,\mathbf{t}_K]\;,\text{ and } \mathbf{t}_j=\mathbf{X}\boldsymbol{\alpha}_j=\sqrt{n}\boldsymbol{\gamma}_j, \quad 1\le j\le K, \label{70019}
\end{align}
are new orthogonal predictors.  

To estimate the decomposition, we first  estimate $\boldsymbol{\alpha}_1,\cdots,\boldsymbol{\alpha}_K$ based on the following theorem, then estimate $\mathbf{t}_1, \cdots,\mathbf{t}_K$ by $\eqref{70019}$. Finally, based on model $\eqref{modeltw.eq}$ and the least squares method, we obtain the  estimates of $\mathbf{w}_1, \cdots,\mathbf{w}_K$.   Define
\begin{align}
&\mathbf{Z}=\frac{1}{\sqrt{n}}\mathbf{X},\quad    \boldsymbol{\Xi}=\frac{1}{n}(\mathbf{X}\mathcal{B}) (\mathbf{X}\mathcal{B})\trans = \mathbf{Z}\mathcal{B}\mathcal{B}\trans\mathbf{Z}\trans,\quad \mathbf{S}=\frac{1}{n}\mathbf{X}\trans\mathbf{X}= \mathbf{Z}\trans\mathbf{Z},\notag\\
 &  \mathbf{B}=\frac{1}{n^2}\mathbf{X}\trans\left(\mathbf{X}\mathcal{B}\right) \left(\mathbf{X}\mathcal{B}\right)\trans\mathbf{X}=\mathbf{S}\mathcal{B}\mathcal{B}\trans \mathbf{S} =\mathbf{Z}\trans\boldsymbol{\Xi}\mathbf{Z}.\label{10008}
\end{align}

\begin{Theorem}\label{theorem_1}
Suppose that  the $n\times n$ nonnegative definite matrix $\boldsymbol{\Xi}$ has $K$   positive eigenvalues $\mu_1(\boldsymbol{\Xi})\ge \mu_2(\boldsymbol{\Xi})\ge \cdots\ge \mu_K(\boldsymbol{\Xi})>0$. Then 
\begin{itemize}
\item[(a).] $K\le \min\{n, p,q\}$. For any $1\le k\le K$, $\boldsymbol{\alpha}_k$ in $\eqref{70002}$ is the solution to  
\begin{align}
&\max_{\boldsymbol{\alpha}\in \mathbb{R}^p}\quad \boldsymbol{\alpha}\trans \mathbf{B}\boldsymbol{\alpha},\quad \text{\rm subject to}\quad \boldsymbol{\alpha}\trans \mathbf{S}\boldsymbol{\alpha}=1,\quad  \boldsymbol{\alpha}_l\trans \mathbf{S}\boldsymbol{\alpha}=0, \quad 1\le l\le k-1.\label{2220}
\end{align}
Moreover, the maximum value of $\eqref{2220}$ is equal to $\boldsymbol{\alpha}_k\trans \mathbf{B}\boldsymbol{\alpha}_k=\mu_k(\boldsymbol{\Xi})$. 
\item[(b).]In the SVD $\eqref{70006}$ of $\mathbf{X}\mathcal{B}$, the singular values are $\sigma_k=\sqrt{n\mu_k(\boldsymbol{\Xi})}$, $1\le k\le K$.  The left-singular vectors  satisfy
 \begin{align}
\boldsymbol{\gamma}_1=\mathbf{Z}\boldsymbol{\alpha}_1 , \quad\boldsymbol{\gamma}_2=\mathbf{Z}\boldsymbol{\alpha}_2 , \quad\cdots, \quad\boldsymbol{\gamma}_K=\mathbf{Z}\boldsymbol{\alpha}_K ,\label{10074}
\end{align} 
 and they are the  $K$ eigenvectors of $\boldsymbol{\Xi}$ corresponding to the $K$ positive eigenvalues  with $\|\boldsymbol{\gamma}_k\|_2=1$.
\item[(c).] The approximation error of the best rank $k$ approximation to $\mathbf{X}\mathcal{B}$ is 
\begin{align*}
 \|\mathbf{X}\mathcal{B}-\sum_{i=1}^k\mathbf{t}_i\mathbf{w}_i\trans\|_F^2=\|\mathbf{X}\mathcal{B}-\mathbf{X}(\sum_{i=1}^k\boldsymbol{\alpha}_i\mathbf{w}_i\trans)\|_F^2=\|\mathbf{X}\mathcal{B}-\mathbf{X}\mathcal{B}_k\|_F^2=n\sum_{i=k+1}^K\mu_i(\boldsymbol{\Xi}) 
\end{align*}
for any $1\le k\le K$.
\end{itemize}
\end{Theorem}
    By Theorem \ref{theorem_1}(b),  $\mu_k(\boldsymbol{\Xi})=\sigma_k^2/n=\|\sigma_k\boldsymbol{\gamma}_k\mathbf{u}_k\trans\|_F^2/n $. So $\mu_k(\boldsymbol{\Xi})$ can be viewed as a measure of the magnitude of the signal in the $k$-th component of the SVD of $\mathbf{X}\mathcal{B}$. Therefore, our choice of $\mathbf{W}$ and $\mathbf{A}$ makes the signal concentrated in  the first few components as much as possible. Theorem \ref{theorem_1}(c) implies that, even if $K$ is not small, as long as $\mu_k(\boldsymbol{\Xi})$ decreases fast enough, $\mathbf{X}\mathcal{B}$ can be well approximated by the first few components. We call this the {\it{Si}}gnal {\it{E}}xtraction multivariate {\it{R}}egression (SiER) method.

 To estimate $\boldsymbol{\alpha}_k$,  we  estimate  $\mathbf{B}$ by
 \begin{align}
&  \widehat{\mathbf{B}}= \frac{1}{n^2}\mathbf{X}\trans(\mathbf{Y}-\mathbf{1}_n\bar{\mathbf{y}}\trans)(\mathbf{Y}-\mathbf{1}_n\bar{\mathbf{y}}\trans)\trans\mathbf{X}\;, \label{10009}
\end{align} 
 where $\bar{\mathbf{y}}$ is the sample mean of $\mathbf{y}_1,\cdots,\mathbf{y}_n$, and $\mathbf{1}_n$ is an $n$-dimensional vector with all elements equal to one. In  the classic setting of small $p$ and large $n$, the estimates $\widehat{\boldsymbol{\alpha}}_1$, $\cdots$, $\widehat{\boldsymbol{\alpha}}_{K}$ can be sequentially obtained by solving 
  \begin{align}
 \max_{\boldsymbol{\alpha}\in \mathbb{R}^p}\quad \boldsymbol{\alpha}\trans \widehat{\mathbf{B}}\boldsymbol{\alpha},\quad \text{\rm subject to}\quad \boldsymbol{\alpha}\trans \mathbf{S}\boldsymbol{\alpha}=1,\quad  \widehat{\boldsymbol{\alpha}}_l\trans \mathbf{S}\boldsymbol{\alpha}=0, \quad 1\le l\le k-1,\label{10096}
\end{align} 
 where $1\le k\le K$.  Once we obtain $\widehat{\boldsymbol{\alpha}}_k$,  we have $\widehat{\mathbf{t}}_k=\mathbf{X}\widehat{\boldsymbol{\alpha}}_k$, which has zero sample mean. With the constraints in $\eqref{10096}$, $\widehat{\mathbf{t}}_1, \widehat{\mathbf{t}}_2, \ldots, \widehat{\mathbf{t}}_K$ are orthogonal and satisfy $\|\widehat{\mathbf{t}}_k\|_2^2/n=1$. Let
 $ \widehat{\mathbf{T}}=[\widehat{\mathbf{t}}_1, \cdots, \widehat{\mathbf{t}}_K]$. 
Then $\widehat{\mathbf{T}}\trans \widehat{\mathbf{T}}=n\mathbf{I}_K$, where  $\mathbf{I}_K$ is the $K$-dimensional identity matrix. 
The matrix  
 $\mathbf{W}$ is estimated by regressing $\mathbf{Y}$ on $\widehat{\mathbf{t}}_1, \widehat{\mathbf{t}}_2, \ldots, \widehat{\mathbf{t}}_K$ with the usual least squares method. That is, 
 \begin{align}
&\widehat{\mathbf{W}}\trans=(\widehat{\mathbf{T}}\trans \widehat{\mathbf{T}})^{-1}\widehat{\mathbf{T}}\trans (\mathbf{Y}-\mathbf{1}_n\bar{\mathbf{y}}\trans)=\frac{1}{n}\widehat{\mathbf{T}}\trans (\mathbf{Y}-\mathbf{1}_n\bar{\mathbf{y}}\trans), \notag\\
\text{ and }\quad &\widehat{\mathcal{B}}=\widehat{\boldsymbol{\alpha}}_1\widehat{\mathbf{w}}_1\trans+\widehat{\boldsymbol{\alpha}}_2\widehat{\mathbf{w}}_2\trans+\cdots+\widehat{\boldsymbol{\alpha}}_k\widehat{\mathbf{w}}_K\trans, \label{10116}
\end{align} 
is the estimate of $\mathcal{B}$, and  $\widehat{\mathcal{B}}_k=\widehat{\boldsymbol{\alpha}}_1\widehat{\mathbf{w}}_1\trans+\widehat{\boldsymbol{\alpha}}_2\widehat{\mathbf{w}}_2\trans+\cdots+\widehat{\boldsymbol{\alpha}}_k\widehat{\mathbf{w}}_k\trans$ is  the estimate of $\mathcal{B}_k$.  Due to the orthogonality of $\widehat{\mathbf{t}}_1, \widehat{\mathbf{t}}_2, \ldots, \widehat{\mathbf{t}}_K$,   $\widehat{\mathcal{B}}_k$ does not depend on the number of selected components in practice. 
The following lemma shows that in the special case of scalar response ($q=1$), our estimate $\widehat{\mathcal{B}}$ in $\eqref{10116}$ is equivalent to the least squares method.  But if $q>1$, this method may not be the same as the least squares method.
\begin{Lemma}\label{lemma_15}
 Suppose that $\mathbf{S}$ is full rank. If $q=1$, the estimate in $\eqref{10116}$ is exactly the same as the least squares estimate.
\end{Lemma}

We will introduce the sparse method for the high-dimensional setting in the next section. 

\section{Sparse estimates and oracle inequalities  in high-dimensional settings}

\subsection{Sparse estimates}
We  make the following sparsity assumption: only a small number of the coefficient vectors, $\boldsymbol{\beta}_1,\cdots,\boldsymbol{\beta}_p$, are nonzero vectors. Since these vectors are the row vectors of $\mathcal{B}$, this assumption is just the row-wise sparsity of $\mathcal{B}$. The definition $\eqref{70002}$ implies that $\boldsymbol{\alpha}_k$ is a sparse vector and the number of its nonzero coordinates is less than or equal to the number of nonzero vectors among $\boldsymbol{\beta}_1,\cdots,\boldsymbol{\beta}_p$. Motivated by the sparsity of $\boldsymbol{\alpha}_k$, we  propose the following penalized optimization problem whose solution is the sparse estimate $\widehat{\boldsymbol{\alpha}}_k$ of  ${\boldsymbol{\alpha}}_{k}$:
 \begin{align}
 \max_{\boldsymbol{\alpha}\in \mathbb{R}^p}\quad & \frac{\boldsymbol{\alpha}\trans \widehat{\mathbf{B}}\boldsymbol{\alpha}}{\boldsymbol{\alpha}\trans \mathbf{S}\boldsymbol{\alpha}+\tau\|\boldsymbol{\alpha}\|^2_{\lambda}}, \label{1240}\\ 
 \text{\rm subject  to}\quad & \boldsymbol{\alpha}\trans \mathbf{S}\boldsymbol{\alpha} =1,\quad  \widehat{\boldsymbol{\alpha}}_l\trans \mathbf{S}\boldsymbol{\alpha}=0, \quad 1\le l\le k-1,\notag
\end{align} 
where $\|\boldsymbol{\alpha}\|^2_{\lambda}=(1-\lambda)\|\boldsymbol{\alpha}\|_2^2+\lambda\|\boldsymbol{\alpha}\|_1^2$, and both $\tau\ge 0$ and $0\le \lambda < 1$ are tuning parameters. In the penalty $\tau\|\boldsymbol{\alpha}\|^2_{\lambda}$,  the $l_2$ term is used to overcome the singularity problem of $\mathbf{S}$ and the $l_1$ term encourages the sparsity of $\widehat{\boldsymbol{\alpha}}_k$. The penalty $\tau\|\boldsymbol{\alpha}\|^2_{\lambda}$ was  introduced in \citet*{spca-qi} for sparse principal component analysis and utilized in \citet*{Qi-2013} for sparse regression and sparse discriminant analysis. The  main reason that we use the squared $l_1$ norm instead of the $l_1$ norm itself as in the elastic-net is to make the objective function in $\eqref{1240}$  scale-invariant, that is, if we  replace $\boldsymbol{\alpha}$ by $t\boldsymbol{\alpha}$, where $t$ is any nonzero number, the value of the objective function is unchanged. This property plays an important role in our theoretical development and algorithms. Due to the scale-invariant property, $\eqref{1240}$ is equivalent to
\begin{align}
& \max {\boldsymbol{\alpha}\trans \widehat{\mathbf{B}}\boldsymbol{\alpha}} 
\quad\text{subject to}\quad \boldsymbol{\alpha}\trans \mathbf{S}\boldsymbol{\alpha}+\tau \|\boldsymbol{\alpha}\|^2_{\lambda }\le 1  \text{ and } \boldsymbol{\alpha}\trans \mathbf{S}\widehat{\boldsymbol{\alpha}}_l=0,\quad 1\le l\le k-1.
\label{alg.eq}
\end{align}
In fact, the solutions of $\eqref{1240}$  and $\eqref{alg.eq}$ differ only by a scale factor. We have proposed   algorithms to solve a  more general optimization problem (see the problem (4.3) in \citet{Qi-2013}) than $\eqref{alg.eq}$.   By a proper rescaling of the solution to $\eqref{alg.eq}$, we obtain  $\widehat{\boldsymbol{\alpha}}_{k}$. With the constraints in $\eqref{1240}$, the estimates $\widehat{\mathbf{t}}_k=\mathbf{X}\widehat{\boldsymbol{\alpha}}_k$, $1\le k \le K$,  are still orthogonal to each other and satisfy $\|\widehat{\mathbf{t}}_k\|_2^2/n=1$. Therefore, we can use $\eqref{10116}$ to get the estimates  $\widehat{\mathbf{W}}$ and $\widehat{\mathcal{B}}$. In the special case of scalar response, the proposed method is just the sparse regression by projection method proposed in \citet{Qi-2013}.

\subsection{Oracle inequalities }\label{sect.4}
 In this section,  we provide  oracle inequalities for the estimates of $\boldsymbol{\alpha}_{k}$, $\mathbf{w}_{k}$ and $\mathcal{B}$  in high-dimensional settings. These oracle inequalities hold for any $n$ and $p$.  

We follow the notations in \citet{bickel2009simultaneous}. For any $p$-dimensional vector $\mathbf{a}=(a_1,\cdots,a_p)\trans$, let $J(\mathbf{a})=\{j\in\{1,\cdots,p\}: a_j\neq 0\}$ denote the collection of indices of nonzero coordinates of $\mathbf{a}$ and $\mathcal{M}(\mathbf{a})=|J(\mathbf{a})|$ denote the number of nonzero coordinates of $\mathbf{a}$, where $|J(\mathbf{a})|$ is the cardinality of $J(\mathbf{a})$. $\mathcal{M}(\mathbf{a})$ is a measure of the sparsity of $\mathbf{a}$. Similarly, for any $p\times q$ matrix $\widetilde{\mathcal{B}}$ with row vectors, $\widetilde{\boldsymbol{\beta}}_1$, $\cdots$, $\widetilde{\boldsymbol{\beta}}_p$,  we define $J(\widetilde{\mathcal{B}})=\{j\in\{1,\cdots,p\}: \widetilde{\boldsymbol{\beta}}_j\neq \mathbf{0}\}$ and $\mathcal{M}(\widetilde{\mathcal{B}})=|J(\widetilde{\mathcal{B}})|$ which is a measure of the row-wise sparsity of $\widetilde{\mathcal{B}}$.   It follows from  $\eqref{70002}$ that 
\begin{align}
J(\boldsymbol{\alpha}_k)\subset J(\mathcal{B}),\quad\text{ and hence}\quad  \mathcal{M}(\boldsymbol{\alpha}_k)\le \mathcal{M}(\mathcal{B}), \quad \forall 1\le k\le K.\label{10012}
\end{align}   

 Before we provide the main results, we first show that $\mathcal{B}_k$  has nearly the smallest expected prediction error among all rank $k$ coefficient matrix estimations when $n$ is large. Let $\mathbf{x}^{\rm new}$ be a new observation of the predictor vector and $\boldsymbol{\Sigma}$ the covariance  matrix of $\mathbf{x}^{\rm new}$. The corresponding new response is $(\mathbf{y}^{\rm new})\trans=(\mathbf{x}^{\rm new})\trans\mathcal{B}+(\boldsymbol{\varepsilon}^{\rm new})\trans$, where $\boldsymbol{\varepsilon}^{\rm new}$ is independent of $\mathbf{x}^{\rm new}$.
\begin{Theorem}\label{theorem_8}
  Suppose that $\|\mathbf{S}-\boldsymbol{\Sigma}\|_\infty\le C\sqrt{\frac{\log p}{n}}$, where $C$ is a constant which does not depend on $n$ and $p$. Then we have
 \begin{align*}
 &E\left[\|(\mathbf{x}^{\rm new})\trans\mathcal{B}_k-(\mathbf{y}^{\rm new})\trans\|_2^2\right]\le \min_{\widetilde{\mathcal{B}_k}}E\left[\|(\mathbf{x}^{\rm new})\trans\widetilde{\mathcal{B}}_k-(\mathbf{y}^{\rm new})\trans\|_2^2\right]+2 C\|\mathcal{B}\|_F^2\mathcal{M}(\mathcal{B})\sqrt{\frac{\log p}{n}},
\end{align*} 
where the minimum is taken over all possible $\widetilde{\mathcal{B}}_k$ of the forms $ \sum_{j=1}^k\mathbf{b}_j\mathbf{v}_j\trans$ with arbitrary $\mathbf{b}_j\in\mathbb{R}^p$ and $\mathbf{v}_j\in\mathbb{R}^q$.
\end{Theorem} 
Note that when $q=1$, $\mathcal{M}(\mathcal{B})\sqrt{\log p/n}$ is the convergence rate of the LASSO and the Dantzig selector \citep{bickel2009simultaneous}. Under the  sparsity assumption  that when $n$, $p\to \infty$,  $\|\mathcal{B}\|_F^2\mathcal{M}(\mathcal{B})\sqrt{\log{p}/n}\to 0$,  the expected prediction error of $\mathcal{B}_k$ is close to the smallest one among all possible rank $k$ approximation to $\mathcal{B}$ when $n$ and $p$ are large. We assume that $\|\mathbf{S}-\boldsymbol{\Sigma}\|_\infty\le C\sqrt{\log{p}/n}$, because it has been shown (Equation (A14) in \citet{bickel2008regularized}) that $\sqrt{\log{p}/n}$ is the order of the  max norm of the difference between the sample covariance matrix and the population covariance matrix of $p$-dimensional multivariate normal distribution.

Now we  state three regularity conditions for the main theorems.  In the setting of large $p$ and small $n$,  the identification problem exists for the model $\eqref{100003}$. That is, there exists $\widetilde{\mathcal{B}} \neq \mathcal{B}$ such that $\mathbf{X}\widetilde{\mathcal{B}} =\mathbf{X}\mathcal{B}$. \citet{bickel2009simultaneous} imposed the following restricted eigenvalue assumptions  on $\mathbf{X}$, which we will also adopt here. 
\begin{Condition}\label{condition_3}
Let $s=\mathcal{M}(\mathcal{B})$ and 
\begin{align*}
 \kappa=\min_{\substack{J_0\subset\{1,\cdots,p\},\\ |J_0|\le  s}}\min_{\substack{\mathbf{0}\neq \boldsymbol{\delta}\in\mathbf{R}^p,\\ \left\|\boldsymbol{\delta}_{J_0^c}\right\|_1\le c\left\|\boldsymbol{\delta}_{J_0}\right\|_1}}\frac{\|\mathbf{X}\boldsymbol{\delta}\|_2}{\sqrt{n}\|\boldsymbol{\delta}_{J_0}\|_2},
\end{align*} 
where $c>1$ and $\kappa>0$ are two constants, $\boldsymbol{\delta}_{J_0}$ and $\boldsymbol{\delta}_{J_0^c}$ are the subvectors of $\boldsymbol{\delta}$ consisted of the coordinates with indices belonging to $J_0$ and $J_0^c$, respectively. 
 \end{Condition}
A consequence of this restricted eigenvalue assumption is that  for any two $p$-dimensional vectors, $\boldsymbol{\alpha}$ and $\boldsymbol{\alpha}^\prime$ with sparsity $\mathcal{M}(\boldsymbol{\alpha})\le s$ and $\mathcal{M}(\boldsymbol{\alpha}^\prime)\le s$, if $\mathbf{X}\boldsymbol{\alpha}=\mathbf{X}\boldsymbol{\alpha}^\prime$, then we have $\boldsymbol{\alpha}= \boldsymbol{\alpha}^\prime$ (see the second remark after Theorem 7.3 in \citet{bickel2009simultaneous}). Therefore, the model $\eqref{100003}$ is identifiable among all coefficient matrices with row-wise sparsity less than or equal to  $\mathcal{M}(\mathcal{B})$. Therefore, if  $\mathbf{X}\mathcal{B}^\prime=\mathbf{X}\mathcal{B}$ and $\mathcal{M}(\mathcal{B}^\prime)\le s=\mathcal{M}(\mathcal{B})$, then $\mathcal{B}^\prime=\mathcal{B}$.
 
 The next regularity condition is on the distribution of the $q$-dimensional noise vector $\boldsymbol{\varepsilon}_i$.  
\begin{Condition}\label{condition_1}
 The random error vectors $\boldsymbol{\varepsilon}_i$, $1\le i\le n$, are i.i.d. mean zero $\mathbb{R}^q$-valued Gaussian variables with  median $M_{\varepsilon}$ and  variance $\sigma^2$, where $M_{\varepsilon}$ is defined as the median of the real-valued random variable $\|\boldsymbol{\varepsilon}_i\|_2$ and the variance is defined as $\sigma^2=\sup_{\mathbf{u}\in\mathbb{R}^q, \|\mathbf{u}\|_2= 1}E[(\mathbf{u}\trans \boldsymbol{\varepsilon}_i)^2]$ (Section 3.1 in \citet{ledoux2011probability}).
\end{Condition}
Note that $(\mathbf{u}\trans \boldsymbol{\varepsilon}_i)$ is the projection of $\boldsymbol{\varepsilon}_i$ onto the direction of $\mathbf{u}$, and has a normal distribution. Hence, $\sigma^2$ is the maximum of the variances of the projections of $\boldsymbol{\varepsilon}_i$ along all possible directions in $\mathbb{R}^q$. In the special case $q=1$, $\sigma^2$ is the just usual variance. In our theoretical development, we need to estimate the tail probabilities of $\|\boldsymbol{\varepsilon}_i\|_2$ which can be controlled by  $M_{\varepsilon}$ and $\sigma^2$ (Section 3.1 in \citet{ledoux2011probability}).   \cite{bunea2012joint} assumes that the coordinates $\{\varepsilon_{ij}: 1 \le j\le q\}$   of $\boldsymbol{\varepsilon}_i$ have independent and identical normal distributions. Condition \ref{condition_1} is weaker and allows correlations among ${\varepsilon}_{i1}, \ldots, {\varepsilon}_{iq}$.
 
\begin{Condition}\label{condition_2}
All the diagonal elements of $\mathbf{S}=\mathbf{X}\trans\mathbf{X}/n$ are equal to 1 and there exist positive constants, $c_2$ and $c_3$, such that 
\begin{itemize}
\item[(a).]  
$ \min\left\{\frac{\mu_1 (\boldsymbol{\Xi})-\mu_2 (\boldsymbol{\Xi})}{\mu_1 (\boldsymbol{\Xi})}, \quad\frac{\mu_2 (\boldsymbol{\Xi})-\lambda_{3}(\boldsymbol{\Xi})}{\mu_2 (\boldsymbol{\Xi})},\cdots, \frac{\mu_{K-1}(\boldsymbol{\Xi})-\mu_K(\boldsymbol{\Xi})}{\mu_{K-1}(\boldsymbol{\Xi})} \right\}\ge c_2$,
\item[(b).]   $\mu_1 (\boldsymbol{\Xi})/\mu_K(\boldsymbol{\Xi})\le c_3$.
\end{itemize}
\end{Condition}
\citet{bickel2009simultaneous} assumed that the diagonal elements of $\mathbf{S}=\mathbf{X}\trans\mathbf{X}/n$ are equal to 1, which can be achieved by scaling $\mathbf{X}$. Condition \ref{condition_2} (a) prevents the cases where the spacing between adjacent eigenvalues is so small that the eigenvalues cannot be well separated based on noisy observations. Condition \ref{condition_2} (b) excludes the situations where the  magnitudes of the higher order components are too small compared to those of lower order components.  

As mentioned in Section \ref{section.2}, in the classic setting, if  $q>1$, our method $\eqref{10009}$ may not be the same as the least squares method. For the integrity of this paper, below we first provide the asymptotic result for our method in Theorem \ref{theorem_2} under the setting when $p$ is fixed and $n$ goes to infinity, and then provide the theoretical property of our sparse estimates for high dimensional setting afterward.
\begin{Theorem}\label{theorem_2}
 Suppose that  Conditions \ref{condition_1}-\ref{condition_2} hold with $M_{\varepsilon}$, $\sigma^2$ and $c_3$ bounded above and $c_2$ bounded below as $n\to\infty$. Moreover, we assume that there exists $c_0>0$ independent of $n$ such that $\mathbf{S}$ is positive definite and its smallest eigenvalue, $\lambda_{min}(\mathbf{S})\ge c_0$, for all $n$ large enough. Then we have
 \begin{align}
&  \|\widehat{\mathcal{B}}-\mathcal{B}\|_F =O_p(1/\sqrt{n}), \quad \|\widehat{\boldsymbol{\alpha}}_k-\boldsymbol{\alpha}_k\|_2=    \mu_1(\boldsymbol{\Xi})^{-1/2}O_p(1/\sqrt{n}),\label{10121}
\end{align} 
for any $1\le k\le K$.
\end{Theorem}
   
In the rest of this  section, we  study the theoretical property of our sparse estimates $\widehat{\boldsymbol{\alpha}}_k$ and $\widehat{\mathcal{B}}$ when both $n, p\to \infty$. We first provide upper bounds on the $l_1$ sparsity of $\widehat{\boldsymbol{\alpha}}_k$, $1\le k\le K$, and the oracle inequalities for them in the following theorem. Although we use the same tuning parameters $(\tau, \lambda)$ in $\eqref{1240}$ for all components in practice for the computational efficiency, in our theoretical results, we allow different tuning parameters for different components. We use $(\tau^{(k)}, \lambda^{(k)})$ to denote the tuning parameters for the $k$-th component, $1\le k\le K$. Let $\widehat{\boldsymbol{\gamma}}_k=\mathbf{Z}\widehat{\boldsymbol{\alpha}}_k=\widehat{\mathbf{t}}_k/\sqrt{n}$, $1\le k\le K$, which are estimates of $\boldsymbol{\gamma}_k=\mathbf{Z}\boldsymbol{\alpha}_k=\mathbf{t}_k/\sqrt{n}$. Define
 \begin{align*}
 \varpi=\sqrt{\frac{\log{p}}{n}}, \quad \text{\rm and recall that }\quad  s=\mathcal{M}(\mathcal{B}).
\end{align*} 
\begin{Theorem}\label{theorem_3}
Assume that Conditions \ref{condition_3}-\ref{condition_2} hold.  Suppose that  
\begin{align}
 \mu_1(\boldsymbol{\Xi})\ge  \hbar^2C_0^2\varpi^2s/\kappa^2, \quad \text{where }C_0=2\max\{M_{\varepsilon}/\sqrt{\log{2}}, 2\sigma\},\label{50020}
\end{align} 
 $\kappa$ is the constant in Condition \ref{condition_3} and $\hbar$ is a constant. Let the tuning parameters $(\tau^{(k)}, \lambda^{(k)})$, $1\le k\le K$, satisfy conditions
 \begin{align}
\tau^{(k)}=\frac{A^{(k)}C_0\varpi}{\|\boldsymbol{\alpha}_1\|_1\sqrt{\mu_1(\boldsymbol{\Xi})}},\qquad c^{-1}+\delta_0<\lambda^{(k)} \le 1,\label{10216}
\end{align} 
where $A^{(k)}$ and $\delta_0$ are positive constants such that $c^{-1}+\delta_0< 1$, and $c$ is the constant in Condition \ref{condition_3}.
\begin{itemize}
\item[(a).] For the first component ($k=1$), there exist constants $A_1^{L}$ and $\hbar_0$ which only depend on $c$, $c_2$ and $\delta_0$, where $c_2$ is the constant in Condition \ref{condition_2} (a), such that with probability at least $ 1-2e^{ M_{\varepsilon}^2/2\sigma^2}p^{1-C_0^2/4\sigma^2}$, if $A^{(1)}\ge A_1^{L} $ and $\hbar\ge \hbar_0$,  we have
 \begin{align}
&\|\widehat{\boldsymbol{\alpha}}_1\|_1\le \sqrt{6c}\|\boldsymbol{\alpha}_1\|_1\le \sqrt{6cs}/\kappa,\label{10205}\\
&\|\widehat{\boldsymbol{\alpha}}_1-\boldsymbol{\alpha}_1\|_1 \le 4(1+c)(1+\sqrt{6c})c_2^{-1}A^{(1)}C_0\kappa^{-2} \mu_1(\boldsymbol{\Xi})^{-1/2}\varpi s,\notag\\
  &\frac{1}{n}\|\mathbf{X}(\widehat{\boldsymbol{\alpha}}_1-\boldsymbol{\alpha}_1)\|_2^2 =\|\widehat{\boldsymbol{\gamma}}_1-\boldsymbol{\gamma}_1\|_2^2 \le 16c_2^{-2}(1+\sqrt{6c})^2(A^{(1)}C_0/\kappa)^2\mu_1(\boldsymbol{\Xi})^{-1}\varpi^2s.  \notag
\end{align} 
\item[(b).] For the higher order components ($1<k\le K$), we further assume that 
 \begin{align}
\max_{1\le k\le K}\|\boldsymbol{\alpha}_k\|_1\le c_4 \min_{1\le k\le K}\|\boldsymbol{\alpha}_k\|_1, \quad \kappa^{-2}\mu_1(\boldsymbol{\Xi})^{-1/2}C_0\varpi s\le c_5\|\boldsymbol{\alpha}_1\|_1, \label{359}
\end{align} 
where $c_4$ and $c_5$ are two constants. Then there exist constants $\hbar_0$, $A^{L}_j < A^{U}_j$, $1\le j\le K$, which only depend on $\delta_0$, $c$, $c_2\sim c_5$, such that with probability at least $ 1-2e^{ M_{\varepsilon}^2/2\sigma^2}p^{1-C_0^2/4\sigma^2}$, for any $1\le k\le K$, if $A^{L}_j\le A^{(j)}\le A^{U}_j$, $1\le j<k$, $A^{(k)}\ge A^{L}_k$ and $\hbar\ge \hbar_0$,  we have
 \begin{align}
   &\|\widehat{\boldsymbol{\alpha}}_k\|_1\le   D_{k,1}\sqrt{s}/\kappa,\quad \|\widehat{\boldsymbol{\alpha}}_k-\boldsymbol{\alpha}_k\|_1\le  D_{k,4}A^{(k)}C_0 \kappa^{-2}\mu_1(\boldsymbol{\Xi})^{-1/2}\varpi s, ,\label{2053}\\
	&  \frac{1}{n}\|\mathbf{X}(\widehat{\boldsymbol{\alpha}}_k-\boldsymbol{\alpha}_k)\|_2^2=\|\widehat{\boldsymbol{\gamma}}_k-\boldsymbol{\gamma}_k\|^2_2 \le D_{k,2} (A^{(k)}C_0/\kappa)^2\mu_1(\boldsymbol{\Xi})^{-1}\varpi^2 s\;,\notag
\end{align} 
 where $D_{k,1}$, $D_{k,2}$ and $D_{k,4}$ are constants only depending on $\delta_0$, $c$, $c_2\sim c_5$.
\end{itemize}
\end{Theorem}
  In the upper bounds above, only $C_0$  depends on the distribution of the noise vector $\boldsymbol{\varepsilon}_i$ and can be regarded as a measure of the magnitude of noise.
  Although we have two tuning parameters, $\tau^{(k)}$ and $\mu^{(k)}$, for each $k$, by Theorem \ref{theorem_3}, $\lambda^{(k)}$ is not essential for the convergence rates. Actually, it can  be any number in a subinterval of the interval $(c^{-1},1]$ and does not affect the convergence rates.  However, the choice of $\lambda^{(k)}$ does affect the predictive performance in the finite-sampling situations. We will propose methods to choose these two parameters in the following section.  Now we provide the oracle inequalities for $\widehat{\mathbf{W}}$, $\widehat{\mathcal{B}}$ and $\mathbf{X}\widehat{\mathcal{B}}$. We will consider the case $q=1$ first and then $q>1$. When $q=1$,  we use $\boldsymbol{\beta}$ to denote the coefficient vector $\mathcal{B}$ and $\mathbf{W}=\mathbf{w}_1$ is a scalar. 
 
\begin{Theorem}\label{theorem_5}
 Let $q=1$ and $\boldsymbol{\varepsilon}_i\sim N(0,\sigma^2)$. Suppose that Condition \ref{condition_3} holds and $\|\mathbf{X}\boldsymbol{\beta}\|_2^2/n\ge  \hbar^2\sigma^2\varpi^2s/\kappa^2$, where $\kappa$ is the constant in Condition \ref{condition_3}. Let the tuning parameters $(\tau, \lambda)
$ satisfy  
 \begin{align*}
\tau=\frac{A\sigma\varpi}{\|\boldsymbol{\alpha}_1\|_1\|\mathbf{X}\boldsymbol{\beta}\|_2/\sqrt{n}},\qquad c^{-1}+\delta_0<\lambda \le 1, 
\end{align*} 
where $A$ is a constant large enough and $\delta_0$ is a constant satisfying  $c^{-1}+\delta_0< 1$. Then with probability at least $1-2\sqrt{e}p^{-3}$, we have
 \begin{align*} 
&|\widehat{\mathbf{w}}_1-\mathbf{w}_1|\le D_1\sigma\kappa^{-1} \varpi \sqrt{s},\quad \|\widehat{\boldsymbol{\beta}}-\boldsymbol{\beta}\|_1 \le  D_3\sigma\kappa^{-2} \varpi s,\quad \|\mathbf{X}(\widehat{\boldsymbol{\beta}}-\boldsymbol{\beta})\|_2\le  \sqrt{n}D_4\sigma\kappa^{-1} \varpi \sqrt{s},
\end{align*} 
where $D_1$, $D_3$ and $D_4$ are constants only depending on $A$,   $c$, $\hbar$ and $c_2$.
\end{Theorem}
The upper bounds of $\|\widehat{\boldsymbol{\beta}}-\boldsymbol{\beta}\|_1$ and $\|\mathbf{X}(\widehat{\boldsymbol{\beta}}-\boldsymbol{\beta})\|_2$ in Theorem \ref{theorem_5} are the same as those  for the Lasso and the Dantzig selector \citep{bickel2009simultaneous} except the constants.  

 When $q>1$,  to measure the difference between $\widehat{\mathcal{B}}$ and $\mathcal{B}$, we consider the $L_{1,2}$ norm which is defined by
\begin{align*} 
\|\mathbf{M}\|_{1,2}=\left[\sum_{j=1}^m\left(\sum_{i=1}^l |M_{ij}|\right)^2\right]^{1/2},
\end{align*} 
where $\mathbf{M}$ is any $l\times m$ matrix and $M_{ij}$ is the $(i,j)$ element. Then for any matrix $\mathbf{M}$, we have $\|\mathbf{M}\|_{1,2}\ge \|\mathbf{M}\|_F$. Therefore, convergence under the $L_{1,2}$ norm is stronger than that under the Frobenius norm. If $m=1$, $L_{1,2}$ norm is just the $l_1$ norm for a vector. 

\begin{Theorem}\label{theorem_7}
   Suppose that all the conditions in Theorem \ref{theorem_3} hold.  Then with probability at least $ 1-2e^{ M_{\varepsilon}^2/2\sigma^2}p^{1-C_0^2/4\sigma^2}$, for any $1\le K_0\le K$, if $A_k^{L}\le A^{(k)}\le A_k^{U}$, $1\le k< K_0$, and $A^{(K_0)}\ge A_{K_0}^{L}$,  we have
	 \begin{align*} 
&\|\widehat{\mathbf{w}}_k-\mathbf{w}_k\|_2\le L_{k,1}C_0\varpi\sqrt{s}/\kappa,\quad \|\widehat{\mathcal{B}}_k- \mathcal{B}_k\|_{1,2} \le  L_{K_0,3}C_0 \varpi s/\kappa^{2},\\
& \| \mathbf{X}\widehat{\mathcal{B}}_k-\mathbf{X}\mathcal{B}_k\|_F\le  \sqrt{n}L_{K_0,4}C_0\varpi \sqrt{s}/\kappa\;, 
\end{align*} 
for $1\le k\le K_0$, where $L_{k,1}$, $L_{k,3}$ and $L_{k,4}$ are constants only depending on $A^{(j)}$ for $1\le j\le k$,  $c$, $\hbar$ and $c_2\sim c_5$. In particular, when $K_0=K$, we have 
 \begin{align*} 
&\|\widehat{\mathcal{B}}-\mathcal{B}\|_{1,2} \le  L_{K,3}C_0 \varpi s/\kappa^{2},\quad \|\mathbf{X}(\widehat{\mathcal{B}}-\mathcal{B})\|_F\le  \sqrt{n}L_{K,4}C_0\varpi \sqrt{s}/\kappa .
\end{align*} 
\end{Theorem}
 \cite{bunea2012joint} provides an upper bound on $E[\|\mathbf{X}(\widehat{\mathcal{B}}-\mathcal{B})\|^2_F/n]$, which is $\{qK/n+Ks\varpi^2\}$ multiplied by a constant (they use different notations). This upper bound depends on $q$,  the dimension of $\mathbf{Y}$, and only if $q/n\to 0$, the upper bound converges to zero. Our bound holds for arbitrary $q$, even if $q$ goes to infinity faster than $n$ and $p$.

 For any $1\le k\le K$, $\widehat{\mathcal{B}}_k$ and $\mathbf{X}\widehat{\mathcal{B}}_k$ are estimates of $\mathcal{B}_k$ and $\mathbf{X}\mathcal{B}_k$, respectively. Theorem  \ref{theorem_7} and Theorem \ref{theorem_1}(c) imply that, even if $K$ is not small, as long as $\mu_k(\boldsymbol{\Xi})$ decreases fast enough, $\mathbf{X}\mathcal{B}$ can be well approximated $\mathbf{X}\widehat{\mathcal{B}}_k$ with a relatively small $k$.

\section{Choice of the number of components and tuning parameters}\label{section_4} 
In this section,  we propose a method to choose the tuning parameters and decide the number of components. We first provide the rationale behind the selection method and then provide the details of the method in Algorithm \ref{algorithm_1}.  In practice, to improve computational efficiency, we use the same tuning parameters for all components and denote them as $(\tau, \lambda)$. The theoretical results in previous section imply that $\lambda$ is not essential for the convergence rates. However, it has effects on the prediction errors in the finite sample situations. With the penalty $\tau\|\boldsymbol{\alpha}\|^2_{\lambda}=\tau(1-\lambda)\|\boldsymbol{\alpha}\|_2^2+\tau\lambda\|\boldsymbol{\alpha}\|_1^2$, the coefficient of the squared $l_1$ term is $\tau\lambda$. Roughly speaking, the effect of $(\tau, \lambda)$ on the sparsity of solutions mainly depends on $\tau\lambda$ and thus a small $\tau$ with a large $\lambda$ has a similar effect on the sparsity of solutions as that of a large $\tau$ with a small $\lambda$. Hence, to improve the computational efficiency, we do not consider all the pairs of $(\tau, \lambda)$ in a two dimensional grid. Instead, we will select the parameters from a sequence of pairs where with the increase of $\tau$, $\lambda$ also increases. Specifically,  in the following simulation studies and applications, we choose $(\tau, \lambda)$ from the following   paired values: $(0.05,0.05)$, $(0.1,0.05)$, $(0.1,0.1)$, $(0.5,0.1)$, $(0.5,0.2)$, $(1,0.2)$, $(1,0.3)$, $(5,0.3)$, $(5,0.4)$, $(10,0.4)$, $(50,0.5)$, and $(100,0.6)$. 

For each of the 12 pairs of tuning parameters,  we first determine $\widehat{K}_i$, $1\le i \le 12$, the maximum number of components to be found. 
As $\mu_k(\boldsymbol{\Xi})=\boldsymbol{\alpha}_k\trans\mathbf{B}\boldsymbol{\alpha}_k$  measures the signal magnitude of the $k$-th component and $\widehat{\mu}_k(\boldsymbol{\Xi})=\widehat{\boldsymbol{\alpha}}_k\trans \widehat{\mathbf{B}} \widehat{\boldsymbol{\alpha}}_k$ is an estimate of $\mu_k(\boldsymbol{\Xi})$, we  only compute the first few components with large values of $\widehat{\mu}_k(\boldsymbol{\Xi})$ and stop when $\widehat{\mu}_k(\boldsymbol{\Xi})$  becomes small enough.  On the other hand, by Theorem \ref{theorem_1} (a), the number of components cannot exceed $\hbox{min}(n, p, q)$. Based on these two considerations, we define
 \begin{align}
 &\widehat{K}_i=\min\{\widehat{K}_i^{(1)}, \widehat{K}_i^{(2)}\},\quad \text{ where } \widehat{K}_i^{(1)}=\hbox{min}(n, p, q), \label{30000}\\
& \text{and } \widehat{K}_i^{(2)}=\min\left\{k>1: \frac{\widehat{\mu}_k(\boldsymbol{\Xi})}{\widehat{\mu}_1(\boldsymbol{\Xi})+\cdots+\widehat{\mu}_{k}(\boldsymbol{\Xi})} \le 0.05\right\}\;. \notag
\end{align} 
The $\widehat{K}_i^{(2)}$ implies that we stop searching for higher order components if the signal magnitude of the $k$th component does not account for more than $5\%$ (by default) of the first $k$ components.
Once we have determined all the $\widehat{K}_i$,  we use the cross-validation method to determine the tuning parameters and the optimal number of components, $\widehat{K}_{opt}$. More specifically, we summarize the procedure in the following algorithm.  
\begin{Algorithm}\label{algorithm_1}
 \begin{enumerate}
\item
For the $i$-th paired value of the tuning parameters, $1\le i\le 12$, we determine $\widehat{K}_i$  using $\eqref{30000}$ and the whole data set. 
\item
 Use the five-fold cross-validation to  determine the number of components and the tuning parameters. We split the whole data set into five subsets and repeat the following procedure. For $1\le l\le 5$, we use the $l$-th subset as the $l$-th validation set and all other observations as the $l$-th training set. Then for the $i$-th pair of tuning parameter values, based on the $l$-th training set,
\begin{enumerate}
\item
 we estimate the first $\widehat{K}_i$ components ($\widehat{\boldsymbol{\alpha}}_1, \ldots, \widehat{\boldsymbol{\alpha}}_{\widehat{K}_i}$) by sequentially solving $\eqref{1240}$.
\item
 For each $j=1,\cdots, \widehat{K}_i$, we define $\widehat{\mathbf{t}}_j=\mathbf{X}\widehat{\boldsymbol{\alpha}}_j$ as the $j$-th new predictor. We use the first $j$ new predictors and $\eqref{10116}$ to get the estimate $\widehat{\mathcal{B}}_{ij}^{(l)}$ of the coefficient matrix, which is the estimate based on  the $l$-th training data set, the $i$-th pair of tuning parameter values and the first $j$ components. 
\item
Then we apply $\widehat{\mathcal{B}}_{ij}^{(l)}$ to the $l$-th validation data set to get the validation error, $e^{(l)}_{ij}$. 
\item
Finally, we calculate the average validation error, $\bar{e}_{ij}=(e^{(1)}_{ij}+\cdots+e^{(5)}_{ij})/5$,  for the $i$-th pair of tuning parameter values and the first $j$ components ($j=1,\cdots, \widehat{K}_i$).
\end{enumerate}
 Let $\bar{e}_{i_0j_0}=\min_{i,j}\bar{e}_{ij}$. Then the $i_0$-th paired value of tuning parameters is chosen, and the optimal number of components is $\widehat{K}_{opt}=j_0$. 
\end{enumerate}
\end{Algorithm}


\section{Simulation studies}
In this section, we compare the performance of the proposed {\bf SiER} method with four related methods on simulated data. The first method is the {\bf SRRR} \citep{chen2012sparse} which assumes the reduced rank structure and estimates the lower rank decomposition of the coefficient matrix by solving a penalized least squares problem   with a group-Lasso type penalty on the first lower rank matrix. The second method is {\bf RemMap} \citep{peng2010regularized} which does not assume the reduced rank structure and solves a penalized least squares problem with both
row-wise and element-wise sparsity imposed on the coefficient matrix. The third method is the {\bf SPLS} \citep{chun} which identifies sparse latent components by maximizing the covariance between them and the responses with sparsity penalty imposed.   The last method is {\bf SepLasso} which fits separate regression models  using Lasso for each individual response. 

  We consider three cases. In each case, we will consider the effects of different factors including  the dimension of the responses ($q$), the number of predictors ($p$) and their correlations ($\rho$), and the magnitude ($\sigma^2$) and correlation ($r$) of noises.  In the first case, $q$ is small. In the last two cases, $q$ is relatively large and we generate the coefficient matrices as the product of two lower rank matrices. 
	For each setting, we repeat the following procedure 50 times. In each replicate, we simulate 590 independent observations among which 90 are the training data and 500 are the test data. Then we apply each of the five methods to the training data to select the tuning parameters and the number of components, and construct the final model which is applied to the test data to obtain the test errors. The test error is defined as $\|\mathbf{Y}_{test}-\mathbf{Y}_{pred}\|^2_F/500$, where $\mathbf{Y}_{test}$ is the $500\times q$ matrix of the test data and $\mathbf{Y}_{pred}$ is the corresponding predicted matrix. The mean squared prediction error (MSPE) is obtained by averaging the 50 test errors.

\subsection{Case 1}\label{section5.1}
We generate data using the model $\mathbf{Y}=\mathbf{X}\mathcal{B}+\boldsymbol{\varepsilon}$. 
We fix $p=500$, $q=3$,
  and set $\mathcal{B}_{j1}=1/\sqrt{15}$ for $j=1,\ldots,15$, $\mathcal{B}_{j2}=0.5/\sqrt{30}$ for $j=16,\ldots,45$,  $\mathcal{B}_{j3}=0.25/\sqrt{60}$ for $j=46,\ldots,105$, and $\mathcal{B}_{jk}=0$ for others.  For each $i=1,\ldots, n$, the first 150 predictors are generated from multivariate normal distribution  $({X}_{i1}, \ldots, {X}_{i,150})\trans \sim N_{150}(0, \Sigma)$, where $\Sigma$ has the $(i,j)$-th element $\Sigma_{jk}=\rho^{|j-k|}$ for $j, k=1,\ldots,150$, and the other predictors are independent normal variables, ${X}_{ij} \sim N(0, 0.1^2)$ for $j=151,\ldots, p$. The noise vector $\boldsymbol{\varepsilon}_i$ is generated from $N_q(0, \sigma^2 R)$, where the correlation matrix $R$ has diagonal elements 1 and off-diagonal elements $r$. In this  and the following cases, we will use $q\sigma^2$, the sum of the diagonal elements of the covariance matrix of the noise vector $\boldsymbol{\varepsilon}_i$, as a measure of the noise level.  We choose $\rho=0.3$ and 0.7, $q\sigma^2=0.1$ and 0.15, $r=0.2$ and 0.9, and consider all their possible combinations.
	
\begin{figure}[h]
\begin{center}
\includegraphics[height=3.3in,width=6.5in, bb=0.5in 0in 8.5in 7in]{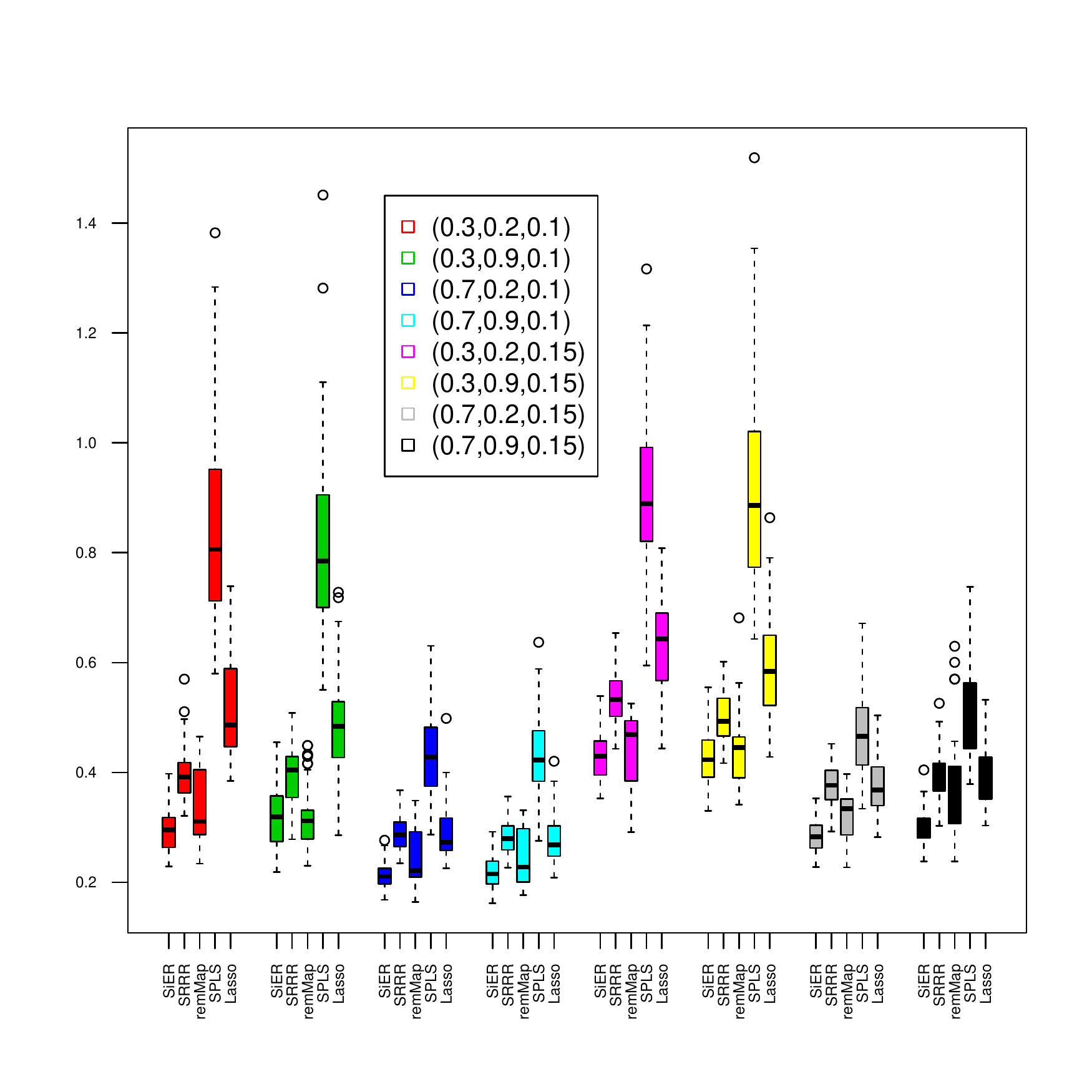}
\end{center}
    \caption{ \baselineskip=10pt Boxplots of MSPE for different methods with different parameter settings $(\rho, r, q\sigma^2)$ in Case 1 of the simulation studies. }
\label{sim1.fig}
\end{figure}

We draw the boxplots of MSPE   in Figure \ref{sim1.fig} and show the average and standard deviation of MSPEs of 50 repeats in Table \ref{sim1.tab}. The SiER has best predictive performance. In this case, $q$ is small, and there is no factor structure and no obvious group sparsity structure, so it is understandable that the SRRR has larger prediction errors than RemMap since the latter takes advantages of both the row-wise and element-wise sparsity.   The SPLS has the largest MSPE because it does not directly target on prediction of the responses and thus has disadvantage in terms of prediction.   The correlation among predictors, $\rho$, has a large effect on MSPE for all methods. A stronger correlation leads to a smaller prediction error in all the methods.  The correlation in random error, $r$, does not  have  obvious effects on MSPE for all methods.  Increasing in the magnitude of random error ($q\sigma^2$) unsurprisingly leads to higher MSPE.

\begin{table}[h]
\caption{\baselineskip=10pt The averages and standard deviations (in parenthesis) of the MSPEs of 50 replicates for Case 1 in the simulation studies.  }
\vspace{5mm}
\label{sim1.tab}\centering
 \small\addtolength{\tabcolsep}{-3pt}
 {
\begin{tabular}{c|c|c|c|c|c|c|c}
\hline
$q\sigma^2$ &  $\rho$ & $r$ &  SiER & SRRR & RemMap & SPLS&SepLasso \\\hline
  \multirow{2}{*}{0.1} & \multirow{2}{*}{0.3} & 0.2 & 0.295(0.039) & 0.395(0.050) & 0.336(0.067) & 0.848(0.180) & 0.504(0.087) \\\cline{3-8}
	&& 0.9 &  0.320(0.053) & 0.395(0.055) & 0.322(0.060) &0.824(0.181) & 0.489(0.088) \\\cline{2-8}
	& \multirow{2}{*}{0.7} & 0.2 & 0.214(0.023) & 0.289(0.031) & 0.244(0.051) &0.432(0.079) & 0.289(0.049) \\\cline{3-8}
	&& 0.9 & 0.217(0.028) & 0.282(0.030) & 0.243(0.050) &0.430(0.075) & 0.282(0.047) \\\hline
  \multirow{2}{*}{0.15} & \multirow{2}{*}{0.3} & 0.2 & 0.429(0.045) & 0.536(0.049) & 0.444(0.063) & 0.918(0.137) & 0.630(0.089)   \\\cline{3-8}
	&& 0.9 & 0.427(0.051) &0.501(0.045) & 0.440(0.068) & 0.923(0.202) & 0.597(0.095) \\\cline{2-8}
	& \multirow{2}{*}{0.7} & 0.2 & 0.283(0.028) & 0.376(0.037) & 0.323(0.044) &0.475(0.079) & 0.377(0.053)\\\cline{3-8}
	&& 0.9 & 0.302(0.033) & 0.394(0.046) & 0.368(0.090)& 0.508(0.089) & 0.392(0.053)\\\hline
\end{tabular}}
\end{table}

\subsection{Case 2}\label{section5.2}

In this case and Case 3, we generate the coefficient matrix as the product of two lower rank matrices,  $\mathcal{B}=CD$, where $C$ is a $p \times K$ matrix and $D$ is $K \times q$. In this case, we fix $K=3$ and choose $(p,q)=(100,20)$ or $(150,30)$. For the matrix $C$, each element in the first $p_0=40$ rows is independently generated from  $N(0,1)$,   the rest $p-p_0$ rows are set to be zero. Then each column of $C$ is scaled to unit norm. All elements in $D$ are first independently generated from the uniform distribution between $-1$ and $1$, and then each row of $D$ is scaled to unit norm. The first 50 predictors in $\mathbf{X}$ are generated from $N_{50}(0, \Sigma)$, where $\Sigma$ has diagonal elements 1 and off-diagonal elements $\rho$. All the other predictors are generated independently from $N(0, 0.1^2)$. The noise vector is generated from $N_{q}(0, \sigma^2 R)$, where $R$ has diagonal elements 1 and off-diagonal elements $r$. 
 We choose $\rho=0.3$ and 0.7, $q\sigma^2=0.015$ and $0.030$, $r=0$ and 0.5, and consider their all possible combinations. We  show the average and standard deviation of MSPE in Table \ref{sim2.tab}.
The SiER has the smallest prediction errors in all the situations. The SRRR has a better prediction performance than RemMap because in this case, there are obvious reduced rank structure and row-wise sparsity structure.   The MSPE of the first three methods are mostly affected by  the magnitude of random error ($q\sigma^2$), and are less sensitive to the correlations of predictors and random errors, and the dimensions $p$ and $q$. 

We also compare the dimension reduction and feature selection of all methods. The number of selected components,  the sensitivity and specificity of variable selection are summarized in Table \ref{sim3.tab}. Only three methods, the SiER, the SRRR and the SPLS, generate low rank latent components. In all settings, the SiER chooses the smallest number of components and is most efficient in dimension reduction.   The SiER, SRRR and RemMap have sensitivity equal to one in all settings, that is, these three methods select all the true features. The SRRR has the highest specificity, and hence performs best in feature selection in this case. The SiER tends to select more features than the SRRR, the RemMap and the SPLS in this simulation setting.

\begin{table}[h]
\caption{\baselineskip=10pt The averages and standard deviations (in parenthesis) of the MSPE for Case 2 in the simulation studies.  }
\vspace{5mm}
\label{sim2.tab}\centering
 \small\addtolength{\tabcolsep}{-3pt}
 {
\begin{tabular}{c|c|c|c|c|c|c|c|c}
\hline
$(p,q)$ & $q\sigma^2$ &  $\rho$ & $r$ &  SiER & SRRR & RemMap & SPLS&SepLasso \\\hline
   \multirow{8}{*}{(100,20)} & \multirow{4}{*}{0.015} & \multirow{2}{*}{0.3} & 0 & 0.020(0.009)  & 0.026(0.003)& 0.038(0.005) & 0.835(0.170) & 0.175(0.054) \\\cline{4-9}
	&&& 0.5 & 0.020(0.005)  & 0.025(0.003) & 0.037(0.005)& 0.839(0.168) & 0.167(0.049)\\\cline{3-9}
	&& \multirow{2}{*}{0.7} & 0 & 0.019(0.001)  & 0.035(0.005)& 0.043(0.004) & 0.473(0.109) & 0.268(0.057)\\\cline{4-9}
	&&& 0.5 & 0.019(0.002)  & 0.034(0.003)& 0.041(0.004) & 0.468(0.096) & 0.272(0.075)\\\cline{2-9}
  &\multirow{4}{*}{0.030} & \multirow{2}{*}{0.3} & 0 & 0.037(0.001) & 0.043(0.003)& 0.063(0.001)& 0.894(0.167) &0.241(0.066) \\\cline{4-9} 
	&&& 0.5 & 0.040(0.017) & 0.044(0.004) & 0.070(0.010) & 0.823(0.138) &0.230(0.056) \\\cline{3-9}
	&& \multirow{2}{*}{0.7} & 0 & 0.038(0.002) & 0.051(0.005) & 0.069(0.005)  & 0.523(0.135) & 0.306(0.068)  \\\cline{4-9}
	&&& 0.5 & 0.037(0.003) & 0.049(0.005) & 0.069(0.007)& 0.527(0.114) &0.298(0.060) \\\hline
	   \multirow{8}{*}{(150,30)} & \multirow{4}{*}{0.015} & \multirow{2}{*}{0.3} & 0 & 0.018(0.001) & 0.025(0.003) & 0.040(0.004) & 1.015(0.172) & 0.475(0.151) \\\cline{4-9}
		&&& 0.5 & 0.021(0.009) & 0.025(0.003) & 0.040(0.004) & 0.953(0.165) & 0.446(0.129) 	\\\cline{3-9}
			&& \multirow{2}{*}{0.7} & 0 & 0.018(0.001) & 0.033(0.004) & 0.047(0.005) & 0.599(0.117) & 0.558(0.101)\\\cline{4-9}
				&&& 0.5 & 0.018(0.001) & 0.032(0.003) & 0.046(0.005) & 0.603(0.110) & 0.575(0.094)\\\cline{2-9}
			& \multirow{4}{*}{0.030} & \multirow{2}{*}{0.3} & 0 & 0.035(0.001) & 0.042(0.003) & 0.070(0.005) & 0.972(0.185) & 0.501(0.104)\\\cline{4-9}
		&&& 0.5 & 0.036(0.002) & 0.042(0.004) & 0.073(0.011) & 1.023(0.183) & 0.581(0.205)	\\\cline{3-9}
			&& \multirow{2}{*}{0.7} & 0 & 0.036(0.001) & 0.050(0.005) &0.073(0.006) & 0.625(0.164) & 0.610(0.098)\\\cline{4-9}
				&&& 0.5 & 0.036(0.002) & 0.049(0.005) & 0.073(0.008) & 0.628(0.171) & 0.597(0.094) \\\hline
\end{tabular}}
\end{table}

\begin{table}[h]
\caption{\baselineskip=10pt The average number ($\hat{K}$) of components selected for the SiER, SRRR and SPLS, and the sensitivity (Se) and specificity (Sp) of all methods in case 2 in the simulation studies. } 
\vspace{5mm}
\label{sim3.tab}\centering
\addtolength{\tabcolsep}{-3pt}
 {
\begin{tabular}{*{18}{c|}c}
\hline
$(p,q)$ & $q\sigma^2$ &  $\rho$ & $r$ &  \multicolumn{3}{c|}{SiER} &  \multicolumn{3}{c|}{SRRR} &  \multicolumn{3}{c|}{RemMap} & \multicolumn{3}{c|}{SPLS} &  \multicolumn{3}{c}{SepLasso}  \\\cline{5-19}
&&&& $\hat{K}$ & Se & Sp & $\hat{K}$ & Se & Sp& $\hat{K}$ & Se & Sp& $\hat{K}$ & Se & Sp& $\hat{K}$ & Se & Sp\\\cline{1-19}
  \multirow{8}{*}{(100,20)} & \multirow{2}{*}{0.015} & \multirow{2}{*}{0.3} 
            	         & 0 & 3 & 1 & 0.58 & 3 & 1 & 0.92 & -- & 1 & 0.83 & 5 & 0.77 & 0.84 & -- & 0.76 & 0.25\\\cline{4-19}
	                   &&&0.5 & 3 & 1 & 0.60 & 3 & 1 & 0.93 & -- & 1 &0.83 & 5 & 0.76 & 0.84 & -- & 0.80 & 0.21\\\cline{3-19}
&	& \multirow{2}{*}{0.7} & 0 & 3 & 1 & 0.33 & 3.22 & 1 & 0.94 & -- & 1&0.83&5 & 0.94 & 0.77 & -- & 0.20 & 0.80\\\cline{4-19}
		                 &&& 0.5 & 3 & 1 & 0.39 & 3.02 & 1 & 0.94 & -- & 1&0.83&5 &  0.92 & 0.77 & --&0.49 & 0.53\\\cline{2-19}
 &\multirow{2}{*}{0.030} & \multirow{2}{*}{0.3} 
                       & 0 & 3 & 1 & 0.59 & 3 & 1 & 0.90 & -- & 1 & 0.83 & 5 & 0.78 & 0.80 & -- & 1.00 & 0.00\\\cline{4-19}
	                   &&&0.5 & 3 & 1 & 0.54 & 3 & 1 & 0.92 & -- & 1 &0.83 & 5 & 0.75 & 0.86 & -- & 1.00 & 0.01\\\cline{3-19}
&& \multirow{2}{*}{0.7} & 0 & 3 & 1 & 0.47 & 3.06 & 1 & 0.93 & --& 1&0.81&4.92&0.94 & 0.74 & -- & 1.00 & 0.02\\\cline{4-19}
		                 &&& 0.5 & 3 & 1 & 0.48 & 3.02 & 1 & 0.96 & --&1&0.82&5  & 0.93 & 0.75 & --&1.00 & 0.02\\\hline  \multirow{8}{*}{(150,30)} & \multirow{2}{*}{0.015} & \multirow{2}{*}{0.3} 
            	          & 0 & 3 & 1 & 0.69 & 3 & 1 & 0.96 & -- & 1 & 0.91 & 5 & 0.69 & 0.90 & -- & 1.00 & 0.04\\\cline{4-19}
	                   &&&0.5 & 3 & 1 & 0.71 & 3 & 1 & 0.96 & -- & 1 &0.91 & 4.98 & 0.70 & 0.89 & -- & 1.00 & 0.04\\\cline{3-19}
&	& \multirow{2}{*}{0.7}& 0 & 3 & 1 & 0.50 & 3.19 & 1 & 0.97 & -- & 1&0.91&4.85 & 0.86 & 0.86 & -- & 0.93 & 0.12\\\cline{4-19}
		                 &&& 0.5 & 3 & 1 & 0.57 & 3 & 1 & 0.97 & -- & 1&0.91&4.85 &  0.83 & 0.87 & --&0.92 & 0.14\\\cline{2-19}
 &\multirow{2}{*}{0.030} & \multirow{2}{*}{0.3} 
                      & 0 & 3 & 1 & 0.75 & 3 & 1 & 0.95 & -- & 1 & 0.91 & 4.98 & 0.65 & 0.91 & -- & 1.00 & 0.03\\\cline{4-19}
	                   &&&0.5 & 3 & 1 & 0.71 & 3 & 1 & 0.95 & -- & 1 &0.91 & 4.98 & 0.73 & 0.89 & -- & 1.00 & 0.05\\\cline{3-19}
&& \multirow{2}{*}{0.7} & 0 & 3 & 1 & 0.54 & 3.04 & 1 & 0.96 & -- & 1&0.90&4.70&0.89 & 0.86 & -- & 0.92 & 0.14\\\cline{4-19}
	                 &&& 0.5 & 3 & 1 & 0.58 & 3 & 1 & 0.96 & -- & 1&0.90&4.77 &  0.80 & 0.86 & --&0.92 & 0.12\\\hline
\end{tabular}}
\end{table}

\subsection{Case 3}\label{section5.2}
In this case, we still generate the coefficient matrix by $\mathcal{B}=CD$. We increase the dimensions $p$ and $q$, and introduce correlation among the coordinates of the row vectors of $D$. We take $K=3$, $p=1000$ or $2000$, $q=100$ or 200.  The matrix $C$ is generated in the same way as in Case 2 with $p_0=100$.  Each row of $D$ is first independently generated from the multivariate normal distribution $\hbox{N}_q(0,\Sigma_D)$ and then scaled to have unit norm. The $(i,j)$-th element in $\Sigma_D$ is defined as $\hbox{exp}[-(|i-j|/100)^{\gamma}]$ for $i,j=1,\ldots, q$. Therefore, the row vectors of $D$ can be viewed as a discretely observed Gaussian process at time points $1/100,2/100,\cdots$. We choose $\gamma=1$ or 2 to represent different correlation levels, where  $\gamma=2$ leads to a smoother Gaussian process than $\gamma=1$. The first 200 predictors in $\mathbf{X}$ are generated from  $N_{200}(0, \Sigma)$, with the $(i,j)$-th element equal $\rho^{|i-j|}$ ($i,j=1,\ldots,200$). The  rest $p-200$ predictors are  independently generated from $N(0, 0.1^2)$. The random noise vector is generated from $N_{q}(0, \sigma^2 R)$, where $R$ have diagonal elements 1 and off-diagonal elements 0.5. We fix $q\sigma^2=0.15$. We summarize the MSPE in Table \ref{sim4.tab}, the dimension reduction and the sensitivity and specificity of variable selection in Table \ref{sim5.tab}.  In this case, the SRRR is not included because the heavy computation load makes it unavailable. As the noise level is fixed, the MSPE is not sensitive to $\gamma$, $p$ and $q$. The SiER has the lowest average MSPE among the four methods and choose less components than the SPLS. In this large dimension case, the SiER has  both high sensitivity and high specificity. 


\begin{table}[h]
\caption{\baselineskip=10pt The averages and standard deviations (in parenthesis) of the MSPE for case 3.  }
\vspace{5mm}
\label{sim4.tab}\centering
 \small\addtolength{\tabcolsep}{-3pt}
 {
\begin{tabular}{c|c|c|c|c|c|c}
\hline
$q$ & $p$ &  $\gamma$ &  SiER  & RemMap & SPLS&SepLasso \\\hline
 \multirow{4}{*}{100} & \multirow{2}{*}{1000} & 1 & 1.739(0.185) & 2.146(0.480) & 2.816(0.370)& 3.103(0.340) \\\cline{3-7}
&  & 2 & 1.844(0.330) & 2.264(0.615) & 2.872(0.462) & 3.117(0.501)  \\\cline{2-7}
  & \multirow{2}{*}{2000} & 1 & 1.748(0.244) & 2.108(0.485) & 3.026(0.432) & 3.202(0.408) \\\cline{3-7}
	& & 2 &1.740(0.337) & 2.185(0.616) & 2.944(0.396) & 3.113(0.385)\\\hline
   \multirow{4}{*}{200} & \multirow{2}{*}{1000} & 1 & 1.763(0.208) & 2.180(0.419) & 2.765(0.235) & 3.131(0.488) \\\cline{3-7}
&	& 2 & 1.774(0.304) & 2.186(0.574) & 2.843(0.320) & 3.070(0.321)\\\cline{2-7}
	& \multirow{2}{*}{2000} & 1 & 1.756(0.224) & 2.109(0.488) & 2.960(0.279) & 3.098(0.266)\\\cline{3-7}
&	& 2 & 1.808(0.280) & 2.243(0.611) & 3.004(0.360) & 3.161(0.354)\\\hline
\end{tabular}}
\end{table}

\begin{table}[h]
\caption{\baselineskip=10pt The average number ($\hat{K}$) of components selected for the SiER and SPLS, and the sensitivity (Se) and specificity (Sp) of all methods in Case 3.  }
\vspace{5mm}
\label{sim5.tab}\centering
\addtolength{\tabcolsep}{-3pt}
 {
\begin{tabular}{*{14}{c|}c}
\hline
$q$ & $p$ &  $\gamma$ &  \multicolumn{3}{c|}{SiER}  & \multicolumn{3}{c|}{RemMap} & \multicolumn{3}{c|}{SPLS} &  \multicolumn{3}{c}{SepLasso}  \\\cline{4-15}
&&& $\hat{K}$ & Se & Sp & $\hat{K}$ & Se & Sp& $\hat{K}$ & Se & Sp& $\hat{K}$ & Se & Sp\\\cline{1-15}
  \multirow{4}{*}{100} & \multirow{2}{*}{1000}  
	                          & 1 & 2.86 & 0.84&0.93&--&0.77& 0.93 & 3.44&0.45&0.77&--&0.42&0.86\\\cline{3-15}
	                          &&2 & 2.32 & 0.75&0.94&--&0.65& 0.94 & 2.76&0.41&0.79&--&0.32&0.89  \\\cline{2-15}
		 & \multirow{2}{*}{2000}& 1 &3     &0.82& 0.97&--&0.77& 0.97 & 3.06&0.35&0.84&--&0.27&0.95 \\\cline{3-15}
		                        &&2 &2.40  &0.76&0.97 &--&0.65& 0.97 & 2.38&0.30&0.85&--&0.23&0.95 \\\cline{1-15}
														
  \multirow{4}{*}{200} & \multirow{2}{*}{1000}  
	                          & 1 &3.04 &0.88 &0.92 &--&0.79& 0.93 & 3.54&0.43&0.81&--&0.51&0.80  \\\cline{3-15}
	                          &&2 &2.34 &0.78 &0.93 &--&0.71& 0.93 & 2.70&0.35&0.83&--&0.39&0.85  \\\cline{2-15}
		 & \multirow{2}{*}{2000}& 1 &2.96 &0.85 &0.97 &--&0.81& 0.96 & 3.26&0.35&0.83&--&0.37&0.91\\\cline{3-15}
		                        &&2 &2.66 &0.77 &0.97 &--&0.68& 0.97 & 2.50&0.32&0.86&--&0.27&0.93\\\cline{1-15}
\end{tabular}}
\end{table}

\section{Application to the communities and crime  data}
 The data set \citep{Bache+Lichman:2013} contains the socio-economic information about a large number of communities over the United States and the corresponding crime data. The goal is to predict different types of crimes of a community based on various socio-economic variables of the community such as the median family income,   per capita number of police officers, and so on.  We first remove the variables with a large proportion of missing. We also remove the response (crime) variables with too many zero values as it is not appropriate to treat them as continuous variables. Due to the high skewness of the variables,  we use the logarithms of each of the original variables plus one as the new variables. 
After excluding samples with extreme values, we get 1514 samples and  102 predictors. 
The response variables are the transformed number of Assault, Burglary, Larceny, Auto Theft, and Arson, the transformed total number of violent crimes, and the transformed total number of non-violent crimes, per 100,000 population. 

\begin{figure}[h]
\begin{center}
\includegraphics[height=2.5in,width=4.5in]{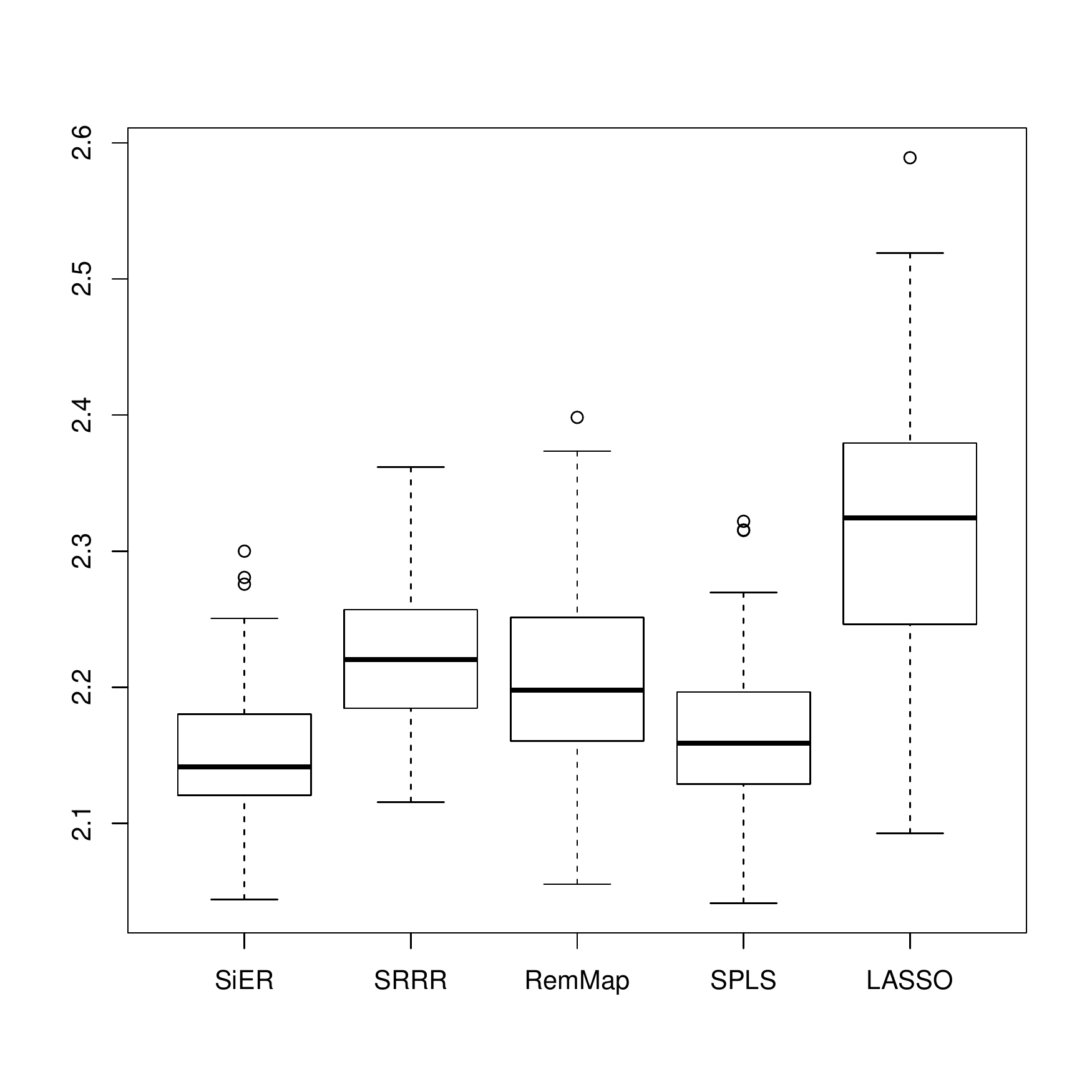}
\end{center}
    \caption{ \baselineskip=10pt Boxplots of prediction errors for different methods in the study for the crime data.}
\label{crime.fig}
\end{figure}

\begin{table}[h]
\caption{\baselineskip=10pt The frequency of the number of selected components ($\hat{K}$)  and the mean (sd) of the number of selected features over 100 simulations for the crime data.} 
\vspace{5mm}
\label{crime.tab}\centering
 \small\addtolength{\tabcolsep}{-3pt}
 {
\begin{tabular}{c|c|c|c|c|c}
\hline
 \multirow{2}{*}{Methods} & \multicolumn{4}{|c|}{ $\hat{K}$ } & \multirow{2}{*}{number of selected features} \\\cline{2-5}
 & 2 & 3 & 4 & 5 &  \\\hline
SiER & 76 & 24 & 0 & 0 & 52.2(23.3)\\
SRRR & 22 & 12 & 17 & 49 & 32.9(18.2) \\
RemMap &--  & -- & -- & -- & 37.6(8.2)\\
SPLS& 1 & 1 & 27 & 71 & 81.5(23.8)\\
SepLasso & -- & -- & -- & -- & 32.7(7.4)\\\hline
\end{tabular}}
\end{table}

To evaluate the prediction performance and the selection of components and features, we repeat the following procedure 100 times: in each replicate, we  randomly take 150 samples as training data, the remaining as test data, and apply all the five methods to the train data to build predictive models and obtain the MSPE by applying the model to the test data. The boxplots of prediction errors of all methods are shown in Figure \ref{crime.fig}. The mean MSPEs of the SiER  and SPLS are close and smaller than others. Table \ref{crime.tab} shows the frequency of the number of selected components and the mean and standard deviation of number of selected features for each method. The SiER selects two components in 76\% of the 100 replicates and three components in all the other replicates. Both the SRRR and the SPLS tend to select more components and they select five components with the highest frequency.

\begin{figure}[h]
\begin{center}
\includegraphics[height=3in,width=5in]{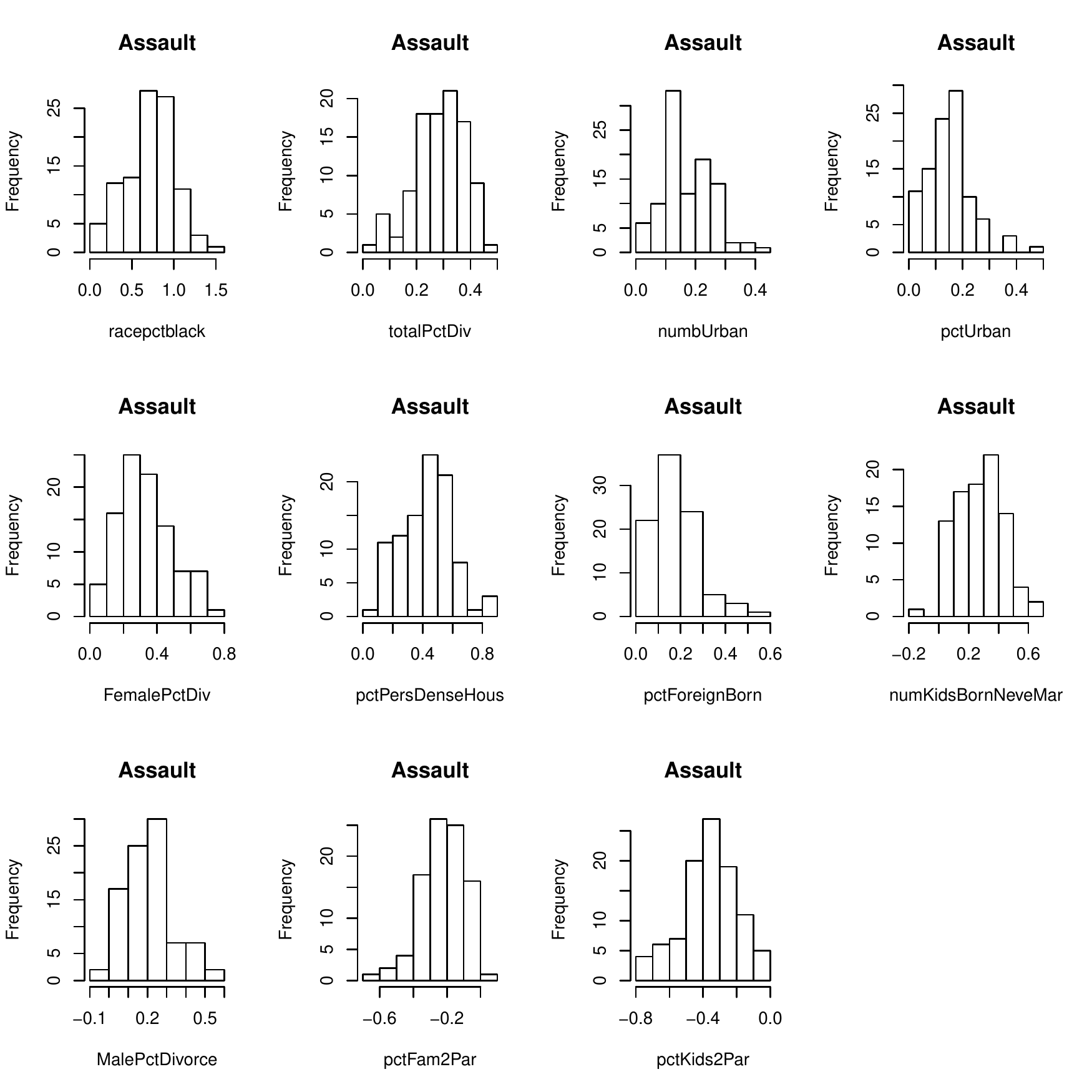}
\end{center}
    \caption{ \baselineskip=10pt Histograms of nonzero coefficients of variables that are selected over 90\% times in predicting the number of assault.}
\label{crimeCoef.fig}
\end{figure}

Among the 100 replicates, 11 features are selected by the SiER over 90 times, such as  the percentage of population who are divorced, 
the number of people living in urban areas, the percent of persons in dense housing, the number of kids born to never married, and so on.
 The histograms of nonzero coefficients of these variables for assault are shown in Figure \ref{crimeCoef.fig}. It can be seen that the percentage of kids in family housing with two parents (pctKids2Par) and the percentage of families with kids that are headed by two parents (pctFam2Par) have protective effects (with  negative coefficients) in crime assault, and the other variables have positive associations with the number of assault. For other crime types, pctKids2Par and pctFam2Par also have protective effects. Percentages for divorce and dense housing, and the number of kids born to never married are positively associated with crimes. Coefficients for variables of population living in urban areas and having foreign born are positive for arson, but negative for auto theft.

\section{Discussion}
In this paper, we propose a signal extracting approach for dimension reduction and regression in multiple response linear model with high-dimensional predictor variables. Under the reduced rank assumptions on the coefficient matrix, we aim to estimate the optimal lower rank decomposition of the coefficient matrix in terms of approximating the regression function. We establish a general eigenvalue problem and its sparse version for high-dimensional settings. The solution of these problems provides the estimate of the first lower rank matrix. Applying  the least squares regression on the response variables and new predictors generated from the estimate of the first lower rank matrix, we obtain the estimate of the second lower rank matrix.  In the high-dimensional setting, we establish the oracle inequalities for the estimation of the lower rank matrices, the coefficient matrix and the estimated regression function, allowing correlation among random errors for different response variables. We do not make restrictions on the dimension of the multivariate response. In the special case of the usual linear regression model with a scalar response, our oracle inequalities provide upper bounds that have the same order as those for the  Lasso and Dantzig selector. The simulation studies  and the application to  real data show that the proposed method has good prediction performance and is efficient in dimension reduction for various reduced rank models.

\section*{Acknowledgments}
The authors would like to thank Professor Jianhua Huang for many constructive comments and suggestions. Xin Qi is supported by NSF DMS 1208786.


\end{document}


\begin{center}
{\LARGE{\bf  Supplementary materials for ``Signal extraction approach for sparse multivariate response regression''}}
\end{center}
\vskip 2mm

\baselineskip=14pt
\vskip 2mm
\begin{center}
 Ruiyan Luo\\
\vskip 2mm
Division of Epidemiology and Biostatistics, Georgia State University School of Public Health, One Park Place, Atlanta, GA 30303\\
rluo@gsu.edu\\
\hskip 5mm\\
Xin Qi\\
\vskip 2mm
Department of Mathematics and Statistics, Georgia State University, 30 Pryor Street, Atlanta, GA 30303-3083\\
xqi3@gsu.edu \\
\end{center}
  
\vskip 20mm
 \renewcommand{\proof}{\noindent\underline{\bf Proof of Theorem}}

\bigskip
This supplementary material is organized as follows.  Appendix A provides additional proofs of theorems and Appendix B provides the proofs of all technical lemmas.

\section*{Appendix A: Proofs of theorems}
 \renewcommand{\thesubsection}{A.\arabic{subsection}}

\renewcommand{\proof}{\underline{\bf Proof of Theorem}}

\setcounter{equation}{0}
\setcounter{table}{0}
\baselineskip=18pt

We first introduce some notations used throughout this supplementary material.    Recall that $\boldsymbol{\varepsilon}=[\boldsymbol{\varepsilon}_1,\cdots, \boldsymbol{\varepsilon}_n]\trans$ is  the $n \times q$ random noise matrix. Let
\begin{align*}
 \bar{\boldsymbol{\varepsilon}}=\frac{1}{n}\sum_{i=1}^n\boldsymbol{\varepsilon}_i,\quad \boldsymbol{\varrho}=\frac{1}{\sqrt{n}}(\boldsymbol{\varepsilon}-\mathbf{1}_n\bar{\boldsymbol{\varepsilon}}\trans).
\end{align*} 
We provide a lemma about the Frobenius norm, the proof of which can be found in Appendix B.
\begin{Lemma}\label{lemma_18}
Let $\mathbf{M}$ be a $p\times p$ symmetric matrix and $\mathbf{D}$ be a $p\times q$ matrix. Then we have  
 \begin{align}
&\|\mathbf{M}\mathbf{D}\|_F\le \|\mathbf{M}\|\|\mathbf{D}\|_F  .\label{50015}
\end{align} 
\end{Lemma}

\subsection{Proof of Theorem $\ref{theorem_1}$}

 By the definition of $\boldsymbol{\Xi}$ in  $\eqref{10008}$, its rank is less than the minimum of the ranks of $\mathbf{X}$ and $\mathcal{B}$. Since $\mathbf{X}$ and $\mathcal{B}$ are $n\times p$ and $p\times q$, respectively, we have $K\le \min\{n, p,q\}$. 

Since $\boldsymbol{\gamma}_k$, $1\le k\le K$  are the left-singular  vectors of  $\mathbf{X}\mathcal{B}$, they are the eigenvectors of the matrix $(\mathbf{X}\mathcal{B}) (\mathbf{X}\mathcal{B})\trans =n \boldsymbol{\Xi}$, and  $\sigma_k=\sqrt{n\mu_k(\boldsymbol{\Xi})}$. Similarly, the right-singular  vectors $\mathbf{u}_k$, $1\le k\le K$, are the eigenvectors of the matrix $(\mathbf{X}\mathcal{B})\trans (\mathbf{X}\mathcal{B})=\mathcal{B} \trans  \mathbf{X}\trans\mathbf{X}\mathcal{B}$ with eigenvalues equal to $n\mu_k(\boldsymbol{\Xi})$.

Consider $k=1$. We first show that the maximum value of $\eqref{2220}$ is equal to $\mu_1(\boldsymbol{\Xi})$. For any $\boldsymbol{\alpha}\in \mathbb{R}^p$ with $\boldsymbol{\alpha}\trans \mathbf{S}\boldsymbol{\alpha}=1$, define $\mathbf{v}=\mathbf{Z}\boldsymbol{\alpha}$. Then we have $\|\mathbf{v}\|_2=1$. By $\eqref{10008}$, 
\begin{align*}
 \boldsymbol{\alpha}\trans \mathbf{B}\boldsymbol{\alpha}=\boldsymbol{\alpha}\trans\mathbf{Z}\trans\boldsymbol{\Xi}\mathbf{Z}\boldsymbol{\alpha}=\mathbf{v}\trans  \boldsymbol{\Xi} \mathbf{v}\le \mu_1(\boldsymbol{\Xi}) \|\mathbf{v}\|_2^2=\mu_1(\boldsymbol{\Xi}).
\end{align*}
Therefore, the maximum value of $\eqref{2220}$ is not greater than $\mu_1(\boldsymbol{\Xi})$. On the other hand, by $\eqref{70002}$, 
\begin{align}
 \boldsymbol{\alpha}_k=\frac{n}{\sigma_k^2}\mathcal{B}\mathbf{w}_k=\frac{\sqrt{n}}{\sigma_k}\mathcal{B}\mathbf{u}_k,\quad 1\le k\le K.\label{70009}
\end{align}
  Then by the definition of $\mathbf{B}$ in $\eqref{10008}$,
 \begin{align*}
&\boldsymbol{\alpha}_1\trans \mathbf{B}\boldsymbol{\alpha}_1=\frac{1}{\sigma^2_1n}\mathbf{u}_1\trans \mathcal{B}\trans\mathbf{X}\trans\left(\mathbf{X}\mathcal{B}\right) \left(\mathbf{X}\mathcal{B}\right)\trans\mathbf{X}\mathcal{B}\mathbf{u}_1\\
=&\frac{1}{\sigma^2_1n}\mathbf{u}_1\trans \left(\mathcal{B}\trans\mathbf{X}\trans\mathbf{X}\mathcal{B}\right) \left(\mathcal{B}\trans\mathbf{X}\trans\mathbf{X}\mathcal{B}\right)\mathbf{u}_1=\frac{n^2\mu_1(\boldsymbol{\Xi})^2}{\sigma^2_1n}\mathbf{u}_1\trans  \mathbf{u}_1=\mu_1(\boldsymbol{\Xi}).
\end{align*} 
Hence, when $k=1$, $\boldsymbol{\alpha}_1$ is the solution to $\eqref{2220}$. 
By similar arguments, we can show that for any $1\le k\le K$, $\boldsymbol{\alpha}_k$ is the solution to $\eqref{2220}$. $\eqref{10074}$ follows from $\eqref{77000}$ and $\eqref{10008}$.
  Part (c) is easily obtained from the following equality.
\begin{align*}
 \|\mathbf{X}\mathcal{B}-\sum_{i=1}^k\sigma_k\boldsymbol{\gamma}_k\mathbf{u}_k\trans\|_F^2=\|\sum_{i=k+1}^K\sigma_k\boldsymbol{\gamma}_k\mathbf{u}_k\trans\|_F^2 =\sum_{i=k+1}^K\sigma^2 =n\sum_{i=k+1}^K\mu_i(\boldsymbol{\Xi}).  
\end{align*}

\subsection{Proof of Theorem $\ref{theorem_8}$}

First, we have
\begin{align}
 &\min_{\widetilde{\mathcal{B}}_k}E\left[\|(\mathbf{x}^{\rm new})\trans\widetilde{\mathcal{B}}_k-(\mathbf{y}^{\rm new})\trans\|_2^2\right]\label{25003}\\
=&\min_{\substack{\mathbf{b}_j\in\mathbb{R}^p, \mathbf{v}_j\in \mathbb{R}^q,\\ 1\le j\le k}}E\left[ \|\sum_{j=1}^k(\mathbf{x}^{\rm new})\trans\mathbf{b}_j\mathbf{v}_j\trans-(\mathbf{x}^{\rm new})\trans \mathcal{B}-(\boldsymbol{\varepsilon}^{\rm new})\trans\|_2^2 \right]\notag\\
=& \min_{\substack{\mathbf{b}_j\in\mathbb{R}^p, \mathbf{v}_j\in \mathbb{R}^q,\\ 1\le j\le k}}E\left[ \|(\mathbf{x}^{\rm new})\trans \mathcal{B}-\sum_{j=1}^k(\mathbf{x}^{\rm new})\trans\mathbf{b}_j\mathbf{v}_j\trans\|_2^2 \right]+E[\|\boldsymbol{\varepsilon}^{\rm new}\|_2^2],\notag
\end{align} 
and similarly,
\begin{align}
 & E\left[\|(\mathbf{x}^{\rm new})\trans\mathcal{B}_k-(\mathbf{y}^{\rm new})\trans\|_2^2\right]=E\left[ \|(\mathbf{x}^{\rm new})\trans \mathcal{B}-(\mathbf{x}^{\rm new})\trans\mathcal{B}_k\|_2^2 \right]+E[\|\boldsymbol{\varepsilon}^{\rm new}\|_2^2],\label{25009}
\end{align} 
Let $\boldsymbol{\Sigma}=\mathbf{Z}\trans\mathbf{Z}$ be a decomposition of $\boldsymbol{\Sigma}$, where $\mathbf{Z}$ is a $p\times p$ matrix. Consider the SVD of $\mathbf{Z}\mathcal{B}$,
\begin{align}
\mathbf{Z}\mathcal{B}=\sum_{j=1}^N\phi^0_j(\mathbf{v}^0_j)\trans, \label{90001}
\end{align} 
where $N$ is the rank of $\mathbf{Z}\mathcal{B}$ and $\mathbf{v}^0_1$, $\cdots$, $\mathbf{v}^0_N$ are orthonormal. Let $\mathbf{b}^0_j=\mathcal{B}\mathbf{v}^0_j$, $1\le j\le N$. Then we have $\phi^0_j=\mathbf{Z}\mathbf{b}^0_j$ and
\begin{equation}
 \mathbf{Z}\mathcal{B}=\sum_{j=1}^N\phi^0_j(\mathbf{v}^0_j)\trans=\sum_{j=1}^N\mathbf{Z}\mathbf{b}^0_j(\mathbf{v}^0_j)\trans, \label{90002}
\end{equation} 
For any square matrix $A$, let $tr(A)$ denote its trace. Then 
\begin{align}
 &  \min_{\substack{\mathbf{b}_j\in\mathbb{R}^p, \mathbf{v}_j\in \mathbb{R}^q,\\ 1\le j\le k}}E\left[ \|(\mathbf{x}^{\rm new})\trans \mathcal{B}-\sum_{j=1}^k(\mathbf{x}^{\rm new})\trans\mathbf{b}_j\mathbf{v}_j\trans\|_2^2 \right]\label{90002}\\
=& \min_{\substack{\mathbf{b}_j\in\mathbb{R}^p, \mathbf{v}_j\in \mathbb{R}^q,\\ 1\le j\le k}} tr\left\{\left(\mathcal{B}-\sum_{j=1}^k\mathbf{b}_j\mathbf{v}_j\trans\right)\trans E\left[\mathbf{x}^{\rm new}  (\mathbf{x}^{\rm new})\trans  \right]\left(\mathcal{B}-\sum_{j=1}^k\mathbf{b}_j\mathbf{v}_j\trans\right)\right\}\notag\\
=& \min_{\substack{\mathbf{b}_j\in\mathbb{R}^p, \mathbf{v}_j\in \mathbb{R}^q,\\ 1\le j\le k}}tr\left\{\left(\mathcal{B}-\sum_{j=1}^k\mathbf{b}_j\mathbf{v}_j\trans\right)\trans \boldsymbol{\Sigma} \left(\mathcal{B}-\sum_{j=1}^k\mathbf{b}_j\mathbf{v}_j\trans\right)\right\}\notag\\
=& \min_{\substack{\mathbf{b}_j\in\mathbb{R}^p, \mathbf{v}_j\in \mathbb{R}^q,\\ 1\le j\le k}}tr\left\{\left(\mathcal{B}-\sum_{j=1}^k\mathbf{b}_j\mathbf{v}_j\trans\right)\trans \mathbf{Z}\trans\mathbf{Z} \left(\mathcal{B}-\sum_{j=1}^k\mathbf{b}_j\mathbf{v}_j\trans\right)\right\}\notag\\
=& \min_{\substack{\mathbf{b}_j\in\mathbb{R}^p, \mathbf{v}_j\in \mathbb{R}^q,\\ 1\le j\le k}}\|\mathbf{Z}\mathcal{B}-\sum_{j=1}^k\mathbf{Z}\mathbf{b}_j\mathbf{v}_j\trans\|_F^2=\|\mathbf{Z}\mathcal{B}-\sum_{j=1}^k\mathbf{Z}\mathbf{b}^0_j(\mathbf{v}^0_j)\trans\|_F^2\notag\\
= &  E\left[ \|(\mathbf{x}^{\rm new})\trans \mathcal{B}-\sum_{j=1}^k(\mathbf{x}^{\rm new})\trans\mathbf{b}^0_j(\mathbf{v}^0_j)\trans\|_2^2 \right],\notag
\end{align} 
where the second last equality is due to the best low rank approximation of the SVD. Let $d_{\ell}$ denote the $\ell$-th column of $\mathcal{B}-\sum_{j=1}^k\mathbf{b}^0_j(\mathbf{v}^0_j)$, $\ell=1, \ldots, q$. Then we have
	\begin{align}
&\left|E\left[ \|(\mathbf{x}^{\rm new})\trans \mathcal{B}-\sum_{j=1}^k(\mathbf{x}^{\rm new})\trans\mathbf{b}^0_j(\mathbf{v}^0_j)\trans\|_2^2 \right]-\frac{1}{n}\|\mathbf{X}\mathcal{B}-\sum_{j=1}^k\mathbf{X}\mathbf{b}^0_j(\mathbf{v}^0_j)\trans\|_F^2\right|\label{25006}\\
=& \left|{\footsize{tr \left\{\left(\mathcal{B}-\sum_{j=1}^k\mathbf{b}^0_j(\mathbf{v}^0_j)\trans\right)\trans\boldsymbol{\Sigma}\left(\mathcal{B}-\sum_{j=1}^k\mathbf{b}^0_j(\mathbf{v}^0_j)\trans\right) -\left(\mathcal{B}-\sum_{j=1}^k\mathbf{b}^0_j(\mathbf{v}^0_j)\trans\right)\trans\mathbf{S}\left(\mathcal{B}-\sum_{j=1}^k\mathbf{b}^0_j(\mathbf{v}^0_j)\trans\right) \right\}}}\right|\notag\\
=&\left|tr\left\{\left(\mathcal{B}-\sum_{j=1}^k\mathbf{b}^0_j(\mathbf{v}^0_j)\trans\right)\trans(\mathbf{S}-\boldsymbol{\Sigma})\left(\mathcal{B}-\sum_{j=1}^k\mathbf{b}^0_j(\mathbf{v}^0_j)\trans\right)  \right\}\right|\notag\\
=&\left|\sum_{\ell=1}^q \mathbf{d}_{\ell}\trans (\mathbf{S}-\boldsymbol{\Sigma})\mathbf{d}_{\ell}\right|
\le \sum_{\ell=1}^q \left|\mathbf{d}_{\ell}\trans (\mathbf{S}-\boldsymbol{\Sigma})\mathbf{d}_{\ell}\right|
\le \|\mathbf{S}-\boldsymbol{\Sigma}\|_\infty\|\sum_{\ell=1}^q \|\mathbf{d}_{\ell}\|_1^2 \notag\\
\le &\|\mathbf{S}-\boldsymbol{\Sigma}\|_\infty\|\sum_{\ell=1}^q \mathcal{M}(\mathbf{d}_{\ell})\|\mathbf{d}_{\ell}\|_2^2 
\le \|\mathbf{S}-\boldsymbol{\Sigma}\|_\infty\mathcal{M}\left(\mathcal{B}-\sum_{j=1}^k \mathbf{b}^0_j(\mathbf{v}^0_j)\right)\sum_{\ell=1}^q \|\mathbf{d}_{\ell}\|_2^2 \notag\\
= & \|\mathbf{S}-\boldsymbol{\Sigma}\|_\infty \mathcal{M}\left(\mathcal{B}-\sum_{j=1}^k \mathbf{b}^0_j(\mathbf{v}^0_j)\trans\right)\|\mathcal{B}-\sum_{j=1}^k\mathbf{b}^0_j(\mathbf{v}^0_j)\trans\|_F^2\notag\\
\le &\|\mathbf{S}-\boldsymbol{\Sigma}\|_\infty \mathcal{M}(\mathcal{B})\|\mathcal{B}-\sum_{j=1}^k\mathbf{b}^0_j(\mathbf{v}^0_j)\trans\|_F^2 \le \|\mathbf{S}-\boldsymbol{\Sigma}\|_\infty \mathcal{M}(\mathcal{B}) \|\mathcal{B}\|_F^2 \le C\|\mathcal{B}\|_F^2\mathcal{M}(\mathcal{B})\sqrt{\frac{\log p}{n}}\;,\notag
\end{align} 
where the first inequality in the fifth line is due to the Cauchy-Schwarz inequality; in the last line, the first inequality  is because $\mathcal{M}(\mathcal{B})\ge \mathcal{M}\left(\mathcal{B}-\sum_{j=1}^k \mathcal{B}\mathbf{v}^0_j(\mathbf{v}^0_j)\trans\right)= \mathcal{M}\left(\mathcal{B}-\sum_{j=1}^k \mathbf{b}^0_j(\mathbf{v}^0_j)\trans\right)$, and the second inequality is because $\mathcal{B}-\sum_{j=1}^k \mathbf{b}^0_j(\mathbf{v}^0_j)\trans=\mathcal{B}\{I-\sum_{j=1}^k \mathbf{v}^0_j(\mathbf{v}^0_j)\trans\}$ and $\{I-\sum_{j=1}^k \mathbf{v}^0_j(\mathbf{v}^0_j)\trans\}$ is an orthogonal projection matrix. Similarly, we have
\begin{align}
\left|E\left[ \|(\mathbf{x}^{\rm new})\trans \mathcal{B}-(\mathbf{x}^{\rm new})\trans\mathcal{B}_k\|_2^2\right]-\frac{1}{n}\|\mathbf{X}\mathcal{B}-\mathbf{X} \mathcal{B}_k\|_F^2\right| \le  C\|\mathcal{B}\|_F^2\mathcal{M}(\mathcal{B})\sqrt{\frac{\log p}{n}}.
\label{25007}
\end{align} 
 By  $\eqref{90002}$,  $\eqref{25006}$ and $\eqref{25007}$, we have
 \begin{align}
&E\left[ \|(\mathbf{x}^{\rm new})\trans \mathcal{B}-(\mathbf{x}^{\rm new})\trans\mathcal{B}_k\|_2^2\right]\le \frac{1}{n}\|\mathbf{X}\mathcal{B}-\mathbf{X} \mathcal{B}_k\|_F^2\label{24048}\\
&\qquad +\left|E\left[ \|(\mathbf{x}^{\rm new})\trans \mathcal{B}-(\mathbf{x}^{\rm new})\trans\mathcal{B}_k\|_2^2\right]-\frac{1}{n}\|\mathbf{X}\mathcal{B}-\mathbf{X} \mathcal{B}_k\|_F^2\right|\notag\\
&\le \frac{1}{n}\|\mathbf{X}\mathcal{B}-\mathbf{X} \mathcal{B}_k\|_F^2+  C\|\mathcal{B}\|_F^2\mathcal{M}(\mathcal{B})\sqrt{\frac{\log p}{n}}\notag\\
&\le \frac{1}{n}\|\mathbf{X}\mathcal{B}-\sum_{j=1}^k\mathbf{X}\mathbf{b}^0_j(\mathbf{v}^0_j)\trans\|_F^2\right|+  C\|\mathcal{B}\|_F^2\mathcal{M}(\mathcal{B})\sqrt{\frac{\log p}{n}}\notag\\
&\le \left|E\left[ \|(\mathbf{x}^{\rm new})\trans \mathcal{B}-\sum_{j=1}^k(\mathbf{x}^{\rm new})\trans\mathbf{b}^0_j(\mathbf{v}^0_j)\trans\|_2^2 \right]-\frac{1}{n}\|\mathbf{X}\mathcal{B}-\sum_{j=1}^k\mathbf{X}\mathbf{b}^0_j(\mathbf{v}^0_j)\trans\|_F^2\right|\notag\\
&+E\left[ \|(\mathbf{x}^{\rm new})\trans \mathcal{B}-\sum_{j=1}^k(\mathbf{x}^{\rm new})\trans\mathbf{b}^0_j(\mathbf{v}^0_j)\trans\|_2^2 \right]+  C\|\mathcal{B}\|_F^2\mathcal{M}(\mathcal{B})\sqrt{\frac{\log p}{n}}\notag\\
&\le E\left[ \|(\mathbf{x}^{\rm new})\trans \mathcal{B}-\sum_{j=1}^k(\mathbf{x}^{\rm new})\trans\mathbf{b}^0_j(\mathbf{v}^0_j)\trans\|_2^2 \right]+ 2 C\|\mathcal{B}\|_F^2\mathcal{M}(\mathcal{B})\sqrt{\frac{\log p}{n}}\notag\\
&= \min_{\substack{\mathbf{b}_j\in\mathbb{R}^p,  \mathbf{v}_j \in \mathbb{R}^q,\\ 1\le j\le k}}E\left[ \|(\mathbf{x}^{\rm new})\trans \mathcal{B}-\sum_{j=1}^k(\mathbf{x}^{\rm new})\trans\mathbf{b}_j\mathbf{v}_j\trans\|_2^2 \right]+2 C\|\mathcal{B}\|_F^2\mathcal{M}(\mathcal{B})\sqrt{\frac{\log p}{n}},\notag
\end{align} 
where the third inequality follows from the facts that $\mathbf{X} \mathcal{B}_k$ is the best rank $k$ approximation to $\mathbf{X} \mathcal{B}$.  $\eqref{25003}$, $\eqref{25009}$, and $\eqref{24048}$ give the theorem.\\

\subsection{Proof of Theorem $\ref{theorem_2}$}

 By Theorem \ref{theorem_1}(a), $K\le p$. Since $p$ is fixed, $K$ is bounded. As in  $\eqref{10202}$, 
 \begin{align}
& \widehat{\mathbf{B}}=  \mathbf{B}+(\mathbf{S}\mathcal{B})(\mathbf{Z}\trans\boldsymbol{\varrho})\trans +(\mathbf{Z}\trans\boldsymbol{\varrho})(\mathbf{S}\mathcal{B})\trans +(\mathbf{Z}\trans\boldsymbol{\varrho})(\mathbf{Z}\trans\boldsymbol{\varrho})\trans . \label{10098}
\end{align} 
Because $\mathbf{S}$ is positive definite, let $\mathbf{S}^{1/2}$ denote the symmetric square root matrix of $\mathbf{S}$ which is positive definite and satisfies $\lambda_{min}(\mathbf{S}^{1/2})=\sqrt{\lambda_{min}(\mathbf{S})}\ge \sqrt{c_0}$ by the assumption of the theorem. Moreover, the inverse $\mathbf{S}^{-1/2}$ satisfies 
\begin{align}
& \|\mathbf{S}^{-1/2}\|\le\lambda_{max}(\mathbf{S}^{-1/2})=\lambda_{min}(\mathbf{S}^{1/2})^{-1}\le c_0^{-1/2}. \label{50011}
\end{align} 
 Because $\boldsymbol{\alpha}_k$, $1\le k\le K$, are the solutions to the generalized eigenvalue problems $\eqref{2220}$ with the maximum value $\mu_k(\boldsymbol{\Xi})$ by Theorem \ref{theorem_1}(a), we have $\mathbf{B}\boldsymbol{\alpha}_k=\mu_k(\boldsymbol{\Xi})\mathbf{S}\boldsymbol{\alpha}_k$. Define 
\begin{align}
& \mathbf{C}=\mathbf{S}^{-1/2}\mathbf{B}\mathbf{S}^{-1/2},\quad \widehat{\mathbf{C}}=\mathbf{S}^{-1/2}\widehat{\mathbf{B}}\mathbf{S}^{-1/2}. \label{50012}
\end{align} 
  Then the first $K$ eigenvectors of $\mathbf{C}$ are 
 \begin{align}
& \boldsymbol{\varphi}_1=\mathbf{S}^{1/2}\boldsymbol{\alpha}_1,\quad \boldsymbol{\varphi}_1=\mathbf{S}^{1/2}\boldsymbol{\alpha}_1,\cdots,\quad  \boldsymbol{\varphi}_K=\mathbf{S}^{1/2}\boldsymbol{\alpha}_K, \label{10101}
\end{align} 
with eigenvalues $\mu_1(\mathbf{C})=\mu_1(\boldsymbol{\Xi}), \cdots, \mu_K(\mathbf{C})=\mu_K(\boldsymbol{\Xi})$. Similarly, the first $K$ eigenvectors of $\widehat{\mathbf{C}}$ are 
\begin{align}
& \widehat{\boldsymbol{\varphi}}_1=\mathbf{S}^{1/2}\widehat{\boldsymbol{\alpha}}_1,\quad \widehat{\boldsymbol{\varphi}}_2=\mathbf{S}^{1/2}\widehat{\boldsymbol{\alpha}}_2, \cdots, \quad \widehat{\boldsymbol{\varphi}}_K=\mathbf{S}^{1/2}\widehat{\boldsymbol{\alpha}}_K, \label{10102}
\end{align}
with eigenvalues  $\widehat{\mu}_1(\widehat{\mathbf{C}}), \cdots, \widehat{\mu}_K(\widehat{\mathbf{C}})$ are also the maximum values of $\eqref{10096}$. $\boldsymbol{\varphi}_1,\cdots,\boldsymbol{\varphi}_K$, are orthogonal to each other with $\|\boldsymbol{\varphi}_k\|_2=1$. $\widehat{\boldsymbol{\varphi}}_1,\cdots,\widehat{\boldsymbol{\varphi}}_K$, are orthogonal to each other with $\|\widehat{\boldsymbol{\varphi}}_k\|_2=1$.
\begin{Lemma}\label{lemma_8}
 \begin{align}
& \|\widehat{\mathbf{C}}-\mathbf{C}\|=\sqrt{\mu_1(\boldsymbol{\Xi})}O_p(1/\sqrt{n}) .\label{10105}
\end{align} 
\end{Lemma}

Since $\widehat{\mu}_1(\widehat{\mathbf{C}})$ and $\mu_1(\mathbf{C})=\mu_1(\boldsymbol{\Xi})$ are the largest eigenvalues of $\widehat{\mathbf{C}}$ and $\mathbf{C}$, respectively, we have
 \begin{align}
 |\widehat{\mu}_1(\widehat{\mathbf{C}})-\mu_1(\boldsymbol{\Xi})|=|\widehat{\mu}_1(\widehat{\mathbf{C}})-\mu_1(\mathbf{C})|=\left|\|\widehat{\mathbf{C}}\|-\|\mathbf{C}\|\right|\notag\\
\le \|\widehat{\mathbf{C}}-\mathbf{C}\|=\sqrt{\mu_1(\boldsymbol{\Xi})}O_p(1/\sqrt{n}). \label{10106}
\end{align} 
Let $a_1=\widehat{\boldsymbol{\varphi}}_1\trans\boldsymbol{\varphi}_1$ and $\boldsymbol{\psi}=\widehat{\boldsymbol{\varphi}}_1-a_1\boldsymbol{\varphi}_1$. Then $\boldsymbol{\psi}$ is orthogonal to $\boldsymbol{\varphi}_1$ (which is the first eigenvector of $\mathbf{C}$). Hence, we have  $\|\mathbf{C}\boldsymbol{\psi}\|_2\le \mu_2(\boldsymbol{\Xi})\| \boldsymbol{\psi}\|_2$, which, together with $1=\|\widehat{\boldsymbol{\varphi}}_1\|^2_2=\|a_1\boldsymbol{\varphi}_1+\boldsymbol{\psi}\|^2_2=a_1^2 +\|\boldsymbol{\psi}\|^2_2$, imply that
 \begin{align}
  \widehat{\boldsymbol{\varphi}}_1\trans\mathbf{C}\widehat{\boldsymbol{\varphi}}_1=( a_1\boldsymbol{\varphi}_1 +\boldsymbol{\psi})\trans( a_1\mathbf{C}\boldsymbol{\varphi}_1 +\mathbf{C}\boldsymbol{\psi})\label{10108}\\
=( a_1\boldsymbol{\varphi}_1 +\boldsymbol{\psi})\trans( a_1\mu_1(\boldsymbol{\Xi})\boldsymbol{\varphi}_1 +\mathbf{C}\boldsymbol{\psi})=a_1^2\mu_1(\boldsymbol{\Xi})\boldsymbol{\varphi}_1\trans\boldsymbol{\varphi}_1+ \boldsymbol{\psi}\trans\mathbf{C}\boldsymbol{\psi}\notag\\
=a_1^2\mu_1(\boldsymbol{\Xi}) +\boldsymbol{\psi}\trans\mathbf{C}\boldsymbol{\psi}\le a_1^2\mu_1(\boldsymbol{\Xi})+\mu_2(\boldsymbol{\Xi})\|\boldsymbol{\psi}\|^2_2=a_1^2\mu_1(\boldsymbol{\Xi})+\mu_2(\boldsymbol{\Xi})(1-a_1^2).\notag
\end{align} 
On the other hand,
 \begin{align}
& \widehat{\boldsymbol{\varphi}}_1\trans\mathbf{C}\widehat{\boldsymbol{\varphi}}_1=\widehat{\boldsymbol{\varphi}}_1\trans\widehat{\mathbf{C}}\widehat{\boldsymbol{\varphi}}_1+\widehat{\boldsymbol{\varphi}}_1\trans(\mathbf{C}-\widehat{\mathbf{C}})\widehat{\boldsymbol{\varphi}}_1\notag\\
&=\widehat{\mu}_1(\widehat{\mathbf{C}})+\widehat{\boldsymbol{\varphi}}_1\trans(\mathbf{C}-\widehat{\mathbf{C}})\widehat{\boldsymbol{\varphi}}_1\ge\widehat{\mu}_1(\widehat{\mathbf{C}})-\|\widehat{\mathbf{C}}-\mathbf{C}\|. \label{10109}
\end{align} 
$\eqref{10108}$-$\eqref{10109}$ imply $\widehat{\mu}_1(\widehat{\mathbf{C}})-\|\widehat{\mathbf{C}}-\mathbf{C}\|\le a_1^2\mu_1(\boldsymbol{\Xi})+\mu_2(\boldsymbol{\Xi})(1-a_1^2)$ which, combined with $\eqref{10106}$ and $\eqref{10105}$, give $  (\mu_1(\boldsymbol{\Xi})-\mu_2(\boldsymbol{\Xi}))(1-a_1^2)\le \mu_1(\boldsymbol{\Xi})-\widehat{\mu}_1(\widehat{\mathbf{C}})+\|\widehat{\mathbf{C}}-\mathbf{C}\|\le \sqrt{\mu_1(\boldsymbol{\Xi})}O_p(1/\sqrt{n})$. Then by Condition \ref{condition_2} (a), 
\begin{align*}
 (1-a_1^2) \le (\mu_1(\boldsymbol{\Xi})-\mu_2(\boldsymbol{\Xi}))^{-1}\sqrt{\mu_1(\boldsymbol{\Xi})}O_p(1/\sqrt{n})\\
\le c_2^{-1}\mu_1(\boldsymbol{\Xi})^{-1}\sqrt{\mu_1(\boldsymbol{\Xi})}O_p(1/\sqrt{n})=\mu_1(\boldsymbol{\Xi})^{-1/2}O_p(1/\sqrt{n}).
 \end{align*} 
 Without loss of generality, we assume that $a_1=\widehat{\boldsymbol{\varphi}}_1\trans\boldsymbol{\varphi}_1\ge 0$. Then
 \begin{align}
& 0\le (1-a_1)\le (1-a_1^2)\le \mu_1(\boldsymbol{\Xi})^{-1/2}O_p(1/\sqrt{n}). \label{10110}
\end{align}
Now  
 \begin{align*}
&   \mu_1(\boldsymbol{\Xi})(\widehat{\boldsymbol{\varphi}}_1-\boldsymbol{\varphi}_1)\\
=&(\mu_1(\boldsymbol{\Xi})-\widehat{\mu}_1(\widehat{\mathbf{C}}))\widehat{\boldsymbol{\varphi}}_1+\widehat{\mu}_1(\widehat{\mathbf{C}})\widehat{\boldsymbol{\varphi}}_1-\mu_1(\boldsymbol{\Xi})(1-a_1)\boldsymbol{\varphi}_1-\mu_1(\boldsymbol{\Xi})a_1\boldsymbol{\varphi}_1  \notag\\
=&(\mu_1(\boldsymbol{\Xi})-\widehat{\mu}_1(\widehat{\mathbf{C}}))\widehat{\boldsymbol{\varphi}}_1+ \widehat{\mathbf{C}} \widehat{\boldsymbol{\varphi}}_1-\mu_1(\boldsymbol{\Xi})(1-a_1)\boldsymbol{\varphi}_1- a_1\mathbf{C}\boldsymbol{\varphi}_1  \notag\\
=&(\mu_1(\boldsymbol{\Xi})-\widehat{\mu}_1(\widehat{\mathbf{C}}))\widehat{\boldsymbol{\varphi}}_1-\mu_1(\boldsymbol{\Xi})(1-a_1)\boldsymbol{\varphi}_1- (\mathbf{C}-\widehat{\mathbf{C}})\widehat{\boldsymbol{\varphi}}_1+\mathbf{C}(\widehat{\boldsymbol{\varphi}}_1-a_1\boldsymbol{\varphi}_1),  
\end{align*} 
which together with $\eqref{10105}$,$\eqref{10106}$ and $\eqref{10110}$, leads to
 \begin{align}
& \quad  \mu_1(\boldsymbol{\Xi})\|\widehat{\boldsymbol{\varphi}}_1-\boldsymbol{\varphi}_1\|_2\label{10112} \\
\le& |\mu_1(\boldsymbol{\Xi})-\widehat{\mu}_1(\widehat{\mathbf{C}})|\|\widehat{\boldsymbol{\varphi}}_1\|_2+\mu_1(\boldsymbol{\Xi})(1-a_1)\|\boldsymbol{\varphi}_1\|_2+\|\mathbf{C}-\widehat{\mathbf{C}}\|\notag\\
 &+\|\mathbf{C}(\widehat{\boldsymbol{\varphi}}_1-a_1\boldsymbol{\varphi}_1)\|_2\le \mu_1(\boldsymbol{\Xi})^{1/2}O_p(1/\sqrt{n})+\mu_2(\boldsymbol{\Xi})\| \widehat{\boldsymbol{\varphi}}_1-a_1\boldsymbol{\varphi}_1\|_2\notag\\
=&\mu_1(\boldsymbol{\Xi})^{1/2}O_p(1/\sqrt{n})+\mu_2(\boldsymbol{\Xi})\| \widehat{\boldsymbol{\varphi}}_1-\boldsymbol{\varphi}_1+(1-a_1)\boldsymbol{\varphi}_1\|_2\notag\\
\le &\mu_1(\boldsymbol{\Xi})^{1/2}O_p(1/\sqrt{n})+\mu_2(\boldsymbol{\Xi})(1-a_1)\|\boldsymbol{\varphi}_1\|_2 +\mu_2(\boldsymbol{\Xi})\| \widehat{\boldsymbol{\varphi}}_1-\boldsymbol{\varphi}_1\|_2\notag\\
\le &\mu_1(\boldsymbol{\Xi})^{1/2}O_p(1/\sqrt{n})+\mu_2(\boldsymbol{\Xi})\| \widehat{\boldsymbol{\varphi}}_1-\boldsymbol{\varphi}_1\|_2.\notag
\end{align} 
It follows from $\eqref{10112}$ that
 \begin{align}
  & \|\widehat{\boldsymbol{\varphi}}_1-\boldsymbol{\varphi}_1\|_2\label{10113}\\
\le&  (\mu_1(\boldsymbol{\Xi})-\mu_2(\boldsymbol{\Xi}))^{-1} \mu_1(\boldsymbol{\Xi})^{1/2}O_p(1/\sqrt{n})= \mu_1(\boldsymbol{\Xi})^{-1/2}O_p(1/\sqrt{n}), \notag
\end{align} 
where the last inequality is due to Condition \ref{condition_2} (a).

To estimate $\|\widehat{\boldsymbol{\varphi}}_2-\boldsymbol{\varphi}_2\|_2$, we define
 \begin{align}
&  \mathbf{C}_1=\mathbf{C}-\mu_1(\boldsymbol{\Xi})\boldsymbol{\varphi}_1\trans \boldsymbol{\varphi}_1, \quad \widehat{\mathbf{C}}_1=\widehat{\mathbf{C}}-\widehat{\mu}_1(\widehat{\mathbf{C}})\widehat{\boldsymbol{\varphi}}_1\trans\widehat{\boldsymbol{\varphi}}_1.\label{10114} 
\end{align} 
Then it can be seen that the first eigenvalues of eigenvectors of $\mathbf{C}_1$ and $\widehat{\mathbf{C}}_1$ are $(\mu_2(\boldsymbol{\Xi}), \boldsymbol{\varphi}_2)$ and $(\widehat{\mu}_2(\widehat{\mathbf{C}}), \widehat{\boldsymbol{\varphi}}_2)$, respectively. Moreover, it follows $\eqref{10105}$,$\eqref{10106}$ and $\eqref{10113}$ that $\|\widehat{\mathbf{C}}_1-\mathbf{C}_1\|=\sqrt{\mu_1(\boldsymbol{\Xi})}O_p(1/\sqrt{n})=\sqrt{\mu_2(\boldsymbol{\Xi})}O_p(1/\sqrt{n})$, where the last equality is due to Condition \ref{condition_2} (b). Hence, by using the similar arguments, we can obtain 
 \begin{align*}
&   |\widehat{\mu}_2(\widehat{\mathbf{C}})-\mu_2(\boldsymbol{\Xi})|\le  \mu_2(\boldsymbol{\Xi})^{-1/2}O_p(1/\sqrt{n})=\mu_1(\boldsymbol{\Xi})^{-1/2}O_p(1/\sqrt{n}),\\
 &\|\widehat{\boldsymbol{\varphi}}_2-\boldsymbol{\varphi}_2\|_2\le   \mu_1(\boldsymbol{\Xi})^{-1/2}O_p(1/\sqrt{n}).
\end{align*} 
By the same arguments, we can iteratively obtain that for all $1\le k\le K$,
 \begin{align}
&   |\widehat{\mu}_k(\widehat{\mathbf{C}})-\mu_k(\boldsymbol{\Xi})|\le \mu_1(\boldsymbol{\Xi})^{-1/2}O_p(1/\sqrt{n}),\notag\\
 &\|\widehat{\boldsymbol{\varphi}}_k-\boldsymbol{\varphi}_k\|_2\le   \mu_1(\boldsymbol{\Xi})^{-1/2}O_p(1/\sqrt{n}).\label{10118}
\end{align} 
  Therefore, for any $1\le k\le K$, by $\eqref{10101}$ and $\eqref{10102}$,
 \begin{align}
&   \|\widehat{\boldsymbol{\alpha}}_k-\boldsymbol{\alpha}_k\|_2=\|\mathbf{S}^{-1/2}\widehat{\boldsymbol{\varphi}}_k-\mathbf{S}^{-1/2}\boldsymbol{\varphi}_k\|_2\notag\\
\le& \|\mathbf{S}^{-1/2}\| \|\widehat{\boldsymbol{\varphi}}_k-\boldsymbol{\varphi}_k\|_2\le   \mu_1(\boldsymbol{\Xi})^{-1/2}O_p(1/\sqrt{n}).\label{10115}
\end{align} 
By $\eqref{10116}$
 \begin{align}
& \widehat{\mathcal{B}}=\widehat{\boldsymbol{\alpha}}_1\widehat{\mathbf{w}}_1\trans+\widehat{\boldsymbol{\alpha}}_2\widehat{\mathbf{w}}_2\trans+\cdots+\widehat{\boldsymbol{\alpha}}_K\widehat{\mathbf{w}}_K\trans\notag\\
=&\frac{1}{n}(\widehat{\boldsymbol{\alpha}}_1\widehat{\mathbf{t}}_1\trans+\widehat{\boldsymbol{\alpha}}_2\widehat{\mathbf{t}}_2\trans+\cdots+\widehat{\boldsymbol{\alpha}}_K\widehat{\mathbf{t}}_K\trans)  (\mathbf{Y}-\mathbf{1}_n\bar{\mathbf{y}}\trans)\notag\\
=&\frac{1}{n} (\widehat{\boldsymbol{\alpha}}_1\widehat{\boldsymbol{\alpha}}_1\trans+\cdots+\widehat{\boldsymbol{\alpha}}_K\widehat{\boldsymbol{\alpha}}_K\trans) \mathbf{X}\trans (\mathbf{X}\mathcal{B}+\boldsymbol{\varepsilon}-\bar{\boldsymbol{\varepsilon}}\mathbf{1}_n)\notag\\
=&(\widehat{\boldsymbol{\alpha}}_1\widehat{\boldsymbol{\alpha}}_1\trans+\cdots+\widehat{\boldsymbol{\alpha}}_K\widehat{\boldsymbol{\alpha}}_K\trans) \mathbf{Z}\trans (\mathbf{Z}\mathcal{B}+\boldsymbol{\varrho})\notag\\
=&(\widehat{\boldsymbol{\alpha}}_1\widehat{\boldsymbol{\alpha}}_1\trans+\cdots+\widehat{\boldsymbol{\alpha}}_K\widehat{\boldsymbol{\alpha}}_K\trans) \mathbf{S}\mathcal{B}+ (\widehat{\boldsymbol{\alpha}}_1\widehat{\boldsymbol{\alpha}}_1\trans+\cdots+\widehat{\boldsymbol{\alpha}}_K\widehat{\boldsymbol{\alpha}}_K\trans)  (\mathbf{Z}\trans\boldsymbol{\varrho}).\label{10117}
\end{align} 
Note that 
 \begin{align}
\widehat{\boldsymbol{\alpha}}_1\widehat{\boldsymbol{\alpha}}_1\trans+\cdots+\widehat{\boldsymbol{\alpha}}_K\widehat{\boldsymbol{\alpha}}_K\trans=\mathbf{S}^{-1/2}(\widehat{\boldsymbol{\varphi}}_1\widehat{\boldsymbol{\varphi}}_1\trans+\cdots+\widehat{\boldsymbol{\varphi}}_K\widehat{\boldsymbol{\varphi}}_K\trans)\mathbf{S}^{-1/2}\notag\\
=\mathbf{S}^{-1/2}\widehat{\mathbf{P}}\mathbf{S}^{-1/2}\label{50013}
\end{align} 
 where $\widehat{\mathbf{P}}=\widehat{\boldsymbol{\varphi}}_1\widehat{\boldsymbol{\varphi}}_1\trans+\cdots+\widehat{\boldsymbol{\varphi}}_K\widehat{\boldsymbol{\varphi}}_K\trans$ is the orthogonal projection matrix onto the subspace spanned by $\{\widehat{\boldsymbol{\varphi}}_1,\cdots, \widehat{\boldsymbol{\varphi}}_K\}$. Similarly, let $ \mathbf{P}=\boldsymbol{\varphi}_1\boldsymbol{\varphi}_1\trans+\cdots+\boldsymbol{\varphi}_K\boldsymbol{\varphi}_K\trans$ be the orthogonal projection matrix onto the subspace spanned by $\{\boldsymbol{\varphi}_1,\cdots, \boldsymbol{\varphi}_K\}$. It follows from $\eqref{10118}$ that  
 \begin{align}
& \|\widehat{\mathbf{P}}-\mathbf{P}\|=\mu_1(\boldsymbol{\Xi})^{-1/2}O_p(1/\sqrt{n}).\label{10119}
\end{align} 
Note that 
\begin{align}
\mathbf{S}^{1/2}\mathcal{B}&=\mathbf{S}^{1/2}\boldsymbol{\alpha}_1\mathbf{w}_1\trans+\mathbf{S}^{1/2}\boldsymbol{\alpha}_2\mathbf{w}_2\trans+\cdots+\mathbf{S}^{1/2}\boldsymbol{\alpha}_K\mathbf{w}_K\trans\notag\\
&=\boldsymbol{\varphi}_1\mathbf{w}_1\trans+\boldsymbol{\varphi}_2\mathbf{w}_2\trans+\cdots+\boldsymbol{\varphi}_K\mathbf{w}_K\trans,\label{55013}
\end{align} 
 which is a linear combination of  $\{\boldsymbol{\varphi}_1,\cdots, \boldsymbol{\varphi}_K\}$. Therefore, we have $\mathbf{P}\mathbf{S}^{1/2}\mathcal{B}=\mathbf{S}^{1/2}\mathcal{B}$, which together with $\eqref{10117}$ and $\eqref{50013}$ lead to
 \begin{align*}
& \widehat{\mathcal{B}}=\mathbf{S}^{-1/2}\widehat{\mathbf{P}}\mathbf{S}^{-1/2} \mathbf{S}\mathcal{B}+ \mathbf{S}^{-1/2}\widehat{\mathbf{P}}\mathbf{S}^{-1/2} (\mathbf{Z}\trans\boldsymbol{\varrho})\notag\\
&=\mathbf{S}^{-1/2}(\widehat{\mathbf{P}}-\mathbf{P})\mathbf{S}^{1/2}\mathcal{B}+ \mathbf{S}^{-1/2}\mathbf{P}\mathbf{S}^{1/2}\mathcal{B}+\mathbf{S}^{-1/2}\widehat{\mathbf{P}}\mathbf{S}^{-1/2} (\mathbf{Z}\trans\boldsymbol{\varrho})\notag\\
&=\mathbf{S}^{-1/2}(\widehat{\mathbf{P}}-\mathbf{P})\mathbf{S}^{1/2}\mathcal{B}+ \mathbf{S}^{-1/2} \mathbf{S}^{1/2}\mathcal{B}+\mathbf{S}^{-1/2}\widehat{\mathbf{P}}\mathbf{S}^{-1/2} (\mathbf{Z}\trans\boldsymbol{\varrho})\notag\\
&=\mathcal{B}+\mathbf{S}^{-1/2}(\widehat{\mathbf{P}}-\mathbf{P})\mathbf{S}^{1/2}\mathcal{B} +\mathbf{S}^{-1/2}\widehat{\mathbf{P}}\mathbf{S}^{-1/2} (\mathbf{Z}\trans\boldsymbol{\varrho}).
\end{align*} 
Therefore, by Lemma \ref{lemma_18} and $\eqref{10119}$, 
 \begin{align}
& \|\widehat{\mathcal{B}}-\mathcal{B}\|_F\label{10120}\\
\le &\|\mathbf{S}^{-1/2}(\widehat{\mathbf{P}}-\mathbf{P})\mathbf{S}^{1/2}\mathcal{B}\|_F +\|\mathbf{S}^{-1/2}\widehat{\mathbf{P}}\mathbf{S}^{-1/2} (\mathbf{Z}\trans\boldsymbol{\varrho})\|_F\notag\\
\le& \|\mathbf{S}^{-1/2}\|\|\widehat{\mathbf{P}}-\mathbf{P}\|\|\mathbf{S}^{1/2}\mathcal{B}\|_F +\|\mathbf{S}^{-1/2}\|\|\widehat{\mathbf{P}}\|\|\mathbf{S}^{-1/2}\|\| (\mathbf{Z}\trans\boldsymbol{\varrho})\|_F\notag\\
\le &\mu_1(\boldsymbol{\Xi})^{-1/2}O_p(1/\sqrt{n})\mu_1(\boldsymbol{\Xi})^{1/2}+O_p(1/\sqrt{n})=O_p(1/\sqrt{n}),\notag
\end{align} 
where we use, by $\eqref{55013}$,
\begin{align*}
& \|\mathbf{S}^{1/2}\mathcal{B}\|_F=\|\boldsymbol{\varphi}_1\mathbf{w}_1\trans+\boldsymbol{\varphi}_2\mathbf{w}_2\trans+\cdots+\boldsymbol{\varphi}_K\mathbf{w}_K\trans\|_F\notag\\
\le& \sum_{i=1}^K\|\boldsymbol{\varphi}_i\mathbf{w}_i\trans\|_F\le \sum_{i=1}^K\|\boldsymbol{\varphi}_i\|_2\|\mathbf{w}_i \|_2=\sum_{i=1}^K\mu_i(\boldsymbol{\Xi})^{1/2}\le K\mu_1(\boldsymbol{\Xi})^{1/2}
\end{align*}
and  by  $\eqref{10104}$,
 \begin{align*}
& \| (\mathbf{Z}\trans\boldsymbol{\varrho})\|_F^2=\sum_{j=1}^p\|(\mathbf{Z}\trans\boldsymbol{\varrho})_j\|_2^2\le p\max_{1\le j\le p}\|(\mathbf{Z}\trans\boldsymbol{\varrho})_j\|_2^2=O_p(1/n).
\end{align*} 

\vspace{10mm}

\subsection{Proof of Theorem $\ref{theorem_3}$}
 
\subsubsection{Proof of Part (a) in Theorem $\ref{theorem_3}$}
 The proof  of Part (a) is broken into several steps. \\

\noindent {\bf  $\bullet$ Step 1: Provide two key inequalities which play the key roles.}\\

 Due to the constraints of $\eqref{2220}$ and $\eqref{1240}$, we have  
\begin{align*}
 \boldsymbol{\alpha}_k\trans \mathbf{S}\boldsymbol{\alpha}_k=1, \quad \widehat{\boldsymbol{\alpha}}_k\trans \mathbf{S}\widehat{\boldsymbol{\alpha}}_k=1,\quad 1\le k\le K.
\end{align*}
  Since $\widehat{\boldsymbol{\alpha}}_1$ is solution to 
 \begin{align*}
\max_{ \boldsymbol{\alpha}\trans \mathbf{S}\boldsymbol{\alpha}=1}\frac{\boldsymbol{\alpha}\trans \widehat{\mathbf{B}}\boldsymbol{\alpha}}{\boldsymbol{\alpha}\trans \mathbf{S}\boldsymbol{\alpha}+\tau^{(1)}\|\boldsymbol{\alpha}\|^2_{\lambda^{(1)}}}=\max_{ \boldsymbol{\alpha}\trans \mathbf{S}\boldsymbol{\alpha}=1}\frac{\boldsymbol{\alpha}\trans \widehat{\mathbf{B}}\boldsymbol{\alpha}}{1+\tau^{(1)}\|\boldsymbol{\alpha}\|^2_{\lambda^{(1)}}},
\end{align*} 
 we have
 \begin{align}
\frac{\boldsymbol{\alpha}_1\trans \widehat{\mathbf{B}}\boldsymbol{\alpha}_1}{1+\tau^{(1)}\|\boldsymbol{\alpha}_1\|^2_{\lambda^{(1)}}}\le \frac{\widehat{\boldsymbol{\alpha}}_1\trans \widehat{\mathbf{B}}\widehat{\boldsymbol{\alpha}}_1}{1+\tau^{(1)}\|\widehat{\boldsymbol{\alpha}}_1\|^2_{\lambda^{(1)}}}. \label{10018}
\end{align} 
On the other hand, because $\boldsymbol{\alpha}_1$ is the first eigenvector of the generalized eigenvalue problem $\eqref{2220}$, by Theorem \ref{theorem_1}(a), we have
 \begin{align}
\mu_1(\boldsymbol{\Xi})=\boldsymbol{\alpha}_1\trans \mathbf{B}\boldsymbol{\alpha}_1=\frac{\boldsymbol{\alpha}_1\trans \mathbf{B}\boldsymbol{\alpha}_1}{\boldsymbol{\alpha}_1\trans \mathbf{S}\boldsymbol{\alpha}_1}\ge \frac{\widehat{\boldsymbol{\alpha}}_1\trans \mathbf{B}\widehat{\boldsymbol{\alpha}}_1}{\widehat{\boldsymbol{\alpha}}_1\trans \mathbf{S}\widehat{\boldsymbol{\alpha}}_1}=\widehat{\boldsymbol{\alpha}}_1\trans \mathbf{B}\widehat{\boldsymbol{\alpha}}_1. \label{10019}
\end{align} 
Our proofs are built on the inequalities $\eqref{10018}$ and $\eqref{10019}$.\\

\noindent{\bf  $\bullet$ Step 2: Define an event $\Omega$ which has probability converging to one as $n,p\to\infty$. Then we will focus on the elements in this event.}\\

 Define the following $q$-dimensional vector and the $n\times q$ matrix,  
\begin{align}
 \bar{\boldsymbol{\varepsilon}}=\frac{1}{n}\sum_{i=1}^n\boldsymbol{\varepsilon}_i,\quad \boldsymbol{\varrho}=\frac{1}{\sqrt{n}}(\boldsymbol{\varepsilon}-\mathbf{1}_n\bar{\boldsymbol{\varepsilon}}\trans).\label{80010}
\end{align} 
Then by the definition $\eqref{10009}$ of $\widehat{\mathbf{B}}$ and the definition of $\mathbf{B}$ in $\eqref{10008}$, 
 \begin{align}
& \widehat{\mathbf{B}}= \frac{1}{n^2}\mathbf{X}\trans(\mathbf{Y}-\mathbf{1}_n\bar{\mathbf{y}}\trans)(\mathbf{Y}-\mathbf{1}_n\bar{\mathbf{y}}\trans)\trans\mathbf{X}\trans\notag\\
 =& \frac{1}{n^2}\mathbf{X}\trans(\mathbf{X}\mathcal{B}+\boldsymbol{\varepsilon}-\mathbf{1}_n\bar{\boldsymbol{\varepsilon}}\trans)(\mathbf{X}\mathcal{B}+\boldsymbol{\varepsilon}- \mathbf{1}_n\bar{\boldsymbol{\varepsilon}}\trans)\trans \mathbf{X}\trans \notag\\
=&\left(\mathbf{Z}\trans\mathbf{Z}\mathcal{B}+\mathbf{Z}\trans\boldsymbol{\varrho}\right) (\mathbf{Z}\trans\mathbf{Z}\mathcal{B}+\mathbf{Z}\trans\boldsymbol{\varrho})\trans \notag\\
=&\mathbf{B}+(\mathbf{S}\mathcal{B})(\mathbf{Z}\trans\boldsymbol{\varrho})\trans +(\mathbf{Z}\trans\boldsymbol{\varrho})(\mathbf{S}\mathcal{B})\trans +(\mathbf{Z}\trans\boldsymbol{\varrho})(\mathbf{Z}\trans\boldsymbol{\varrho})\trans . \label{10202}
\end{align} 
\begin{Lemma}\label{lemma_2}
Suppose that Condition \ref{condition_1} holds and $p\ge 2$. Then for any $C>M_{\epsilon}/\sqrt{\log{p}}$, we have 
 \begin{align*}
&P\left(\max_{1\le j\le p}\|(\mathbf{Z}\trans\boldsymbol{\varrho})_j\|_2> C\sqrt{\frac{\log{p}}{n}}\right)\le pe^{ M_{\epsilon}^2/2\sigma^2}p^{-C^2/4\sigma^2},\notag\\
\text{and }\qquad &P\left(\bigcup_{k=1}^K\left\{\|(\mathbf{Z}\boldsymbol{\alpha}_k)\trans\boldsymbol{\varrho}\|_2>C\sqrt{\frac{\log{p}}{n}}\right\}\right)\le Ke^{ M_{\epsilon}^2/2\sigma^2}p^{-C^2/4\sigma^2},
\end{align*} 
where $(\mathbf{Z}\trans\boldsymbol{\varrho})_j$ denotes the $j$-th row of the $p\times q$ matrix $ \mathbf{Z}\trans\boldsymbol{\varrho} $.
\end{Lemma}
Note that $\varpi=\sqrt{\log{p}/n}$. Define the event
\begin{align}
\Omega=\left\{\max_{1\le j\le p}\|(\mathbf{Z}\trans\boldsymbol{\varrho})_j\|_2\le  C_0\varpi\right\}\bigcap_{k=1}^K\left\{\|(\mathbf{Z}\boldsymbol{\alpha}_k)\trans\boldsymbol{\varrho}\|_2\le  C_0\varpi\right\},  \label{1033}
\end{align}
where $C_0=2\max\{M_{\epsilon}/\sqrt{\log{2}}, 2\sigma\}$. Hence, by Lemma \ref{lemma_2}, we have 
$$P(\Omega)\ge 1-(p+K)e^{ M_{\epsilon}^2/2\sigma^2}p^{-C_0^2/4\sigma^2}\ge 1-2pe^{ M_{\epsilon}^2/2\sigma^2}p^{-C_0^2/4\sigma^2},$$
where we use the fact that $K\le p$.\\

\noindent{\bf  $\bullet$ Step 3: Provide upper bounds for  $\|\widehat{\boldsymbol{\alpha}}_1\|_1$ and   $\widehat{\boldsymbol{\alpha}}_1\trans \widehat{\mathbf{B}}\widehat{\boldsymbol{\alpha}}_1$.}\\

We first give a technical lemma, Lemma \ref{lemma_4}, and then provide the upper bounds in Lemma \ref{lemma_3}.
\begin{Lemma}\label{lemma_4}
Under Condition \ref{condition_3},  for any $1\le k\le K$, we have $\|\boldsymbol{\alpha}_k\|_2\le 1/\kappa$. Under Condition \ref{condition_2},  $\|\boldsymbol{\alpha}_k\|_1\ge 1$ and $\|\widehat{\boldsymbol{\alpha}}_k\|_1\ge 1$ for any $1\le k\le K$. Moreover, if both the two conditions are satisfied, we have $\|\boldsymbol{\alpha}_k\|_1^2\le   s/\kappa^2$.
\end{Lemma}
By the condition $\mu_1(\boldsymbol{\Xi})\ge  \hbar^2C_0^2\varpi^2s/\kappa^2$ and Lemma \ref{lemma_4}, we have
 \begin{align}
&  C_0\varpi\|\boldsymbol{\alpha}_1\|_1\le C_0\varpi\sqrt{s}/\kappa\le \hbar^{-1}  \mu_1(\boldsymbol{\Xi})^{1/2}, \quad \text{and hence,}\notag\\
& \tau^{(1)}\|\boldsymbol{\alpha}_1\|_{\lambda^{(1)}}^2\le \tau^{(1)}\|\boldsymbol{\alpha}_1\|_1^2= \frac{A^{(1)}C_0\varpi\|\boldsymbol{\alpha}_1\|_1^2}{\|\boldsymbol{\alpha}_1\|_1\sqrt{\mu_1(\boldsymbol{\Xi})}}\le   A^{(1)} \hbar^{-1}\;. \label{10203}
\end{align} 

\begin{Lemma}\label{lemma_3}
In the event $\Omega$,  there exist $(A_1^{L})^\prime$ and $(\hbar_0)^\prime$ only depending on $c$ such that for any $A^{(1)}\ge (A_1^{L})^\prime $ and $\hbar\ge (\hbar_0)^\prime$,  we have  
   \begin{align*}
 &\|\widehat{\boldsymbol{\alpha}}_1\|_1\le D_1\|\boldsymbol{\alpha}_1\|_1,\quad \widehat{\boldsymbol{\alpha}}_1\trans \widehat{\mathbf{B}}\widehat{\boldsymbol{\alpha}}_1 \le(1+\hbar^{-1}D_1)^2\mu_1(\boldsymbol{\Xi}),\\
&\text{ and }\quad \boldsymbol{\alpha}_1\trans \widehat{\mathbf{B}}\boldsymbol{\alpha}_1-\widehat{\boldsymbol{\alpha}}_1\trans \widehat{\mathbf{B}}\widehat{\boldsymbol{\alpha}}_1\notag\\
&\ge \frac{1}{2}c_2\mu_1(\boldsymbol{\Xi})\|\widehat{\boldsymbol{\gamma}}_1-\boldsymbol{\gamma}_1\|_2^2-[4+\hbar^{-1} (1+D_1)]\sqrt{\mu_1(\boldsymbol{\Xi})}C_0\varpi\|\boldsymbol{\alpha}_1-\widehat{\boldsymbol{\alpha}}_1\|_1,
\end{align*} 
where $D_1=\sqrt{6c}$.
\end{Lemma}

\noindent{\bf  $\bullet$ Step 4: Derive the oracle inequalities in Part (a)}.\\

Now by $\eqref{10018}$, we have $\boldsymbol{\alpha}_1\trans \widehat{\mathbf{B}}\boldsymbol{\alpha}_1(1+\tau^{(1)}\|\widehat{\boldsymbol{\alpha}}_1\|^2_{\lambda^{(1)}})\le  \widehat{\boldsymbol{\alpha}}_1\trans \widehat{\mathbf{B}}\widehat{\boldsymbol{\alpha}}_1(1+\tau^{(1)}\|\boldsymbol{\alpha}_1\|^2_{\lambda^{(1)}})$ which leads to,  
 \begin{align}
  &\widehat{\boldsymbol{\alpha}}_1\trans \widehat{\mathbf{B}}\widehat{\boldsymbol{\alpha}}_1(\tau^{(1)}\|\boldsymbol{\alpha}_1\|^2_{\lambda^{(1)}}-\tau^{(1)}\|\widehat{\boldsymbol{\alpha}}_1\|^2_{\lambda^{(1)}})\label{10029}\\
	\ge& (\boldsymbol{\alpha}_1\trans \widehat{\mathbf{B}}\boldsymbol{\alpha}_1-\widehat{\boldsymbol{\alpha}}_1\trans \widehat{\mathbf{B}}\widehat{\boldsymbol{\alpha}}_1)(1+ \tau^{(1)}\|\widehat{\boldsymbol{\alpha}}_1\|^2_{\lambda^{(1)}})\notag\\
\ge &\left(\frac{1}{2}c_2\mu_1(\boldsymbol{\Xi})\|\widehat{\boldsymbol{\gamma}}_1-\boldsymbol{\gamma}_1\|_2^2-[4+\hbar^{-1} (1+D_1)]C_0\sqrt{\mu_1(\boldsymbol{\Xi})}\varpi\|\boldsymbol{\alpha}_1-\widehat{\boldsymbol{\alpha}}_1\|_1\right)\notag\\
&\times (1+ \tau^{(1)}\|\widehat{\boldsymbol{\alpha}}\|^2_{\lambda^{(1)}})\ge \frac{1}{2}c_2\mu_1(\boldsymbol{\Xi})\|\widehat{\boldsymbol{\gamma}}_1-\boldsymbol{\gamma}_1\|_2^2\notag\\
&-[4+\hbar^{-1} (1+D_1)]C_0\sqrt{\mu_1(\boldsymbol{\Xi})}\varpi\|\boldsymbol{\alpha}_1-\widehat{\boldsymbol{\alpha}}_1\|_1(1+ D_1^2A^{(1)}\hbar^{-1})\notag,
\end{align} 
where the third inequality follows from the last inequality in Lemma \ref{lemma_3} and the last one is due to the first inequality in Lemma \ref{lemma_3} and $\eqref{10203}$. We  provide upper bounds for $\widehat{\boldsymbol{\alpha}}_1\trans \widehat{\mathbf{B}}\widehat{\boldsymbol{\alpha}}_1(\tau^{(1)}\|\boldsymbol{\alpha}_1\|^2_{\lambda^{(1)}}-\tau^{(1)}\|\widehat{\boldsymbol{\alpha}}_1\|^2_{\lambda^{(1)}})$ in the following lemma.
\begin{Lemma}\label{lemma_1}
In the event $\Omega$,  we have two alternative cases:
\begin{itemize}
\item \text{\bf Case (a)}: $\|(\boldsymbol{\alpha}_1-\widehat{\boldsymbol{\alpha}}_1)_{J_1}\|_1<\lambda^{(1)} \|(\widehat{\boldsymbol{\alpha}}_1)_{J_1^c}\|_1$. In this case, we have 
 \begin{align}
 &\widehat{\boldsymbol{\alpha}}_1\trans \widehat{\mathbf{B}}\widehat{\boldsymbol{\alpha}}_1(\tau^{(1)}\|\boldsymbol{\alpha}_1\|^2_{\lambda^{(1)}}-\tau^{(1)}\|\widehat{\boldsymbol{\alpha}}_1\|^2_{\lambda^{(1)}})\notag\\
\le& (1-2\hbar^{-1}) A^{(1)}C_0\mu_1(\boldsymbol{\Xi})^{1/2}\varpi(\|(\boldsymbol{\alpha}_1-\widehat{\boldsymbol{\alpha}}_1)_{J_1}\|_1-\lambda^{(1)}\|(\widehat{\boldsymbol{\alpha}}_1)_{J_1^c}\|_1) .\notag
\end{align} 
\item \text{\bf Case (b)}: $\|(\boldsymbol{\alpha}_1-\widehat{\boldsymbol{\alpha}}_1)_{J_1}\|_1\ge \lambda^{(1)} \|(\widehat{\boldsymbol{\alpha}}_1)_{J_1^c}\|_1$. In this case, we have  
   \begin{align*}
 &  \widehat{\boldsymbol{\alpha}}_1\trans \widehat{\mathbf{B}}\widehat{\boldsymbol{\alpha}}_1(\tau^{(1)}\|\boldsymbol{\alpha}_1\|^2_{\lambda^{(1)}}-\tau^{(1)}\|\widehat{\boldsymbol{\alpha}}_1\|^2_{\lambda^{(1)}})\notag\\
\le &  (1+\hbar^{-1}D_1)^2(1 +D_1) A^{(1)}C_0\mu_1(\boldsymbol{\Xi})^{1/2}\varpi\|(\boldsymbol{\alpha}_1-\widehat{\boldsymbol{\alpha}}_1)_{J_1}\|_1,\notag
\end{align*} 
\end{itemize}
where $J_1=J(\boldsymbol{\alpha}_1)$ is the collection of the nonzero coordinates of $\boldsymbol{\alpha}_1$ and $J_1^c$ is its complement set in the set $\{1,2,\cdots, p\}$.
\end{Lemma}

In the following, we will consider the two cases separately.

\vspace{5mm}

\noindent {\bf Case (a)}.  By $\eqref{10029}$ and Lemma \ref{lemma_1},
 \begin{align*}
 & \frac{1}{2}c_2\mu_1(\boldsymbol{\Xi})\|\widehat{\boldsymbol{\gamma}}_1-\boldsymbol{\gamma}_1\|_2^2\notag\\
\le& [4+\hbar^{-1} (1+D_1)] (1+ D_1^2A^{(1)}\hbar^{-1})C_0 \sqrt{\mu_1(\boldsymbol{\Xi})}\varpi\|\boldsymbol{\alpha}_1-\widehat{\boldsymbol{\alpha}}_1\|_1 \notag\\
&+ (1-2\hbar^{-1}) A^{(1)}C_0\sqrt{\mu_1(\boldsymbol{\Xi})}\varpi\|(\widehat{\boldsymbol{\alpha}}_1-\boldsymbol{\alpha}_1)_{J_1}\|_1  \notag\\
&-(1-2\hbar^{-1}) A^{(1)}C_0\sqrt{\mu_1(\boldsymbol{\Xi})}\varpi\lambda^{(1)}\|(\widehat{\boldsymbol{\alpha}}_1)_{J_1^c}\|_1,
\end{align*} 
 which together  the equality $\|\widehat{\boldsymbol{\alpha}}_1-\boldsymbol{\alpha}_1\|_1=\|(\widehat{\boldsymbol{\alpha}}_1-\boldsymbol{\alpha}_1)_{J_1}\|_1+\|(\widehat{\boldsymbol{\alpha}}_1)_{J_1^c}\|_1$,  leads to
 \begin{align}
    0\le& \frac{1}{2}c_2\mu_1(\boldsymbol{\Xi})\|\widehat{\boldsymbol{\gamma}}_1-\boldsymbol{\gamma}_1\|_2^2\notag\\
\le&  \left[(1-2\hbar^{-1}) A^{(1)}+[4+\hbar^{-1} (1+D_1)] (1+ D_1^2A^{(1)}\hbar^{-1}) \right]\notag\\
&\times \sqrt{\mu_1(\boldsymbol{\Xi})}C_0\varpi\|(\widehat{\boldsymbol{\alpha}}_1-\boldsymbol{\alpha}_1)_{J_1}\|_1\notag\\
&  -\left[(1-2\hbar^{-1}) A^{(1)}\lambda^{(1)}-[4+\hbar^{-1} (1+D_1)] (1+ D_1^2A^{(1)}\hbar^{-1})\right]\notag\\
&\times\sqrt{\mu_1(\boldsymbol{\Xi})}C_0\varpi\|(\widehat{\boldsymbol{\alpha}}_1)_{J_1^c}\|_1.\label{10034}
\end{align} 
 Therefore,  it follows from $\eqref{10034}$ that 
 \begin{align}
 &\left[(1-2\hbar^{-1}) A^{(1)}+[4+\hbar^{-1} (1+D_1)] (1+ D_1^2A^{(1)}\hbar^{-1}) \right]\notag\\
&\times \sqrt{\mu_1(\boldsymbol{\Xi})}C_0\varpi\|(\widehat{\boldsymbol{\alpha}}_1-\boldsymbol{\alpha}_1)_{J_1}\|_1\notag\\
\ge  &\left[(1-2\hbar^{-1}) A^{(1)}\lambda^{(1)}-[4+\hbar^{-1} (1+D_1)] (1+ D_1^2A^{(1)}\hbar^{-1})\right]\notag\\
&\times\sqrt{\mu_1(\boldsymbol{\Xi})}C_0\varpi\|(\widehat{\boldsymbol{\alpha}}_1)_{J_1^c}\|_1\;.\label{400} 
\end{align} 
Note that
 \begin{align*}
&\lim_{\substack{ A^{(1)}\to\infty\\\hbar\to\infty}}\frac{(1-2\hbar^{-1}) A^{(1)} +[4+\hbar^{-1} (1+D_1)] (1+ D_1^2A^{(1)}\hbar^{-1}) }{(1-2\hbar^{-1}) A^{(1)} \lambda^{(1)}-[4+\hbar^{-1} (1+D_1)] (1+ D_1^2A^{(1)}\hbar^{-1}) } \to ( \lambda^{(1)})^{-1} , 
 \end{align*}
and by $\eqref{10216}$, $( \lambda^{(1)})^{-1}<(c^{-1}+\delta_0)^{-1}<c$. Then by $\eqref{400}$, there exist $(A_1^{L})^{\prime\prime}$ and $(\hbar_0)^{\prime\prime}$ only depending on $D_1=\sqrt{6c}$ (Lemma \ref{lemma_3}) and $\delta_0$ such that for any $A^{(1)}\ge (A_1^{L})^{\prime\prime} $ and $\hbar\ge (\hbar_0)^{\prime\prime}$,   
 \begin{align}
 \|(\widehat{\boldsymbol{\alpha}}_1-\boldsymbol{\alpha}_1)_{J_1^c}\|_1=\|(\widehat{\boldsymbol{\alpha}}_1)_{J_1^c}\|_1< c\|(\widehat{\boldsymbol{\alpha}}_1-\boldsymbol{\alpha}_1)_{J_1}\|_1.\label{10035}
\end{align} 
 By Condition \ref{condition_3},  $\eqref{10035}$ and the inequality $|J_1|=|J(\boldsymbol{\alpha}_1)|\le |J(\mathcal{B})|=\mathcal{M}(\mathcal{B})=s$, we have 
{\small\begin{align}
 \kappa\|(\widehat{\boldsymbol{\alpha}}_1-\boldsymbol{\alpha}_1)_{J_1}\|_2\le \|\frac{\mathbf{X}}{\sqrt{n}}(\widehat{\boldsymbol{\alpha}}_1-\boldsymbol{\alpha}_1)\|_2=\|\mathbf{Z}(\widehat{\boldsymbol{\alpha}}_1-\boldsymbol{\alpha}_1)\|_2=\|\widehat{\boldsymbol{\gamma}}_1-\boldsymbol{\gamma}_1\|_2,\label{10036}
\end{align}}
which, together with $\eqref{10034}$ and the  Cauchy-Schwarz inequality, leads to
 \begin{align}
 &  \frac{1}{2}c_2\mu_1(\boldsymbol{\Xi})\|\widehat{\boldsymbol{\gamma}}_1-\boldsymbol{\gamma}_1\|_2^2\notag\\
\le& \left[(1-2\hbar^{-1}) A^{(1)}+[4+\hbar^{-1} (1+D_1)] (1+ D_1^2A^{(1)}\hbar^{-1}) \right]\notag\\
&\qquad \times\sqrt{\mu_1(\boldsymbol{\Xi})}C_0\varpi\|(\widehat{\boldsymbol{\alpha}}_1-\boldsymbol{\alpha}_1)_{J_1}\|_1\notag\\
\le& \left[(1-2\hbar^{-1}) A^{(1)}+[4+\hbar^{-1} (1+D_1)] (1+ D_1^2A^{(1)}\hbar^{-1}) \right]\notag\\
&\qquad \times\sqrt{\mu_1(\boldsymbol{\Xi})}C_0\varpi\sqrt{|J_1|}\|(\widehat{\boldsymbol{\alpha}}_1-\boldsymbol{\alpha}_1)_{J_1}\|_2\notag\\
\le &\left[(1-2\hbar^{-1}) A^{(1)}+[4+\hbar^{-1} (1+D_1)] (1+ D_1^2A^{(1)}\hbar^{-1}) \right]\notag\\
&\qquad \times\sqrt{\mu_1(\boldsymbol{\Xi})}C_0\varpi\sqrt{s}\|\widehat{\boldsymbol{\gamma}}_1-\boldsymbol{\gamma}_1\|_2/\kappa.\notag
\end{align} 
Then it follows that 
 \begin{align}
 & \|\widehat{\boldsymbol{\gamma}}_1-\boldsymbol{\gamma}_1\|_2\le D_2C_0\kappa^{-1}\mu_1(\boldsymbol{\Xi})^{-1/2}\varpi\sqrt{s},\notag\\
& \|\mathbf{Z}(\widehat{\boldsymbol{\alpha}}_1-\boldsymbol{\alpha}_1)\|_2^2=\|\widehat{\boldsymbol{\gamma}}_1-\boldsymbol{\gamma}_1\|_2^2\le D_2^2C_0^2\kappa^{-2}\mu_1(\boldsymbol{\Xi})^{-1}\varpi^2s,\notag\\
 &\|\mathbf{X}(\widehat{\boldsymbol{\alpha}}_1-\boldsymbol{\alpha}_1)\|_2^2=n\|\widehat{\boldsymbol{\gamma}}_1-\boldsymbol{\gamma}_1\|_2^2\le nD_2^2C_0^2\kappa^{-2}\mu_1(\boldsymbol{\Xi})^{-1}\varpi^2s, \label{10038}
\end{align} 
where $D_2=2c_2^{-1}\left[(1-2\hbar^{-1}) A^{(1)} +[4+\hbar^{-1} (1+D_1)] (1+ D_1^2A^{(1)}\hbar^{-1}) \right]$. By $\eqref{10035}$-$\eqref{10038}$,
 \begin{align}
 &\|\widehat{\boldsymbol{\alpha}}_1-\boldsymbol{\alpha}_1\|_1=\|(\widehat{\boldsymbol{\alpha}}_1-\boldsymbol{\alpha}_1)_{J_1^c}\|_1+\|(\widehat{\boldsymbol{\alpha}}_1-\boldsymbol{\alpha}_1)_{J_1}\|_1\notag\\
 <& (1+c)\|(\widehat{\boldsymbol{\alpha}}_1-\boldsymbol{\alpha}_1)_{J_1}\|_1\le (1+c)\sqrt{s}\|(\widehat{\boldsymbol{\alpha}}_1-\boldsymbol{\alpha}_1)_{J_1}\|_2\notag\\
 \le &(1+c)\sqrt{s}\kappa^{-1}\|\widehat{\boldsymbol{\gamma}}_1-\boldsymbol{\gamma}_1\|_2\le (1+c)D_2C_0\kappa^{-2} \mu_1(\boldsymbol{\Xi})^{-1/2}\varpi s.\label{10055}
\end{align} 
 Note that
\begin{align}
 & \lim_{\substack{ A^{(1)}\to\infty\\\hbar\to\infty}}\frac{D_2}{A^{(1)}}=2c_2^{-1}\le 2c_2^{-1}(1+D_1)=2c_2^{-1}(1+\sqrt{6c}).\label{50033}
\end{align}
 Hence, there exist $(A_1^{L})^{\prime\prime\prime}$ and $(\hbar_0)^{\prime\prime\prime}$ only depending on $\delta_0$, $c$ and $c_2$ such that for any $A^{(1)}\ge (A_1^{L})^{\prime\prime\prime} $ and $\hbar\ge (\hbar_0)^{\prime\prime\prime}$, we have  
 \begin{align*}
 & D_2\le 4c_2^{-1}(1+\sqrt{6c})A^{(1)},
\end{align*} 
and the inequalities in Theorem \ref{theorem_3} (a) follow from $\eqref{10038}$ and $\eqref{10055}$ in this case.
\vspace{5mm}

\noindent {\bf Case (b)}. The arguments are similar to Case (a). We summarize the results in the following lemma and provide the proof of the lemma in supplementary materials. 
\begin{Lemma}\label{lemma_20}
In the event $\Omega$ and Case (b), there exist $(A_1^{L})^{\prime\prime\prime\prime}$ and $(\hbar_0)^{\prime\prime\prime\prime}$ only depending on $\delta_0$, $c$ and $c_2$ such that for any $A^{(1)}\ge (A_1^{L})^{\prime\prime\prime\prime}$ and $\hbar\ge (\hbar_0)^{\prime\prime\prime\prime}$,  the inequalities in Theorem \ref{theorem_3} (a) hold.  
\end{Lemma}

Finally, we choose 
$$A_1^{L}=\max\{(A_1^{L})^{\prime}, (A_1^{L})^{\prime\prime}, (A_1^{L})^{\prime\prime\prime},(A_1^{L})^{\prime\prime\prime\prime}\},\quad h_0=\max\{(h_0)^{\prime}, (h_0)^{\prime\prime}, (h_0)^{\prime\prime\prime}, (h_0)^{\prime\prime\prime\prime}\}.$$
 Then it can be seen that $A_1^{L}$ and $h_0$ only depend on $c$, $c_2$ and $\delta_0$ and the inequalities in Theorem \ref{theorem_3} (a) hold if $A^{(1)}\ge A_1^{L} $ and $\hbar\ge \hbar_0$. The proof of Theorem \ref{theorem_3} (a) is completed.

\subsubsection{Proof of Part (b) in Theorem $\ref{theorem_3}$}
In this proof, we will proceed by induction to prove a set of inequalities from which Part (b) follows.  The basis step of the induction (that is, $k=1$) will follow from Part (a). Then we will assume that the inequalities hold for all $1\le i\le k-1$, based on which we will use the similar ideas as in the proof of Part (a) to show that these inequalities are also true for $i=k$. In this proof, we only consider the elements in the event $\Omega$ defined in the proof of Part (a). \\

\noindent {\bf  $\bullet$ Step 1: Introduction of some notations and inequalities.}\\

By the first inequality in the condition $\eqref{359}$, for any $1\le k\le K$, we have 
 \begin{align}
   &c_4^{-1} \| \boldsymbol{\alpha}_1\|_1\le \| \boldsymbol{\alpha}_k\|_1\le c_4 \| \boldsymbol{\alpha}_1\|_1,\label{360}
\end{align} 
which together with the condition $\eqref{50020}$ and Lemma \ref{lemma_4} lead to
 \begin{align}
&  C_0\varpi\|\boldsymbol{\alpha}_k\|_1\le C_0\varpi\sqrt{s}/\kappa\le \hbar^{-1}  \mu_1(\boldsymbol{\Xi})^{1/2}, \label{10247}\\
& \tau^{(k)}\|\boldsymbol{\alpha}_k\|_{\lambda^{(k)}}^2\le \tau^{(k)}\|\boldsymbol{\alpha}_k\|_1^2\le c_4^2\tau^{(k)}\|\boldsymbol{\alpha}_1\|_1^2\notag\\
&\qquad = \frac{c_4^2A^{(k)}C_0\varpi\|\boldsymbol{\alpha}_1\|_1^2}{\|\boldsymbol{\alpha}_1\|_1\sqrt{\mu_1(\boldsymbol{\Xi})}}\le c_4^2  A^{(k)} \hbar^{-1}, \notag
\end{align} 
for any $1\le k\le K$. Note that 
 \begin{align}
&  \boldsymbol{\alpha}_k\trans \mathbf{S}\boldsymbol{\alpha}_k=\|\boldsymbol{\gamma}_k\|_2^2=1,\qquad \widehat{\boldsymbol{\alpha}}_k\trans \mathbf{S}\widehat{\boldsymbol{\alpha}}_k=\|\widehat{\boldsymbol{\gamma}}_k\|_2^2=1,\notag\\
&\boldsymbol{\alpha}_k\trans \mathbf{S}\boldsymbol{\alpha}_l= \boldsymbol{\gamma}_k\trans\boldsymbol{\gamma}_l=0,\qquad \widehat{\boldsymbol{\alpha}}_k\trans \mathbf{S}\widehat{\boldsymbol{\alpha}}_l= \widehat{\boldsymbol{\gamma}}_k\trans\widehat{\boldsymbol{\gamma}}_l=0, \label{50036}
\end{align} 
  for any $1\le k\neq l\le K$. Define the following subspaces which are spanned by different sets of vectors. The first two are subspaces in $\mathbb{R}^n$ and the last two in  $\mathbb{R}^p$. 
\begin{align*}
& \mathbf{V}_k=\text{span}\{\boldsymbol{\gamma}_1,\boldsymbol{\gamma}_2, \cdots, \boldsymbol{\gamma}_k\},\quad \widehat{\mathbf{V}}_k=\text{span}\{\widehat{\boldsymbol{\gamma}}_1,\widehat{\boldsymbol{\gamma}}_2, \cdots, \widehat{\boldsymbol{\gamma}}_k\}, \notag\\
	&\mathbf{W}_k={\rm span}\{\mathbf{S}\boldsymbol{\alpha}_1,\mathbf{S}\boldsymbol{\alpha}_2, \cdots, \mathbf{S}\boldsymbol{\alpha}_k\},\quad \widehat{\mathbf{W}}_k={\rm span}\{\mathbf{S}\widehat{\boldsymbol{\alpha}}_1,\mathbf{S}\widehat{\boldsymbol{\alpha}}_2, \cdots, \mathbf{S}\widehat{\boldsymbol{\alpha}}_k\},
\end{align*}
and 
\begin{align*}
& \mathbf{P}_k=\sum_{i=1}^{k} \boldsymbol{\gamma}_i\boldsymbol{\gamma}_i\trans,\quad \widehat{\mathbf{P}}_k=\sum_{i=1}^{k}\widehat{\boldsymbol{\gamma}}_i\widehat{\boldsymbol{\gamma}}_i\trans
\end{align*}
  be the orthogonal projection matrices onto   $\mathbf{V}_k$ and $\widehat{\mathbf{V}}_k$, respectively. Let
 \begin{align}
&\boldsymbol{\delta}_k=\sum_{i=1}^{k-1}(\boldsymbol{\gamma}_k\trans\widehat{\boldsymbol{\gamma}}_i)\widehat{\boldsymbol{\alpha}}_i,\quad \boldsymbol{\beta}_k=\boldsymbol{\alpha}_k-\boldsymbol{\delta}_k, \notag\\
 &\widehat{\boldsymbol{\delta}}_k= \sum_{i=1}^{k-1}(\widehat{\boldsymbol{\gamma}}_k\trans\boldsymbol{\gamma}_i)\boldsymbol{\alpha}_i,\quad  \widehat{\boldsymbol{\beta}}_k=\widehat{\boldsymbol{\alpha}}_k- \widehat{\boldsymbol{\delta}}_k,\label{378}
\end{align} 
which play an important role in the proof. We have $\mathbf{Z}\boldsymbol{\delta}_k=\sum_{i=1}^{k-1}(\boldsymbol{\gamma}_k\trans\widehat{\boldsymbol{\gamma}}_i)\widehat{\boldsymbol{\gamma}}_i=\widehat{\mathbf{P}}_{k-1}\boldsymbol{\gamma}_k$ and similarly,
 \begin{align}
  \mathbf{Z}\boldsymbol{\beta}_k=(\mathbf{I}-\widehat{\mathbf{P}}_{k-1})\boldsymbol{\gamma}_k, \quad \mathbf{Z}\widehat{\boldsymbol{\delta}}_k= \mathbf{P}_{k-1}\widehat{\boldsymbol{\gamma}}_k,\quad  \mathbf{Z}\widehat{\boldsymbol{\beta}}_k=(\mathbf{I}-\mathbf{P}_{k-1})\widehat{\boldsymbol{\gamma}}_k,\label{362}
\end{align} 
for all $1\le k\le K$. For any $1\le i\le k-1$, $(\mathbf{S}\widehat{\boldsymbol{\alpha}}_i)\trans \boldsymbol{\beta}_k=\widehat{\boldsymbol{\alpha}}_i\trans \mathbf{Z}\trans\mathbf{Z}\boldsymbol{\beta}_k=\widehat{\boldsymbol{\gamma}}_i\trans(\mathbf{I}-\widehat{\mathbf{P}}_{k-1})\boldsymbol{\gamma}_k=0$, therefore, we have 
\begin{align}
  \boldsymbol{\beta}_k\perp \widehat{\mathbf{W}}_{k-1}, \quad \text{\rm and similarly,}\quad  \widehat{\boldsymbol{\beta}}_k\perp \mathbf{W}_{k-1}, \label{50037}
\end{align}
 for any $1\le k\le K$.\\

 \noindent {\bf  $\bullet$ Step 2: Induction hypothesis.}\\

We will prove that we can find constants $\hbar_0$, $A^{L}_j < A^{U}_j$, $1\le j\le K$, which only depend on $\delta_0$, $c$, $c_2\sim c_5$, such that for any $1\le i\le K$, if $A^{L}_j\le A^{(j)}\le A^{U}_j$, $1\le j<i$,  $\hbar\ge \hbar_0$ and $A^{(i)}\ge A^{L}_i$, we have 
 \begin{align}
   &\|\widehat{\boldsymbol{\alpha}}_i\|_1\le   D_{i,1}\kappa^{-1}\sqrt{s} ,\label{2053}\\
	&\|\widehat{\boldsymbol{\gamma}}_i-\boldsymbol{\gamma}_i\|^2_2\le (A^{(i)})^2D_{i,2} C_0^2\kappa^{-2}\mu_1(\boldsymbol{\Xi})^{-1}\varpi^2 s, \notag\\ 
	&\|\widehat{\mathbf{P}}_i-\mathbf{P}_i\|^2_2\le (A^{(i)})^2D_{i,3} C_0^2\kappa^{-2}\mu_1(\boldsymbol{\Xi})^{-1}\varpi^2 s, \notag\\
	&\|\widehat{\boldsymbol{\alpha}}_i-\boldsymbol{\alpha}_i\|_1\le  D_{i,4}A^{(i)}\mu_1(\boldsymbol{\Xi})^{-1/2}\kappa^{-2}C_0 \varpi s\notag
\end{align} 
 where $D_{i,1}\sim D_{i,4}$ are constants only depending on $\delta_0$, $c$ and $c_2\sim c_5$. We will proceed by induction. \\

\noindent {\bf  $\bullet$ Step 3: Proof of $\eqref{2053}$ for $i=1$.}\\

When $i=1$, by Theorem \ref{theorem_3} (a), we can find constants $A_1^{L}$ and $\hbar_0^{(1)}$ such that the inequalities in $\eqref{2053}$ except the third one hold for any $A^{(1)}\ge A_1^{L} $ and $\hbar\ge \hbar_0^{(1)}$ with $D_{1,1}=\sqrt{6c}$, $D_{1,2}=16c_2^{-2}(1+\sqrt{6c})^2$ and $D_{1,4}= 4(1+c)(1+\sqrt{6c})c_2^{-1}$. The third inequality follows from 
 \begin{align*}
    &\|\widehat{\mathbf{P}}_1-\mathbf{P}_1\|^2_2=\|\widehat{\boldsymbol{\gamma}}_1\widehat{\boldsymbol{\gamma}}_1\trans-\boldsymbol{\gamma}_1\boldsymbol{\gamma}_1\trans\|^2\\
		\le& 2\|\widehat{\boldsymbol{\gamma}}_1-\boldsymbol{\gamma}_1\|^2_2\|\widehat{\boldsymbol{\gamma}}_1\|^2_2+2\|\widehat{\boldsymbol{\gamma}}_1-\boldsymbol{\gamma}_1\|^2_2\| \boldsymbol{\gamma}_1\|^2_2\le 4\|\widehat{\boldsymbol{\gamma}}_1-\boldsymbol{\gamma}_1\|^2_2
\end{align*} 
with $D_{1,3}=4D_{1,4}$. Therefore, $D_{1,1}\sim D_{1,4}$ only depends on $c$ and $c_2$. We arbitrarily choose a number $A^{U}_1$ such that $A^{U}_1> A^{L}_1$. \\

\noindent {\bf  $\bullet$ Step 4: Inductive step, the proof of $\eqref{2053}$ for $i=k$.}\\

 Now we assume that we have found constants $A^{L}_j<  A^{U}_j$, $1\le j\le k-1$ and  $\hbar_0^{(k-1)}$ which only depend on $\delta_0$, $c$, $c_2\sim c_5$, such that $\eqref{2053}$ hold for any $i<k$ if $A^{L}_j\le A^{(j)}\le A^{U}_j$, $1\le j\le i-1$,  $\hbar\ge \hbar_0^{(k-1)}$ and $A^{(i)}\ge A^{L}_i$. Based on these assumptions,  we will find  $A^{L}_{k}$ and $\hbar_0^{(k)}\ge \hbar_0^{(k-1)}$ such that $\eqref{2053}$ are true for $i=k$ when $A^{L}_j\le A^{(j)}\le A^{U}_j$, $1\le j\le k-1$,  $\hbar\ge \hbar_0^{(k)}$ and $A^{(k)}\ge A^{L}_k$. We choose an arbitrary $A^{U}_{k}>A^{L}_{k}$.  Then by induction the claims $\eqref{2053}$ are true for all $1\le k\le K$ with $\hbar_0=\hbar_0^{(K-1)}$.  \\

 Define
  \begin{align}
   &b_{i,1}= D_{i,1},\quad b_{i,2}= D_{i,2}(A^{U}_i)^2,\quad b_{i,3}=(A^{U}_i)^2D_{i,3},\quad 1\le i\le k-1. \label{388}
\end{align} 
Then $b_{i,1}$, $b_{i,2}$ and $b_{i,3}$ only depend on $\delta_0$, $c$ and $c_2\sim c_5$.   It follows from $\eqref{2053}$ and $\eqref{388}$ that for any $1\le i\le k-1$,
   \begin{align}
   &\|\widehat{\boldsymbol{\alpha}}_i\|_1\le   b_{i,1}\kappa^{-1}\sqrt{s} ,\quad \|\widehat{\boldsymbol{\gamma}}_i-\boldsymbol{\gamma}_i\|^2_2\le b_{i,2}C_0^2\kappa^{-2}\mu_1(\boldsymbol{\Xi})^{-1}\varpi^2 s, \notag\\ &\|\widehat{\mathbf{P}}_i-\mathbf{P}_i\|^2_2\le b_{i,3}C_0^2\kappa^{-2}\mu_1(\boldsymbol{\Xi})^{-1}\varpi^2 s. \label{1053}
\end{align} 
for all $\hbar\ge \hbar_0^{(k-1)}$ and all  $A^{L}_j\le A^{(j)}\le A^{U}_j$, $1\le j\le i$. \\

We provide several inequalities related to $\boldsymbol{\delta}_k$ and $\boldsymbol{\beta}_k$. By $\eqref{360}$, $\eqref{50036}$, $\eqref{1053}$ and the definitions of $\boldsymbol{\delta}_k$ and $\boldsymbol{\beta}_k$ in $\eqref{378}$,
 \begin{align}
&  \|\boldsymbol{\delta}_k\|_1\le  \sum_{i=1}^{k-1}|\boldsymbol{\gamma}_k\trans\widehat{\boldsymbol{\gamma}}_i|\|\widehat{\boldsymbol{\alpha}}_i\|_1=\sum_{i=1}^{k-1}|\boldsymbol{\gamma}_k\trans\widehat{\boldsymbol{\gamma}}_i-\boldsymbol{\gamma}_k\trans \boldsymbol{\gamma}_i|\|\widehat{\boldsymbol{\alpha}}_i\|_1\notag\\
\le&\sum_{i=1}^{k-1}\|\widehat{\boldsymbol{\gamma}}_i- \boldsymbol{\gamma}_i\|_2\|\widehat{\boldsymbol{\alpha}}_i\|_1\le\sum_{i=1}^{k-1}b_{i,2}^{1/2}C_0\kappa^{-1}\mu_1(\boldsymbol{\Xi})^{-1/2}\varpi \sqrt{s}b_{i,1}\kappa^{-1}\sqrt{s}\notag\\
=&M_{k,0}\kappa^{-2}\mu_1(\boldsymbol{\Xi})^{-1/2}C_0\varpi s\le M_{k,0}c_5 \|\boldsymbol{\alpha}_1\|_1, \label{50043}\\
 \text{ and }\quad &\|\boldsymbol{\beta}_k\|_1\le \|\boldsymbol{\alpha}_k\|_1+\|\boldsymbol{\delta}_k\|_1\le c_4\|\boldsymbol{\alpha}_1\|_1+\|\boldsymbol{\delta}_k\|_1\notag\\
&\le M_{k,1} \|\boldsymbol{\alpha}_1\|_1\le M_{k,1} \sqrt{s}/\kappa, \label{10245}
\end{align} 
where 
\begin{align}
M_{k,0}=\sum_{i=1}^{k-1}b_{i,2}^{1/2} b_{i,1},\quad M_{k,1}=\left(c_4+M_{k,0}c_5\right),\label{50048}
\end{align} 
the inequality in the third line is due to the second inequality in the condition $\eqref{359}$ and the last inequality follows from Lemma \ref{lemma_4}.  $\eqref{10245}$ and $\eqref{10247}$ lead to
 \begin{align}
 &\tau^{(k)}\|\boldsymbol{\beta}_k\|_1^2 \le \frac{A^{(k)}C_0\varpi M_{k,1}^2\|\boldsymbol{\alpha}_1\|_1^2}{\|\boldsymbol{\alpha}_1\|_1\sqrt{\mu_1(\boldsymbol{\Xi})}}=A^{(k)}M_{k,1}^2\frac{C_0 \varpi\|\boldsymbol{\alpha}_1\|_1}{\sqrt{\mu_1(\boldsymbol{\Xi})}}\notag\\
\le& A^{(k)}M_{k,1}^2\kappa^{-1}\frac{C_0 \varpi\sqrt{s}}{\sqrt{\mu_1(\boldsymbol{\Xi})}}  \le A^{(k)}M_{k,1}^2\hbar^{-1}.  \label{10257}
\end{align} 

Similar to the proof of Part (a), we next provide a key inequality $\eqref{10252}$ based on which the proof is. $\widehat{\boldsymbol{\alpha}}_k$ is the solution to the optimization problem $\eqref{1240}$. Because  $\widehat{\boldsymbol{\alpha}}_k$ is   the solution to $\eqref{1240}$ which has a scale-invariant objective function, it is also the solution to   
	 \begin{align*}
\max_{\boldsymbol{\alpha}\neq \mathbf{0},\quad  \boldsymbol{\alpha}\perp \widehat{\mathbf{W}}_{k-1}}\frac{\boldsymbol{\alpha}\trans \widehat{\mathbf{B}}\boldsymbol{\alpha}}{\boldsymbol{\alpha}\trans \mathbf{S}\boldsymbol{\alpha}+\tau^{(k)}\|\boldsymbol{\alpha}\|^2_{\lambda^{(k)}}},
\end{align*} 
where we do not imose the constraint, $\boldsymbol{\alpha}\trans \mathbf{S}\boldsymbol{\alpha}=1$. Therefore, since $\boldsymbol{\beta}_k\perp \widehat{\mathbf{W}}_{k-1}$ by $\eqref{50037}$, we have
 \begin{align}
\frac{\boldsymbol{\beta}_k\trans \widehat{\mathbf{B}}\boldsymbol{\beta}_k}{\boldsymbol{\beta}_k\trans \mathbf{S}\boldsymbol{\beta}_k+\tau^{(k)}\|\boldsymbol{\beta}_k\|^2_{\lambda^{(k)}}}\le \frac{\widehat{\boldsymbol{\alpha}}_k\trans \widehat{\mathbf{B}}\widehat{\boldsymbol{\alpha}}_k}{1+\tau^{(k)}\|\widehat{\boldsymbol{\alpha}}_k\|^2_{\lambda^{(k)}}}, \label{10241}
\end{align} 
  which is one of the key inequalities in this proof.
\begin{Lemma}\label{lemma_10}
 \begin{align*}
&\boldsymbol{\beta}_k\trans  \mathbf{S}\boldsymbol{\beta}_k \le 1,\\
&\boldsymbol{\beta}_k\trans  \widehat{\mathbf{B}}\boldsymbol{\beta}_k  \ge  \mu_k(\boldsymbol{\Xi})-  N_{k,3}C_0\kappa^{-1}\mu_1(\boldsymbol{\Xi})^{1/2}\varpi \sqrt{s},\\
&\widehat{\boldsymbol{\alpha}}_k\trans \widehat{\mathbf{B}}\widehat{\boldsymbol{\alpha}}_k \le  \mu_k(\boldsymbol{\Xi})+(b_{k-1,3}\hbar^{-1}+1)\sqrt{\mu_1(\boldsymbol{\Xi})}\kappa^{-1}C_0\varpi\sqrt{s}\notag\\
&\qquad\qquad +\frac{(1+\hbar^{-1})}{\sqrt{s}}\kappa\sqrt{\mu_1(\boldsymbol{\Xi})}C_0\varpi \|\widehat{\boldsymbol{\alpha}}_k\|_1^2,
\end{align*} 
where $N_{k,3}=2(b_{k-1,3}\hbar^{-1}+ M_{k,1})$.
\end{Lemma}

 By $\eqref{10241}$ and the first inequality in Lemma \ref{lemma_10}, 
 \begin{align}
&(\boldsymbol{\beta}_k\trans \widehat{\mathbf{B}}\boldsymbol{\beta}_k)(1+\tau^{(k)}\|\widehat{\boldsymbol{\alpha}}_k\|^2_{\lambda^{(k)}})\le (\widehat{\boldsymbol{\alpha}}_k\trans \widehat{\mathbf{B}}\widehat{\boldsymbol{\alpha}}_k)(\boldsymbol{\beta}_k\trans \mathbf{S}\boldsymbol{\beta}_k+\tau^{(k)}\|\boldsymbol{\beta}_k\|^2_{\lambda^{(k)}}) \label{10252}\\
\le &(\widehat{\boldsymbol{\alpha}}_k\trans \widehat{\mathbf{B}}\widehat{\boldsymbol{\alpha}}_k)(1+\tau^{(k)}\|\boldsymbol{\beta}_k\|^2_{\lambda^{(k)}}).\notag
\end{align} 
Based on $\eqref{10252}$ and Lemma \ref{lemma_10},  we provide an upper bound for $\|\widehat{\boldsymbol{\alpha}}_k\|_1$ in the following lemma.
\begin{Lemma}\label{lemma_12}
there exist $(A_k^{L})^\prime$ and $(\hbar_0)^\prime$ only depending on $\delta_0$, $c$ and $c_2\sim c_5$ such that for any $A^{(k)}\ge (A_k^{L})^\prime $ and $\hbar\ge (\hbar_0)^\prime$,  we have 
 \begin{align}
\|\widehat{\boldsymbol{\alpha}}_k\|_1\le D_{k,1}\kappa^{-1}\sqrt{s}.\label{10258}
\end{align} 
where $D_{k,1}=2M_{k,1}\sqrt{c_3c}$.
\end{Lemma}

Next,  by $\eqref{10252}$, we have 
 \begin{align}
&(\boldsymbol{\beta}_k\trans \widehat{\mathbf{B}}\boldsymbol{\beta}_k-\widehat{\boldsymbol{\alpha}}_k\trans \widehat{\mathbf{B}}\widehat{\boldsymbol{\alpha}}_k)(1+\tau^{(k)}\|\widehat{\boldsymbol{\alpha}}_k\|^2_{\lambda^{(k)}}) \notag\\
\le& (\widehat{\boldsymbol{\alpha}}_k\trans \widehat{\mathbf{B}}\widehat{\boldsymbol{\alpha}}_k)(\tau^{(k)}\|\boldsymbol{\beta}_k\|^2_{\lambda^{(k)}}-\tau^{(k)}\|\widehat{\boldsymbol{\alpha}}_k\|^2_{\lambda^{(k)}}),\label{20252}
\end{align} 
which leads to the following lemma.

\begin{Lemma}\label{lemma_11}
We have either
 \begin{align}
& \frac{1}{2}c_2\mu_k(\boldsymbol{\Xi})\|\boldsymbol{\gamma}_k-\widehat{\boldsymbol{\gamma}}_k\|_2^2\notag\\
\le& N_{k,4}C_0^2\kappa^{-2} \varpi^2 s+N_{k,5}C_0\sqrt{\mu_1(\boldsymbol{\Xi})}\varpi\|\boldsymbol{\beta}_k-\widehat{\boldsymbol{\alpha}}_k\|_1\notag\\
&+ N_{k,6}\mu_1(\boldsymbol{\Xi})[\tau^{(k)}\|\boldsymbol{\beta}_k\|^2_{\lambda^{(k)}}-\tau^{(k)}\|\widehat{\boldsymbol{\alpha}}_k\|^2_{\lambda^{(k)}}].\label{383}
\end{align} 
or
 \begin{align}
& \frac{1}{2}c_2\mu_k(\boldsymbol{\Xi})\|\boldsymbol{\gamma}_k-\widehat{\boldsymbol{\gamma}}_k\|_2^2\notag\\
\le& N_{k,4}C_0^2\kappa^{-2} \varpi^2 s+N_{k,5}C_0\sqrt{\mu_1(\boldsymbol{\Xi})}\varpi\|\boldsymbol{\beta}_k-\widehat{\boldsymbol{\alpha}}_k\|_1\notag\\
&+ N_{k,7}\mu_1(\boldsymbol{\Xi})[\tau^{(k)}\|\boldsymbol{\beta}_k\|^2_{\lambda^{(k)}}-\tau^{(k)}\|\widehat{\boldsymbol{\alpha}}_k\|^2_{\lambda^{(k)}}].\label{384}
\end{align} 
where 
 \begin{align*}
&N_{k,4}=3b_{k-1,3}N_{k,6}N_{k,7}^{-1}(1+A^{(k)}M_{k,1}^2\hbar^{-1}),\\
& N_{k,5}=[4+ (M_{k,1}+D_{k,1})\hbar^{-1}]N_{k,6}N_{k,7}^{-1}(1+A^{(k)}M_{k,1}^2\hbar^{-1})\notag\\
 &N_{k,6}=1+(b_{k-1,3}\hbar^{-1}+1)\hbar^{-1}+(1+\hbar^{-1}) D_{k,1}^2\hbar^{-1},\\
& N_{k,7}=c_3-2(b_{k-1,3}\hbar^{-1}+ M_{k,1})\hbar^{-1}.
\end{align*} 
\end{Lemma}

The two inequalities $\eqref{383}$ and $\eqref{384}$ leads to the same upper bounds except the constants. Hence, we only prove the results for the $\eqref{383}$ and exactly the same arguments can be applied to $\eqref{384}$.  
 
\begin{Lemma}\label{lemma_13}
 By $\eqref{383}$, we have 
 \begin{align}
  \frac{1}{2}c_2\mu_k(\boldsymbol{\Xi})\|\boldsymbol{\gamma}_k-\widehat{\boldsymbol{\gamma}}_k\|_2^2\le& N_{k,1}C_0^2\kappa^{-2} \varpi^2 s\notag\\
&+ (N_{k,2}+N_{k,5})\mu_1(\boldsymbol{\Xi})^{1/2}C_0\varpi \|(\boldsymbol{\alpha}_k-\widehat{\boldsymbol{\alpha}}_k)_{J_k}\|_1\notag\\
&-(\lambda^{(k)}N_{k,2}-N_{k,5})\mu_1(\boldsymbol{\Xi})^{1/2}C_0\varpi \|(\widehat{\boldsymbol{\alpha}}_k)_{J_k^c}\|_1 ,\label{399}
\end{align} 
where 
\begin{align*}
&N_{k,1}=N_{k,4}+N_{k,5} M_{k,0} +2A^{(k)}N_{k,6}M_{k,1} M_{k,0}\notag\\
 &N_{k,2}=A^{(k)}N_{k,6}(2\|\boldsymbol{\alpha}_k\|_1/\|\boldsymbol{\alpha}_1\|_1).
\end{align*} 
\end{Lemma}
 We will consider the following two cases separately, 
$$N_{k,1}C_0^2\kappa^{-2} \varpi^2 s\le \nu_0(N_{k,2}+N_{k,5})\mu_1(\boldsymbol{\Xi})^{1/2}C_0\varpi \|(\boldsymbol{\alpha}_k-\widehat{\boldsymbol{\alpha}}_k)_{J_k}\|_1$$ 
and 
$$N_{k,1}C_0^2\kappa^{-2} \varpi^2 s> \nu_0(N_{k,2}+N_{k,5})\mu_1(\boldsymbol{\Xi})^{1/2}C_0\varpi \|(\boldsymbol{\alpha}_k-\widehat{\boldsymbol{\alpha}}_k)_{J_k}\|_1,$$  
  where
$$\nu_0=(\delta_0/2)(c^{-1}+\delta_0/2)^{-1}.$$

\vspace{10mm}

\noindent {\bf  $\bullet$ Case 1:} $N_{k,1}C_0^2\kappa^{-2} \varpi^2 s\le \nu_0(N_{k,2}+N_{k,5})\mu_1(\boldsymbol{\Xi})^{1/2}C_0\varpi \|(\boldsymbol{\alpha}_k-\widehat{\boldsymbol{\alpha}}_k)_{J_k}\|_1$.\\

  In this case, $\eqref{399}$ leads to
 \begin{align}
 &\frac{1}{2}c_2\mu_k(\boldsymbol{\Xi})\|\boldsymbol{\gamma}_k-\widehat{\boldsymbol{\gamma}}_k\|_2^2\label{10280}\\
\le &(1+\nu_0)[N_{k,2}+N_{k,5}]\mu_1(\boldsymbol{\Xi})^{1/2}C_0\varpi \|(\boldsymbol{\alpha}_k-\widehat{\boldsymbol{\alpha}}_k)_{J_k}\|_1 \notag\\
 & -(\lambda^{(k)}N_{k,2}-N_{k,5})\mu_1(\boldsymbol{\Xi})^{1/2}C_0\varpi \|(\widehat{\boldsymbol{\alpha}}_k)_{J_k^c}\|_1.\notag
\end{align} 
Then we have 
 \begin{align}
&  (\lambda^{(k)}N_{k,2}-N_{k,5})\mu_1(\boldsymbol{\Xi})^{1/2}C_0\varpi \|(\widehat{\boldsymbol{\alpha}}_k)_{J_k^c}\|_1\notag\\
\le &(1+\nu_0)[N_{k,2}+N_{k,5}]\mu_1(\boldsymbol{\Xi})^{1/2}C_0\varpi \|(\boldsymbol{\alpha}_k-\widehat{\boldsymbol{\alpha}}_k)_{J_k}\|_1.\label{10281}
\end{align} 
By the definitions of $N_{k,5}$ and $N_{k,2}$ in Lemmas \ref{lemma_11} and \ref{lemma_13}, the facts that $b_{i, 1}$, $b_{i, 2}$ $b_{i, 3}$, $M_{k,0}$,  $M_{k,1}$,  and $D_{k,1}$, $1\le i\le k-1$, only depend on $\delta_0$, $c$ and $c_2\sim c_5$ (see $\eqref{388}$, $\eqref{50048}$ and Lemma \ref{lemma_12}) and the inequality $\eqref{360}$, we have 
 \begin{align}
&\lim_{\substack{ A^{(k)}\to\infty\\\hbar\to\infty}}\frac{(1+\nu_0)[N_{k,2}+N_{k,5}]}{\lambda^{(k)}N_{k,2}-N_{k,5}}=(1+\nu_0)(\lambda^{(k)})^{-1}\notag\\
\le &(1+\nu_0)(c+\delta_0)^{-1}=(c^{-1}+\delta_0/2)^{-1}<c .\label{402}
 \end{align} 
Therefore, by $\eqref{10281}$ and $\eqref{402}$, as $A^{(k)} $ and $\hbar $ are large enough, we have  $\|(\widehat{\boldsymbol{\alpha}}_k)_{J_k^c}\|_1\le c\|(\widehat{\boldsymbol{\alpha}}_k-\boldsymbol{\alpha}_k)_{J_k}\|_1$. Note that $(\boldsymbol{\alpha}_k)_{J_k^c}=\mathbf{0}$, hence
\begin{align}
& \|(\widehat{\boldsymbol{\alpha}}_k-\boldsymbol{\alpha}_k)_{J_k^c}\|_1\le c\|(\widehat{\boldsymbol{\alpha}}_k-\boldsymbol{\alpha}_k)_{J_k}\|_1,\label{50049}
 \end{align} 
  which together with $|J_k|=|J(\boldsymbol{\alpha}_k)|\le |J(\mathcal{B})|=\mathcal{M}(\mathcal{B})=s$ and Condition \ref{condition_3}, imply 
 \begin{align}
 &\kappa\|(\widehat{\boldsymbol{\alpha}}_k-\boldsymbol{\alpha}_k)_{J_k}\|_2\le \|\frac{\mathbf{X}}{\sqrt{n}}(\widehat{\boldsymbol{\alpha}}_k-\boldsymbol{\alpha}_k)\|_2\notag\\
=&\|\mathbf{Z}(\widehat{\boldsymbol{\alpha}}_k-\boldsymbol{\alpha}_k)\|_2=\|\widehat{\boldsymbol{\gamma}}_k-\boldsymbol{\gamma}_k\|_2.\label{10282}
\end{align} 
By $\eqref{10282}$, $\eqref{10280}$ and the  Cauchy-Schwarz inequality lead to
 \begin{align}
 &  \frac{1}{2}c_2\mu_k(\boldsymbol{\Xi})\|\boldsymbol{\gamma}_k-\widehat{\boldsymbol{\gamma}}_k\|_2^2 \notag\\
\le& (1+\nu_0)[N_{k,2}+N_{k,5}]\mu_1(\boldsymbol{\Xi})^{1/2}C_0\varpi \|(\boldsymbol{\alpha}_k-\widehat{\boldsymbol{\alpha}}_k)_{J_k}\|_1\notag\\
\le &(1+\nu_0)[N_{k,2}+N_{k,5}]\mu_1(\boldsymbol{\Xi})^{1/2}C_0\varpi \sqrt{|J_k|}\|(\widehat{\boldsymbol{\alpha}}_k-\boldsymbol{\alpha}_k)_{J_k}\|_2\notag\\
\le &(1+\nu_0)[N_{k,2}+N_{k,5}]\mu_1(\boldsymbol{\Xi})^{1/2}C_0\varpi \sqrt{s}\kappa^{-1}\|\widehat{\boldsymbol{\gamma}}_k-\boldsymbol{\gamma}_k\|_2\notag
 \end{align} 
Hence, by the inequality above and $\mu_k(\boldsymbol{\Xi})\ge c_3\mu_1(\boldsymbol{\Xi})$ (Condition \ref{condition_2}(b)), we have  $\|\widehat{\boldsymbol{\gamma}}_1-\boldsymbol{\gamma}_1\|_2\le 2c_2^{-1}c_3(1+\nu_0)[N_{k,2}+N_{k,5}]\kappa^{-1}\mu_1(\boldsymbol{\Xi})^{-1/2}C_0\varpi\sqrt{s}$. Because
 \begin{align*}
 &  \lim_{\substack{ A^{(k)}\to\infty\\\hbar\to\infty}}\frac{2c_2^{-1}c_3(1+\nu_0)[N_{k,2}+N_{k,5}]}{A^{(k)}}\notag\\
= &2c_2^{-1}c_3 (1+\nu_0) (2\|\boldsymbol{\alpha}_k\|_1/\|\boldsymbol{\alpha}_1\|_1)\notag\\
< & 2c_2^{-1}c_3 (1+2\nu_0) (2\|\boldsymbol{\alpha}_k\|_1/\|\boldsymbol{\alpha}_1\|_1)\le 4 c_2^{-1}c_3 (1+2\nu_0) c_4  ,
\end{align*} 
where the last inequality is due to the inequality $\eqref{360}$. Therefore, as $A^{(k)} $ and $\hbar $ are large enough, we have 
 \begin{align}
& \|\widehat{\boldsymbol{\gamma}}_k-\boldsymbol{\gamma}_k\|_2\le  4 c_2^{-1}c_3 (1+2\nu_0) c_4A^{(k)}\kappa^{-1}\mu_1(\boldsymbol{\Xi})^{-1/2}C_0\varpi\sqrt{s},\label{10283}\\
 \text{and hence} \quad & \|\widehat{\boldsymbol{\gamma}}_k-\boldsymbol{\gamma}_k\|_2^2\le 16 c_2^{-2}c_3^{2} (1+2\nu_0)^2 c_4^2(A^{(k)})^2\kappa^{-2}\mu_1(\boldsymbol{\Xi})^{-1}C_0^2\varpi^2s. \notag
\end{align} 
Now by $\eqref{50049}$, $\eqref{10282}$ and $\eqref{10283}$,
 \begin{align}
 &\|\widehat{\boldsymbol{\alpha}}_k-\boldsymbol{\alpha}_k\|_1=\|(\widehat{\boldsymbol{\alpha}}_k)_{J_k^c}\|_1+\|(\widehat{\boldsymbol{\alpha}}_k-\boldsymbol{\alpha}_k)_{J_k}\|_1\label{10284}\\
<& (1+c)\|(\widehat{\boldsymbol{\alpha}}_k-\boldsymbol{\alpha}_k)_{J_k}\|_1\notag\\
\le& (1+c)\sqrt{s}\|(\widehat{\boldsymbol{\alpha}}_k-\boldsymbol{\alpha}_k)_{J_k}\|_2\le (1+c)\sqrt{s}\kappa^{-1}\|\widehat{\boldsymbol{\gamma}}_k-\boldsymbol{\gamma}_k\|_2\notag\\
\le& 4(1+c)c_2^{-1}c_3c_4 (1+2\nu_0)A^{(k)}\kappa^{-2} \mu_k(\boldsymbol{\Xi})^{-1/2}C_0\varpi s.\notag
\end{align} 

\vspace{10mm}

\noindent {\bf  $\bullet$ Case 2:} $N_{k,1}C_0^2\kappa^{-2} \varpi^2 s> \nu_0(N_{k,2}+N_{k,5})\mu_1(\boldsymbol{\Xi})^{1/2}C_0\varpi \|(\boldsymbol{\alpha}_k-\widehat{\boldsymbol{\alpha}}_k)_{J_k}\|_1$.\\

 In this case, the inequality above leads to 
 \begin{align}
& \|(\boldsymbol{\alpha}_k-\widehat{\boldsymbol{\alpha}}_k)_{J_k}\|_1\le \frac{N_{k,1}}{\nu_0(N_{k,2}+N_{k,5})}\mu_1(\boldsymbol{\Xi})^{-1/2}\kappa^{-2}C_0\varpi s.\label{405}
\end{align} 
Because
\begin{align}
&\lim_{\substack{ A^{(k)}\to\infty\\\hbar\to\infty}}\frac{ N_{k,5}}{N_{k,2}}= 0,\notag\\
 &  \lim_{\substack{ A^{(k)}\to\infty\\\hbar\to\infty}}\frac{N_{k,1}}{\nu_0(N_{k,2}+N_{k,5})}=\nu_0^{-1} M_{k,1} M_{k,0}(2\|\boldsymbol{\alpha}_k\|_1/\|\boldsymbol{\alpha}_1\|_1)^{-1}\notag\\
& <\nu_0^{-1} M_{k,1} M_{k,0}c_4,\notag
\end{align}
we have, as $A^{(k)} $ and $\hbar $ are large enough, 
 \begin{align}
& \lambda^{(k)}N_{k,2}-N_{k,5}>0,\notag\\
& \|(\boldsymbol{\alpha}_k-\widehat{\boldsymbol{\alpha}}_k)_{J_k}\|_1\le \nu_0^{-1} M_{k,1} M_{k,0}c_4\mu_1(\boldsymbol{\Xi})^{-1/2}\kappa^{-2}C_0\varpi s,\label{406}
\end{align} 
In this case, $\eqref{399}$ gives
 \begin{align}
& \frac{1}{2}c_2\mu_k(\boldsymbol{\Xi})\|\boldsymbol{\gamma}_k-\widehat{\boldsymbol{\gamma}}_k\|_2^2+(\lambda^{(k)}N_{k,2}-N_{k,5})\mu_1(\boldsymbol{\Xi})^{1/2}C_0\varpi \|(\widehat{\boldsymbol{\alpha}}_k)_{J_k^c}\|_1\label{10285}\\
\le &  (1+\nu_0^{-1}) N_{k,1}C_0^2\kappa^{-2} \varpi^2 s,\notag
\end{align} 
which together with the first inequality in $\eqref{406}$ lead to
 \begin{align}
& \|\boldsymbol{\gamma}_k-\widehat{\boldsymbol{\gamma}}_k\|_2^2 \le 2c_2^{-1}  \mu_k(\boldsymbol{\Xi})^{-1}(1+\nu_0^{-1}) N_{k,1}C_0^2\kappa^{-2} \varpi^2 s\notag\\
\le&  2c_2^{-1} c_3 \mu_1(\boldsymbol{\Xi})^{-1}(1+\nu_0^{-1}) N_{k,1}C_0^2\kappa^{-2} \varpi^2 s, \label{408}
\end{align} 
 Because
 \begin{align}
 &  \lim_{\substack{ A^{(k)}\to\infty\\\hbar\to\infty}}\frac{ N_{k,1}}{A^{(k)}}= 2M_{k,1} M_{k,0}<4M_{k,1} M_{k,0},\notag
\end{align} 
as $A^{(k)} $ and $\hbar $ are large enough, we have 
 \begin{align}
& \|\boldsymbol{\gamma}_k-\widehat{\boldsymbol{\gamma}}_k\|_2^2 \le   8c_2^{-1} c_3 (1+\nu_0^{-1})M_{k,1} M_{k,0}A^{(k)}\mu_1(\boldsymbol{\Xi})^{-1}C_0^2\kappa^{-2} \varpi^2 s\notag\\
\le &8c_2^{-1} c_3 (1+\nu_0^{-1})M_{k,1} M_{k,0}(A^{(k)})^2\mu_1(\boldsymbol{\Xi})^{-1}C_0^2\kappa^{-2} \varpi^2 s,\label{10287}
\end{align} 
By $\eqref{10285}$,
 \begin{align}
& (\lambda^{(k)}N_{k,2}-N_{k,5})\mu_1(\boldsymbol{\Xi})^{1/2}C_0\varpi \|(\widehat{\boldsymbol{\alpha}}_k)_{J_k^c}\|_1\notag\\
\le & (1+\nu_0^{-1}) N_{k,1}C_0^2\kappa^{-2} \varpi^2 s.\label{10288}
\end{align} 
Because
 \begin{align}
    \lim_{\substack{ A^{(k)}\to\infty\\\hbar\to\infty}}\frac{(1+\nu_0^{-1}) N_{k,1}}{(\lambda^{(k)}N_{k,2}-N_{k,5})}&=(1+\nu_0^{-1})(\lambda^{(k)})^{-1}M_{k,1} M_{k,0}(2\|\boldsymbol{\alpha}_k\|_1/\|\boldsymbol{\alpha}_1\|_1)^{-1}\notag\\
&<(1+\nu_0^{-1})cM_{k,1} M_{k,0}c_4 ,
\end{align} 
as $A^{(k)} $ and $\hbar $ are large enough, we have 
 \begin{align*}
&  \|(\widehat{\boldsymbol{\alpha}}_k)_{J_k^c}\|_1\le  (1+\nu_0^{-1})cM_{k,1} M_{k,0}c_4\mu_1(\boldsymbol{\Xi})^{-1/2}\kappa^{-2}C_0 \varpi s
\end{align*} 
which together with $\eqref{406}$ lead to
 \begin{align}
&   \|\widehat{\boldsymbol{\alpha}}_k-\boldsymbol{\alpha}_k\|_1=\|(\widehat{\boldsymbol{\alpha}}_k-\boldsymbol{\alpha}_k)_{J_k}\|_1+\|(\widehat{\boldsymbol{\alpha}}_k)_{J_k^c}\|_1\notag\\
<&(1+2\nu_0^{-1})cM_{k,1} M_{k,0}c_4\mu_1(\boldsymbol{\Xi})^{-1/2}\kappa^{-2}C_0 \varpi s\notag\\
\le& (1+2\nu_0^{-1})cM_{k,1} M_{k,0}c_4A^{(k)}\mu_1(\boldsymbol{\Xi})^{-1/2}\kappa^{-2}C_0 \varpi s.\label{407}
\end{align}

Now we combine the results for the two cases. By $\eqref{10283}$ and $\eqref{10287}$, in both the two cases,  we have 
{\small\begin{align}
& \|\boldsymbol{\gamma}_k-\widehat{\boldsymbol{\gamma}}_k\|_2^2 \le D_{k,2} (A^{(k)})^2\kappa^{-2}\mu_1(\boldsymbol{\Xi})^{-1}C_0^2\varpi^2s, \label{409}
\end{align}}
where $D_{k,2}=16 c_2^{-2}c_3^{2} (1+2\nu_0)^2 c_4^2+8c_2^{-1} c_3 (1+\nu_0^{-1})M_{k,1} M_{k,0}$ which only depends on $\delta_0$, $c$ and $c_2\sim c_5$.
 Now by $\eqref{409}$ and $\eqref{1053}$, as $A^{(k)} $ and $\hbar $ are large enough,
 \begin{align}
    &\|\widehat{\mathbf{P}}_k-\mathbf{P}_{k-1}\|^2=\|(\widehat{\mathbf{P}}_{k-1}+\widehat{\boldsymbol{\gamma}}_k\trans \widehat{\boldsymbol{\gamma}}_k)-(\mathbf{P}_{k-1}-\boldsymbol{\gamma}_k\trans \boldsymbol{\gamma}_k)\|^2\label{10292}\\
		\le& 2\|\widehat{\mathbf{P}}_{k-1}-\mathbf{P}_{k-1}\|^2+2\| \widehat{\boldsymbol{\gamma}}_k\trans \widehat{\boldsymbol{\gamma}}_k -\boldsymbol{\gamma}_k\trans \boldsymbol{\gamma}_k \|^2\notag\\
		\le & 2\|\widehat{\mathbf{P}}_{k-1}-\mathbf{P}_{k-1}\|^2+4\| \widehat{\boldsymbol{\gamma}}_k\|_2^2\| \widehat{\boldsymbol{\gamma}}_k -\boldsymbol{\gamma}_k\|_2^2+ 4\| \widehat{\boldsymbol{\gamma}}_k -\boldsymbol{\gamma}_k\|_2^2\|\boldsymbol{\gamma}_k\|_2^2\notag\\
	=&	2\|\widehat{\mathbf{P}}_{k-1}-\mathbf{P}_{k-1}\|^2+4 \| \widehat{\boldsymbol{\gamma}}_k -\boldsymbol{\gamma}_k\|_2^2+ 4\| \widehat{\boldsymbol{\gamma}}_k -\boldsymbol{\gamma}_k\|_2^2 \notag\\
	\le &2b_{k-1,3} \kappa^{-2}\mu_1(\boldsymbol{\Xi})^{-1}C_0^2\varpi^2s+8 \| \widehat{\boldsymbol{\gamma}}_k -\boldsymbol{\gamma}_k\|_2^2\notag\\
		\le &2b_{k-1,3} (A^{(k)})^2\kappa^{-2}\mu_1(\boldsymbol{\Xi})^{-1}C_0^2\varpi^2s+8D_{k,2} (A^{(k)})^2\kappa^{-2}\mu_1(\boldsymbol{\Xi})^{-1}C_0^2\varpi^2s\notag\\
		=& D_{k,3} (A^{(k)})^2\kappa^{-2}\mu_1(\boldsymbol{\Xi})^{-1}C_0^2\varpi^2s,\notag
\end{align} 
where $ D_{k,3}=2b_{k-1,3}+8D_{k,2}$. Finally, by $\eqref{10284}$ and $\eqref{407}$, in both the two cases,  we have 
 \begin{align}
  \|\widehat{\boldsymbol{\alpha}}_k-\boldsymbol{\alpha}_k\|_1\le& 4(1+c)c_2^{-1}c_3c_4 (1+2\nu_0)A^{(k)}\kappa^{-2} \mu_k(\boldsymbol{\Xi})^{-1/2}C_0\varpi s\notag\\
&+(1+2\nu_0^{-1})cM_{k,1} M_{k,0}c_4A^{(k)}\mu_1(\boldsymbol{\Xi})^{-1/2}\kappa^{-2}C_0 \varpi s\notag\\
=&D_{k,4}A^{(k)}\mu_1(\boldsymbol{\Xi})^{-1/2}\kappa^{-2}C_0 \varpi s.\notag
\end{align} 
where $D_{k,4}=4(1+c)c_2^{-1}c_3c_4 (1+2\nu_0)+(1+2\nu_0^{-1})cM_{k,1} M_{k,0}c_4$. 

Hence, we have proved the claims $\eqref{2053}$ for $i=k$. By induction, the claims are true for all $1\le i\le K$.\\

\subsection{Proof of Theorem $\ref{theorem_5}$}

In this particular case, Conditions \ref{condition_1}-\ref{condition_2} are automatically satisfied and $\mu_1(\boldsymbol{\Xi})=\|\mathbf{X}\boldsymbol{\beta}\|_2^2/n$. Therefore, all the conditions for Theorem \ref{theorem_3} (a) are satisfied and the tuning parameters are the same as those in Theorem \ref{theorem_3}. Therefore, the conclusions in Theorem \ref{theorem_3} (a) hold. On the other hand, by the paragraph below Condition \ref{condition_1}, we have 
\begin{align*}
  M_{\epsilon}\le \sigma\sqrt{d\Phi^{-1}(1-1/(4d))} =\sigma\sqrt{\Phi^{-1}(3/4)}=0.6745\sigma\le \sigma.
\end{align*} 
 Therefore, we have $C_0= 4\sigma$, where $C_0$ is the constant in $\eqref{50020}$. The probability in Theorem \ref{theorem_3} (a)
\begin{align*}
  1-2pe^{ M_{\epsilon}^2/2\sigma^2}p^{-C_0^2/4\sigma^2}\ge 1-2p\sqrt{e}p^{-4}=1-2\sqrt{e}p^{-3}.
\end{align*} 
Hence, with probability at least $1-2\sqrt{e}p^{-3}$, the oracle inequalities in Theorem \ref{theorem_3} (a) hold. In this special case, $K=1$, $\mathbf{w}_1$ is a scalar and $\boldsymbol{\alpha}_1$ is proportional to $\boldsymbol{\beta}$. Due to the constraint $\boldsymbol{\alpha}_1\trans\mathbf{S}\boldsymbol{\alpha}_1=1$, we have
\begin{align*}
  &\boldsymbol{\alpha}_1= \boldsymbol{\beta}/\sqrt{\boldsymbol{\beta}\trans\mathbf{S}\boldsymbol{\beta}}=\boldsymbol{\beta}/\sqrt{\|\mathbf{X}\boldsymbol{\beta}\|_2^2/n}=\mu_1(\boldsymbol{\Xi})^{-1/2}\boldsymbol{\beta},\notag\\
	&\mathbf{t}_1=\mathbf{X}\boldsymbol{\alpha}_1=\mu_1(\boldsymbol{\Xi})^{-1/2}\mathbf{X}\boldsymbol{\beta}
\end{align*} 
  On the other hand, 
	\begin{align*}
  \mathbf{t}_1=\sqrt{n}\boldsymbol{\gamma}_1, \quad \widehat{\mathbf{t}}_1 =\sqrt{n}\widehat{\boldsymbol{\gamma}}_1.
\end{align*} 
Therefore, we have $\mathbf{X}\boldsymbol{\beta}=\mu_1(\boldsymbol{\Xi})^{1/2}\sqrt{n}\boldsymbol{\gamma}_1$  and by $\eqref{10116}$,
 \begin{align*} 
 \widehat{\mathbf{w}}_1= &\frac{1}{n}\widehat{\mathbf{t}}_1\trans (\mathbf{Y} -\mathbf{1}_n\bar{\mathbf{y}}\trans)=\frac{1}{n}\widehat{\mathbf{t}}_1\trans (\mathbf{X}\boldsymbol{\beta}+\boldsymbol{\varepsilon}-\mathbf{1}_n\bar{\boldsymbol{\varepsilon}}\trans)\notag\\
&= \frac{1}{n}\sqrt{n}\widehat{\boldsymbol{\gamma}}_1\trans (\mu_1(\boldsymbol{\Xi})^{1/2}\sqrt{n}\boldsymbol{\gamma}_1+\boldsymbol{\varepsilon}-\mathbf{1}_n\bar{\boldsymbol{\varepsilon}}\trans)=\widehat{\boldsymbol{\gamma}}_1\trans (\mu_1(\boldsymbol{\Xi})^{1/2} \boldsymbol{\gamma}_1+\varrho)\notag\\
&=\mu_1(\boldsymbol{\Xi})^{1/2}\widehat{\boldsymbol{\gamma}}_1\trans\boldsymbol{\gamma}_1+\widehat{\boldsymbol{\gamma}}_1\trans\varrho=\mu_1(\boldsymbol{\Xi})^{1/2}\widehat{\boldsymbol{\gamma}}_1\trans\boldsymbol{\gamma}_1+\widehat{\boldsymbol{\alpha}}_1\trans(\mathbf{Z}\trans\varrho).
\end{align*} 
Because by $\eqref{70002}$, $\mathbf{w}_1=\mu_1(\boldsymbol{\Xi})^{1/2}$ and hence we have 
 \begin{align} 
 |\widehat{\mathbf{w}}_1-\mathbf{w}_1|&\le \mu_1(\boldsymbol{\Xi})^{1/2}(1-\widehat{\boldsymbol{\gamma}}_1\trans\boldsymbol{\gamma}_1)+\max_{1\le j\le p}|(\mathbf{Z}\trans\boldsymbol{\varrho})_j|\|\widehat{\boldsymbol{\alpha}}_1\|_1\notag\\
&=\frac{1}{2}\mu_1(\boldsymbol{\Xi})^{1/2}(\|\widehat{\boldsymbol{\gamma}}_1\|_2^2+\| \boldsymbol{\gamma}_1\|_2^2-2\widehat{\boldsymbol{\gamma}}_1\trans\boldsymbol{\gamma}_1)+\max_{1\le j\le p}|(\mathbf{Z}\trans\boldsymbol{\varrho})_j|\|\widehat{\boldsymbol{\alpha}}_1\|_1\notag\\
&=\frac{1}{2}\mu_1(\boldsymbol{\Xi})^{1/2}\|\widehat{\boldsymbol{\gamma}}_1-\boldsymbol{\gamma}_1\|_2^2 +\max_{1\le j\le p}|(\mathbf{Z}\trans\boldsymbol{\varrho})_j|\|\widehat{\boldsymbol{\alpha}}_1\|_1.\label{334}
\end{align} 
In the event $\Omega$ defined in $\eqref{1033}$ of the proof of Theorem \ref{theorem_3}, we have the inequality $\max_{1\le j\le p}\|(\mathbf{Z}\trans\boldsymbol{\varrho})_j\|\le  C_0\varpi$. By $\eqref{334}$ and Theorem \ref{theorem_3} (a),  
 \begin{align} 
& |\widehat{\mathbf{w}}_1-\mathbf{w}_1|\le \frac{1}{2}\mu_1(\boldsymbol{\Xi})^{-1/2}16c_2^{-2}(1+\sqrt{6c})^2A^2C_0^2\kappa^{-2} \varpi^2 s + C_0\varpi \sqrt{6c}\sqrt{s}\kappa^{-1}\label{335}\\
\le& 8\hbar^{-1}c_2^{-2}(1+\sqrt{6c})^2A^2C_0\kappa^{-1} \varpi \sqrt{s}+ C_0\varpi \sqrt{6c}\sqrt{s}\kappa^{-1}=D_1C_0\kappa^{-1} \varpi \sqrt{s},\notag
\end{align} 
where $D_1=8\hbar^{-1}c_2^{-2}(1+\sqrt{6c})^2A^2+\sqrt{6c}$ and the first inequality in the last line is due to the condition $\mu_1(\boldsymbol{\Xi})\ge  \hbar^2C_0^2\varpi^2s/\kappa^2$. Now by   $\eqref{335}$ and Theorem \ref{theorem_3}(a),
 \begin{align*} 
&\|\widehat{\boldsymbol{\beta}}-\boldsymbol{\beta}\|_1=\|\widehat{\boldsymbol{\alpha}}_1\widehat{\mathbf{w}}_1-\boldsymbol{\alpha}_1\mathbf{w}_1\|_1\le \|\widehat{\boldsymbol{\alpha}}_1\|_1|\widehat{\mathbf{w}}_1-\mathbf{w}_1|+\|\widehat{\boldsymbol{\alpha}}_1-\boldsymbol{\alpha}_1\|_1|\mathbf{w}_1|\notag\\
\le & \sqrt{6c}D_1C_0\kappa^{-2} \varpi s+ 4(1+c)c_2^{-1}(1+\sqrt{6c})AC_0\kappa^{-2}\mu_1(\boldsymbol{\Xi})^{-1/2}\varpi s\mu_1(\boldsymbol{\Xi})^{1/2}\notag\\
=& D_3C_0\kappa^{-2} \varpi s,
\end{align*} 
where $D_3=[\sqrt{6c}D_1+4(1+c)c_2^{-1}(1+\sqrt{6c})A]$. Moreover,
 \begin{align*} 
&\|\mathbf{X}(\widehat{\boldsymbol{\beta}}-\boldsymbol{\beta})\|_2=\|\mathbf{X}\widehat{\boldsymbol{\alpha}}_1\widehat{\mathbf{w}}_1-\mathbf{X}\boldsymbol{\alpha}_1\mathbf{w}_1\|_2\notag\\
\le& \|\mathbf{X}\widehat{\boldsymbol{\alpha}}_1\|_2|\widehat{\mathbf{w}}_1-\mathbf{w}_1|+\|\mathbf{X}(\widehat{\boldsymbol{\alpha}}_1-\boldsymbol{\alpha}_1)\|_2|\mathbf{w}_1|\notag\\
\le & \sqrt{n} D_1C_0\kappa^{-1} \varpi \sqrt{s}+ \sqrt{n} 4c_2^{-1}(1+\sqrt{6c})A C_0 \kappa^{-1}\mu_1(\boldsymbol{\Xi})^{-1/2} \varpi \sqrt{s} \mu_1(\boldsymbol{\Xi})^{1/2} \notag\\
=& \sqrt{n}D_4C_0\kappa^{-1} \varpi \sqrt{s},
\end{align*} 
where $D_4=[D_1+4c_2^{-1}(1+\sqrt{6c})A]$. The theorem follows from the above inequalities and $C_0=4\sigma$.\\

\subsection{Proof of Theorem $\ref{theorem_7}$}

Recall that  
 \begin{align}
&\mathbf{t}_k=\sqrt{n}\boldsymbol{\gamma}_k,\quad \widehat{\mathbf{t}}_k=\sqrt{n}\widehat{\boldsymbol{\gamma}}_k,\notag\\
&\|\mathbf{w}_k\|_2=\mu_k(\boldsymbol{\Xi})^{1/2}\le \mu_1(\boldsymbol{\Xi})^{1/2}, \quad 1\le k\le K\notag\\
&\mathbf{X}\mathcal{B}= \mathbf{t}_1\mathbf{w}_1\trans+\mathbf{t}_2\mathbf{w}_2\trans+\cdots+\mathbf{t}_K\mathbf{w}_K\trans. \label{30001}
\end{align} 
 By $\eqref{10116}$,
 \begin{align*} 
 \widehat{\mathbf{w}}_k\trans&= \frac{1}{n}\widehat{\mathbf{t}}_k\trans ( \mathbf{Y}-\mathbf{1}_n\bar{\mathbf{y}}\trans)=\frac{1}{n}\widehat{\mathbf{t}}_k\trans (\mathbf{X}\mathcal{B}+\boldsymbol{\varepsilon}-\mathbf{1}_n\bar{\boldsymbol{\varepsilon}}\trans)\notag\\
&= \frac{1}{n}\sqrt{n}\widehat{\boldsymbol{\gamma}}_k\trans (\mathbf{t}_1\mathbf{w}_1\trans+\mathbf{t}_2\mathbf{w}_2\trans+\cdots+\mathbf{t}_K\mathbf{w}_K\trans+\boldsymbol{\varepsilon}-\mathbf{1}_n\bar{\boldsymbol{\varepsilon}}\trans)\notag\\
&=\frac{1}{n}\sqrt{n}\widehat{\boldsymbol{\gamma}}_k\trans (\sqrt{n}\boldsymbol{\gamma}_1\mathbf{w}_1\trans+\sqrt{n}\boldsymbol{\gamma}_2\mathbf{w}_2\trans+\cdots+\sqrt{n}\boldsymbol{\gamma}_K\mathbf{w}_K\trans+\sqrt{n}\varrho)\notag\\
&=\widehat{\boldsymbol{\gamma}}_k\trans\boldsymbol{\gamma}_k\mathbf{w}_k\trans+\sum_{i\neq k}\widehat{\boldsymbol{\gamma}}_k\trans\boldsymbol{\gamma}_i\mathbf{w}_i\trans+\widehat{\boldsymbol{\gamma}}_k\trans\varrho \noatg\\
&=\widehat{\boldsymbol{\gamma}}_k\trans\boldsymbol{\gamma}_k\mathbf{w}_k\trans+\sum_{i\neq k}\widehat{\boldsymbol{\gamma}}_k\trans\boldsymbol{\gamma}_i\mathbf{w}_i\trans+\widehat{\boldsymbol{\alpha}}_k\trans(\mathbf{Z}\trans\varrho),
\end{align*} 
where in the last equality is because $\widehat{\boldsymbol{\gamma}}_k=\mathbf{Z}\widehat{\boldsymbol{\alpha}}_k$. Because $1-\widehat{\boldsymbol{\gamma}}_k\trans\boldsymbol{\gamma}_k=\|\widehat{\boldsymbol{\gamma}}_k-\boldsymbol{\gamma}_k\|_2^2/2$ and $\mathbf{w}_1,\cdots,\mathbf{w}_K$ are orthogonal to each other, we have
\begin{align} 
 &\|\sum_{i\neq k}\widehat{\boldsymbol{\gamma}}_k\trans\boldsymbol{\gamma}_i\mathbf{w}_i\|_2^2=\sum_{i\neq k}(\widehat{\boldsymbol{\gamma}}_k\trans\boldsymbol{\gamma}_i)^2\|\mathbf{w}_i\|_2^2=\sum_{i\neq k}(\widehat{\boldsymbol{\gamma}}_k\trans\boldsymbol{\gamma}_i)^2\mu_k(\boldsymbol{\Xi})\notag\\
&\le \mu_1(\boldsymbol{\Xi})\sum_{i\neq k}(\widehat{\boldsymbol{\gamma}}_k\trans\boldsymbol{\gamma}_i)^2\le\mu_1(\boldsymbol{\Xi}) [1-(\widehat{\boldsymbol{\gamma}}_k\trans\boldsymbol{\gamma}_k)^2]\notag\\
&\le 2\mu_1(\boldsymbol{\Xi}) [1- \widehat{\boldsymbol{\gamma}}_k\trans\boldsymbol{\gamma}_k ]=\mu_1(\boldsymbol{\Xi})\|\widehat{\boldsymbol{\gamma}}_k-\boldsymbol{\gamma}_k\|_2^2, 
\end{align} 
where the first inequality in the last line is because $|1+\widehat{\boldsymbol{\gamma}}_k\trans\boldsymbol{\gamma}_k |\le 2$. Then we have 
 \begin{align} 
 \|\widehat{\mathbf{w}}_k-\mathbf{w}_k\|_2&\le  (1-\widehat{\boldsymbol{\gamma}}_k\trans\boldsymbol{\gamma}_k)\|\mathbf{w}_k\|_2+\|\sum_{i\neq k}\widehat{\boldsymbol{\gamma}}_k\trans\boldsymbol{\gamma}_i\mathbf{w}_i\|_2+\max_{1\le j\le p}\|(\mathbf{Z}\trans\boldsymbol{\varrho})_j\|_2\|\widehat{\boldsymbol{\alpha}}_k\|_1 \notag\\
& \le \frac{1}{2}\mu_1(\boldsymbol{\Xi})^{1/2}\|\widehat{\boldsymbol{\gamma}}_k-\boldsymbol{\gamma}_k\|_2^2+ \mu_1(\boldsymbol{\Xi})^{1/2}\|\widehat{\boldsymbol{\gamma}}_k-\boldsymbol{\gamma}_k\|_2 \notag\\
&\quad +\max_{1\le j\le p}\|(\mathbf{Z}\trans\boldsymbol{\varrho})_j\|_2\|\widehat{\boldsymbol{\alpha}}_k\|_1.\label{334}
\end{align} 
In the event $\Omega$, we have $\max_{1\le j\le p}\|(\mathbf{Z}\trans\boldsymbol{\varrho})_j\|_2\le  C_0\varpi$. By $\eqref{334}$ and  Theorem \ref{theorem_3} (b) and the definition of the $L_{1,2}$ norm,
 \begin{align} 
& \|\widehat{\mathbf{w}}_k-\mathbf{w}_k\|_2\label{30003}\\
\le& \frac{1}{2}\mu_1(\boldsymbol{\Xi})^{-1/2}D_{k,2}(A^{(k)})^2C_0^2\kappa^{-2} \varpi^2 s +D_{k,2}^{1/2}(A^{(k)})C_0\kappa^{-1} \varpi \sqrt{s}\notag\\
&\quad + C_0\varpi D_{k,1}\sqrt{s}\kappa^{-1}\notag\\
\le& \frac{1}{2}D_{k,2}(A^{(k)})^2\hbar^{-1}C_0\kappa^{-1} \varpi \sqrt{s}+ D_{k,2}^{1/2}(A^{(k)})C_0\kappa^{-1} \varpi \sqrt{s}\notag\\
&\quad + C_0\varpi D_{k,1}\sqrt{s}\kappa^{-1}\notag\\
=&L_{k,1}C_0\varpi\sqrt{s}\kappa^{-1},\notag
\end{align} 
where $L_{k,1}=\frac{1}{2}D_{k,2}(A^{(k)})^2\hbar^{-1}+D_{k,2}^{1/2}(A^{(k)})+D_{k,1}$ and the second inequality  is due to the condition $\mu_1(\boldsymbol{\Xi})\ge  \hbar^2C_0^2\varpi^2s/\kappa^2$. Now by  $\eqref{10116}$, $\eqref{30003}$ and Theorem \ref{theorem_3}(b),
 \begin{align*} 
& \|\sum_{k=1}^{K_0}\widehat{\boldsymbol{\alpha}}_k\widehat{\mathbf{w}}_k\trans-\sum_{k=1}^{K_0}\boldsymbol{\alpha}_k\mathbf{w}_k\trans\|_{1,2}\le \sum_{k=1}^{K_0}\|\widehat{\boldsymbol{\alpha}}_k\widehat{\mathbf{w}}_k\trans-\boldsymbol{\alpha}_k\mathbf{w}_k\trans\|_{1,2}\notag\\
\le& \sum_{k=1}^{K_0}\|\widehat{\boldsymbol{\alpha}}_k\|_1\|\widehat{\mathbf{w}}_k-\mathbf{w}_k\|_2+\sum_{k=1}^{K_0}\|\widehat{\boldsymbol{\alpha}}_k-\boldsymbol{\alpha}_k\|_1\|\mathbf{w}_k\|_2\notag\\
\le & \sum_{k=1}^{K_0}D_{k,1}L_{k,1}C_0\kappa^{-2} \varpi s+ \sum_{k=1}^{K_0}D_{k,4}(A^{(k)})C_0\kappa^{-2}\mu_1(\boldsymbol{\Xi})^{-1/2}\varpi s\mu_1(\boldsymbol{\Xi})^{1/2}\notag\\
= & L_{K_0,3}C_0\kappa^{-2} \varpi s,
\end{align*} 
where $L_{K_0,3}=\sum_{k=1}^{K_0}D_{k,1}L_{k,1}+\sum_{k=1}^{K_0}D_{k,4}(A^{(k)})$. In particular,
\begin{align*} 
& \|\widehat{\mathcal{B}}-\mathcal{B}\|_{1,2}  \le L_{K,3}C_0\kappa^{-2} \varpi s,
\end{align*} 
 Moreover,
 \begin{align*} 
& \|\sum_{k=1}^{K_0}\mathbf{X}\widehat{\boldsymbol{\alpha}}_k\widehat{\mathbf{w}}_k\trans-\sum_{k=1}^{K_0}\mathbf{X}\boldsymbol{\alpha}_k\mathbf{w}_k\trans\|_F\le \sum_{k=1}^{K_0}\|\mathbf{X}\widehat{\boldsymbol{\alpha}}_k\widehat{\mathbf{w}}_k\trans-\mathbf{X}\boldsymbol{\alpha}_k\mathbf{w}_k\trans\|_F\notag\\
\le& \sum_{k=1}^{K_0}\|\mathbf{X}\widehat{\boldsymbol{\alpha}}_k\|_2\|\widehat{\mathbf{w}}_1-\mathbf{w}_1\|_2+\sum_{k=1}^{K_0}\|\mathbf{X}(\widehat{\boldsymbol{\alpha}}_k-\boldsymbol{\alpha}_k)\|_2\|\mathbf{w}_k\|_2\notag\\
\le & \sum_{k=1}^{K_0}\sqrt{n} L_{k,1}C_0\kappa^{-1} \varpi \sqrt{s}+ \sum_{k=1}^{K_0}\sqrt{n} D_{k,2}^{1/2}(A^{(k)})C_0\kappa^{-1}  \mu_1(\boldsymbol{\Xi})^{-1/2} \varpi \sqrt{s} \mu_1(\boldsymbol{\Xi})^{1/2} \notag\\
=& \sqrt{n}L_{K_0,4}C_0\kappa^{-1} \varpi \sqrt{s},
\end{align*} 
where $L_{K_0,4}=\sum_{k=1}^{K_0}L_{k,1}+\sum_{k=1}^{K_0} D_{k,2}^{1/2}(A^{(k)})$. In particular,
\begin{align*} 
&\|\mathbf{X}(\widehat{\mathcal{B}}-\mathcal{B})\|_F \le \sqrt{n}L_{K,4}C_0\kappa^{-1} \varpi \sqrt{s}.
\end{align*}

\newpage

\section*{Appendix B: Proofs of technical lemmas}

\renewcommand{\proof}{\underline{\bf Proof of Lemma}}

\baselineskip=18pt

 In Appendix B, we provide the proofs for all technical lemmas.\\

\begin{proof} $\ref{lemma_15}$. \vspace{3mm}
 
In this special case, $\mathcal{B}$ is a $p$-dimensional vector and $\mathbf{Y}$ is a $n$-dimensional vector, and $\bar{\mathbf{y}}$ is a scalar. Therefore, for convenience, in this proof, we will use the traditional notation $\boldsymbol{\beta}$ to replace $\mathcal{B}$.  Let $\widehat{\boldsymbol{\beta}}_{LS}$ be  the least squares estimates. In this special case,  $K=1$. Therefore, there is only one component $\boldsymbol{\alpha}_1$. Its estimate $\widehat{\boldsymbol{\alpha}}_1$ is the solution to
 \begin{align}
 &\max_{ \boldsymbol{\alpha}\trans \mathbf{S}\boldsymbol{\alpha}=1} \boldsymbol{\alpha}\trans \widehat{\mathbf{B}}\boldsymbol{\alpha} =\max_{ \|\mathbf{Z}\boldsymbol{\alpha}\|_2=1}\left(\frac{1}{n} (\mathbf{X}\boldsymbol{\alpha})\trans(\mathbf{Y}-\mathbf{1}_n\bar{\mathbf{y}})\right)^2\notag\\
&=\max_{ \|\mathbf{Z}\boldsymbol{\alpha}\|_2=1}\left(\frac{1}{\sqrt{n}} (\mathbf{Z}\boldsymbol{\alpha})\trans(\mathbf{Y}-\mathbf{1}_n\bar{\mathbf{y}})\right)^2,\label{10212}
\end{align} 
and its solution is $\widehat{\boldsymbol{\alpha}}_1$. In the case, $\widehat{\mathbf{w}}_1$ is a scalar and by $\eqref{10116}$, we obtained that our estimate of $\boldsymbol{\beta}$ is equal to
{\small\begin{align*}
&\widehat{\boldsymbol{\beta}}=\widehat{\boldsymbol{\alpha}}_1\widehat{\mathbf{w}}_1\trans =\frac{1}{n}[\widehat{\mathbf{t}}_1\trans (\mathbf{Y}-\mathbf{1}_n\bar{\mathbf{y}})]\widehat{\boldsymbol{\alpha}}_1\notag\\
=&\frac{1}{n} [(\mathbf{X}\widehat{\boldsymbol{\alpha}}_1)\trans(\mathbf{Y}-\mathbf{1}_n\bar{\mathbf{y}})]\widehat{\boldsymbol{\alpha}}_1 
=\frac{1}{\sqrt{n}} \left[(\mathbf{Z}\widehat{\boldsymbol{\alpha}}_1)\trans(\mathbf{Y}-\mathbf{1}_n\bar{\mathbf{y}})\right]\widehat{\boldsymbol{\alpha}}_1.
\end{align*}} 
Hence, $\mathbf{X}\widehat{\boldsymbol{\beta}}=\left[(\mathbf{Z}\widehat{\boldsymbol{\alpha}}_1)\trans(\mathbf{Y}-\mathbf{1}_n\bar{\mathbf{y}})\right]\mathbf{Z}\widehat{\boldsymbol{\alpha}}_1$, which together with the fact that $\|\mathbf{Z}\widehat{\boldsymbol{\alpha}}_1\|_2^2=\widehat{\boldsymbol{\alpha}}_1\trans\mathbf{Z}\trans\mathbf{Z}\widehat{\boldsymbol{\alpha}}_1=\widehat{\boldsymbol{\alpha}}_1\trans\mathbf{S} \widehat{\boldsymbol{\alpha}}_1=1$ imply that $\mathbf{X}\widehat{\boldsymbol{\beta}}$ is the orthogonal projection of $\mathbf{Y}-\mathbf{1}_n\bar{\mathbf{y}}$ along the direction of   $\mathbf{Z}\widehat{\boldsymbol{\alpha}}_1$. Therefore, by $\eqref{10212}$,
 \begin{align}
&\|(\mathbf{Y}-\mathbf{1}_n\bar{\mathbf{y}})-\mathbf{X}\widehat{\boldsymbol{\beta}}\|^2_2=\|(\mathbf{Y}-\mathbf{1}_n\bar{\mathbf{y}})\|^2_2-\|\mathbf{X}\widehat{\boldsymbol{\beta}}\|^2_2\label{10213}\\
=&\|(\mathbf{Y}-\mathbf{1}_n\bar{\mathbf{y}})\|^2_2-\left[(\mathbf{Z}\widehat{\boldsymbol{\alpha}}_1)\trans(\mathbf{Y}-\mathbf{1}_n\bar{\mathbf{y}})\right]^2\notag\\
=&\|(\mathbf{Y}-\mathbf{1}_n\bar{\mathbf{y}})\|^2_2-\max_{ \boldsymbol{\alpha}\trans \mathbf{S}\boldsymbol{\alpha}=1} \left[(\mathbf{Z} \boldsymbol{\alpha} )\trans(\mathbf{Y}-\mathbf{1}_n\bar{\mathbf{y}})\right]^2\notag\\
=&\min_{ \boldsymbol{\alpha}\trans \mathbf{S}\boldsymbol{\alpha}=1} \left\{\|(\mathbf{Y}-\mathbf{1}_n\bar{\mathbf{y}})\|^2_2-\left[(\mathbf{Z} \boldsymbol{\alpha} )\trans(\mathbf{Y}-\mathbf{1}_n\bar{\mathbf{y}})\right]^2\right\}.\notag
\end{align} 
Now let $\widehat{\boldsymbol{\alpha}}_{LS}=\widehat{\boldsymbol{\beta}}_{LS}/\|\mathbf{Z}\widehat{\boldsymbol{\beta}}_{LS}\|_2$. Then $\|\mathbf{Z}\widehat{\boldsymbol{\alpha}}_{LS}\|_2=1$ and $\mathbf{X}\widehat{\boldsymbol{\beta}}_{LS}$ is the orthogonal projection of $\mathbf{Y}-\mathbf{1}_n\bar{\mathbf{y}}$ along the direction of $\mathbf{Z}\widehat{\boldsymbol{\alpha}}_{LS}$. Therefore, by $\eqref{10213}$,
 \begin{align*}
&\|(\mathbf{Y}-\mathbf{1}_n\bar{\mathbf{y}})-\mathbf{X}\widehat{\boldsymbol{\beta}}_{LS}\|^2_2=\|\mathbf{Y}-\mathbf{1}_n\bar{\mathbf{y}}\|_2^2-\|\mathbf{X}\widehat{\boldsymbol{\beta}}_{LS}\|^2_2\notag\\
&= \left\{\|(\mathbf{Y}-\mathbf{1}_n\bar{\mathbf{y}})\|^2_2-\left[(\mathbf{Z}\widehat{\boldsymbol{\alpha}}_{LS})\trans(\mathbf{Y}-\mathbf{1}_n\bar{\mathbf{y}})\right]^2\right\}\notag\\
&\ge \min_{ \boldsymbol{\alpha}\trans \mathbf{S}\boldsymbol{\alpha}=1} \left\{\|(\mathbf{Y}-\mathbf{1}_n\bar{\mathbf{y}})\|^2_2-\left[(\mathbf{Z} \boldsymbol{\alpha} )\trans(\mathbf{Y}-\mathbf{1}_n\bar{\mathbf{y}})\right]^2\right\}\notag\\
&=\|(\mathbf{Y}-\mathbf{1}_n\bar{\mathbf{y}})-\mathbf{X}\widehat{\boldsymbol{\beta}}\|^2_2,
\end{align*} 
which, together with the definition and the uniqueness (due to the full rank of $\mathbf{S}$) of the least squares estimate, implies $\widehat{\boldsymbol{\beta}}=\widehat{\boldsymbol{\beta}}_{LS}$.  \\

\end{proof}

\begin{proof}  $\ref{lemma_2}$. \\

By $\eqref{10200}$ in the proof of Lemma \ref{lemma_8}, for any $t>M_{\epsilon}$,
 \begin{align}
  &P\left(\|(\mathbf{Z}\trans\boldsymbol{\varrho})_j\|_2>\frac{t}{\sqrt{n}}\right)\le e^{-\frac{(t-M_{\epsilon})^2}{2\sigma^2}}\notag\\
	=&e^{ M_{\epsilon}^2/2\sigma^2}e^{-\frac{(t-M_{\epsilon})^2+M_{\epsilon}^2}{2\sigma^2}}\le e^{ M_{\epsilon}^2/2\sigma^2}e^{-\frac{t^2}{4\sigma^2}}.\label{10201}
\end{align} 
For any $C>M_{\epsilon}/\sqrt{\log{p}}$, let $t=C\sqrt{\log{p}}$. Then $t>M_{\epsilon}$ and by $\eqref{10201}$,   we have 
\begin{align}
  &P\left(\max_{1\le j\le p}\|(\mathbf{Z}\trans\boldsymbol{\varrho})_j\|_2> C\sqrt{\frac{\log{p}}{n}}\right)\notag\\
	=&P\left(\max_{1\le j\le p}\|(\mathbf{Z}\trans\boldsymbol{\varrho})_j\|_2>\frac{t}{\sqrt{n}}\right)\le \sum_{j=1}^p P\left( \|(\mathbf{Z}\trans\boldsymbol{\varrho})_j\|_2>\frac{t}{\sqrt{n}}\right)\notag\\
	\le& pe^{ M_{\epsilon}^2/2\sigma^2}e^{ -t^2/4\sigma^2}=pe^{ M_{\epsilon}^2/2\sigma^2}p^{-C^2/4\sigma^2}.\label{10013}
\end{align}
On the other hand, for any $1\le k\le K$,  
\begin{align*}
  &\boldsymbol{\varrho}\trans (\mathbf{Z}\boldsymbol{\alpha}_k)=\frac{1}{\sqrt{n}}(\boldsymbol{\varepsilon}-\mathbf{1}_n\bar{\boldsymbol{\varepsilon}}\trans)\trans(\mathbf{Z}\boldsymbol{\alpha}_k)=\frac{1}{\sqrt{n}}\boldsymbol{\varepsilon}\trans(\mathbf{Z}\boldsymbol{\alpha}_k)-\frac{1}{\sqrt{n}}\bar{\boldsymbol{\varepsilon}}\mathbf{1}_n\trans(\mathbf{Z}\boldsymbol{\alpha}_k)\\
	=& \frac{1}{\sqrt{n}}\sum_{i=1}^n(\mathbf{Z}\boldsymbol{\alpha}_k)_i\boldsymbol{\varepsilon}_i,
\end{align*}
 where $(\mathbf{Z}\boldsymbol{\alpha}_k)_i$ is the $i$-th coordinate of the vector $\mathbf{Z}\boldsymbol{\alpha}_k$ and last equality is because the column means of $\mathbf{Z}=\mathbf{X}/\sqrt{n}$ are all zero. Note that $\sum_{i=1}^n(\mathbf{Z}\boldsymbol{\alpha}_k)_i\boldsymbol{\varepsilon}_i/\sqrt{n}$ has the same distribution as $\|(\mathbf{Z}\boldsymbol{\alpha}_k)/\sqrt{n}\|_2\boldsymbol{\varepsilon}_1=\boldsymbol{\varepsilon}_1/\sqrt{n}$, where we use the equality: $\|\mathbf{Z}\boldsymbol{\alpha}_k\|_2^2=\boldsymbol{\alpha}_k\trans\mathbf{Z}\trans\mathbf{Z}\boldsymbol{\alpha}_k=\boldsymbol{\alpha}_k\trans\mathbf{S}\boldsymbol{\alpha}_k=1$. Similar to $\eqref{10201}$, for $t=C\sqrt{\log{p}}$, we have
\begin{align*}
  &P\left(\|(\mathbf{Z}\boldsymbol{\alpha}_k)\trans\boldsymbol{\varrho}\|_2>\frac{t}{\sqrt{n}}\right)=P\left(\|\frac{\boldsymbol{\varepsilon}_1}{\sqrt{n}}\|_2> \frac{t}{\sqrt{n}}\right)\notag\\
	\le& e^{ M_{\epsilon}^2/2\sigma^2}e^{ -t^2/4\sigma^2}=e^{ M_{\epsilon}^2/2\sigma^2}p^{-C^2/4\sigma^2}.
\end{align*}
Hence
{\small\begin{align*}
  &P\left(\bigcup_{k=1}^K\left\{\|(\mathbf{Z}\boldsymbol{\alpha}_k)\trans\boldsymbol{\varrho}\|_2>C\sqrt{\frac{\log{p}}{n}}\right\}\right)\le Ke^{ M_{\epsilon}^2/2\sigma^2}p^{-C^2/4\sigma^2}.
\end{align*}}

\end{proof}

\begin{proof} $\ref{lemma_4}$.\\

Given $1\le k\le K$, let $J_k=J(\boldsymbol{\alpha}_k)$. Then by $\eqref{10012}$, $|J_k|\le \mathcal{M}(\mathcal{B})=s$ and $(\boldsymbol{\alpha}_k)_{J_k^c}=\mathbf{0}$. By Condition \ref{condition_3}, 
$$\kappa\le\frac{\|\mathbf{X}\boldsymbol{\alpha}_k\|_2}{\sqrt{n}\|(\boldsymbol{\alpha}_k)_{J_k}\|_2}=\frac{\|\mathbf{Z}\boldsymbol{\alpha}_k\|_2}{\|(\boldsymbol{\alpha}_k)_{J_k}\|_2}=\frac{\|\mathbf{Z}\boldsymbol{\alpha}_k\|_2}{\|\boldsymbol{\alpha}_k\|_2}.$$
Therefore, $\kappa^2\|\boldsymbol{\alpha}_k\|_2^2\le \|\mathbf{Z}\boldsymbol{\alpha}_k\|_2^2=\boldsymbol{\alpha}_k\trans\mathbf{Z}\trans\mathbf{Z}\boldsymbol{\alpha}_k=\boldsymbol{\alpha}_k\trans\mathbf{S}\boldsymbol{\alpha}_k=1$, which leads to the first inequality in the lemma.

 Under Condition \ref{condition_2}, all the diagonal elements of $\mathbf{S}=\mathbf{X}\trans\mathbf{X}/n$ are equal to 1. Then it is easy to see that the absolute values of all the elements of $\mathbf{S}$ are less than or equal to 1. That is, $\|\mathbf{S}\|_{\infty}\le 1$. Therefore, we have $1=\boldsymbol{\alpha}_k\trans\mathbf{S}\boldsymbol{\alpha}_k\le \|\mathbf{S}\|_{\infty}\|\boldsymbol{\alpha}_k\|_1^2\le \|\boldsymbol{\alpha}_k\|_1^2$ and $1=\widehat{\boldsymbol{\alpha}}_k\trans\mathbf{S}\widehat{\boldsymbol{\alpha}}_k\le \|\mathbf{S}\|_{\infty}\|\widehat{\boldsymbol{\alpha}}_k\|_1^2\le \|\widehat{\boldsymbol{\alpha}}_k\|_1^2$. 

As for the last inequality, if both Conditions \ref{condition_2} and \ref{condition_3} are satisfied, by the Cauchy-Schwarz inequality, $\|\boldsymbol{\alpha}_k\|_1^2\le s\|\boldsymbol{\alpha}_k\|_2^2\le s/\kappa^2$.\\

\end{proof}

\begin{proof} $\ref{lemma_3}$.\\

In this proof, we only consider the element in $\Omega$. By $\eqref{10202}$,
 \begin{align}
 \boldsymbol{\alpha}_1\trans \widehat{\mathbf{B}}\boldsymbol{\alpha}_1 &=\boldsymbol{\alpha}_1\trans \mathbf{B}\boldsymbol{\alpha}_1 +2 \boldsymbol{\alpha}_1\trans \mathbf{S}\mathcal{B} (\mathbf{Z}\trans\boldsymbol{\varrho})\trans\boldsymbol{\alpha}_1+\boldsymbol{\alpha}_1\trans (\mathbf{Z}\trans\boldsymbol{\varrho})(\mathbf{Z}\trans\boldsymbol{\varrho})\trans\boldsymbol{\alpha}_1\label{10017}\\
 &=\mu_1(\boldsymbol{\Xi})+ 2 \boldsymbol{\alpha}_1 \trans \mathbf{S}\mathcal{B}(\mathbf{Z}\trans\boldsymbol{\varrho})\trans\boldsymbol{\alpha}_1+\|(\mathbf{Z}\trans\boldsymbol{\varrho})\trans\boldsymbol{\alpha}_1\|_2^2\notag\\
&\ge \mu_1(\boldsymbol{\Xi})-2\|\boldsymbol{\alpha}_1\trans \mathbf{S}\mathcal{B}\|_2\|(\mathbf{Z}\trans\boldsymbol{\varrho})\trans\boldsymbol{\alpha}_1\|_2\notag\\
 &=\mu_1(\boldsymbol{\Xi})-2\sqrt{\mu_1(\boldsymbol{\Xi})}\|(\mathbf{Z}\trans\boldsymbol{\varrho})\trans\boldsymbol{\alpha}_1\|_2\notag\\
&\ge \mu_1(\boldsymbol{\Xi}) -2\sqrt{\mu_1(\boldsymbol{\Xi})}\left[\max_{1\le j\le p}\|(\mathbf{Z}\trans\boldsymbol{\varrho})_j\|_2 \right]\|\boldsymbol{\alpha}_1\|_1\notag\\
 &\ge\mu_1(\boldsymbol{\Xi}) -2\sqrt{\mu_1(\boldsymbol{\Xi})}C_0\varpi \|\boldsymbol{\alpha}_1\|_1 ,\notag
\end{align}
where the  equality in the fourth line is because {\small $\|\boldsymbol{\alpha}_1\trans \mathbf{S}\mathcal{B}\|_2=\sqrt{\boldsymbol{\alpha}_1\trans \mathbf{S}\mathcal{B}\mathcal{B}\trans \mathbf{S}\boldsymbol{\alpha}_1}=  \sqrt{\boldsymbol{\alpha}_1\trans \mathbf{B}\boldsymbol{\alpha}_1}=\sqrt{\mu_1(\boldsymbol{\Xi})}$}, and the  inequality in the fifth line is because
\begin{align}
 \|(\mathbf{Z}\trans\boldsymbol{\varrho})\trans\boldsymbol{\alpha}_1\|_2&=\|\sum_{j=1}^p(\mathbf{Z}\trans\boldsymbol{\varrho})_j(\boldsymbol{\alpha}_1)_j\|_2\le \sum_{j=1}^p\|(\mathbf{Z}\trans\boldsymbol{\varrho})_j\|_2|(\boldsymbol{\alpha}_1)_j|\notag\\
&\le \left[\max_{1\le j\le p}\|(\mathbf{Z}\trans\boldsymbol{\varrho})_j\|_2 \right]\|\boldsymbol{\alpha}_1\|_1\le C_0\varpi \|\boldsymbol{\alpha}_1\|_1,\label{17700}
\end{align}
where $(\boldsymbol{\alpha}_1)_j$ is the $j$-th coordinate of the vector $\boldsymbol{\alpha}_1$ and the last inequality is due to the definition of $\Omega$. By  the similar arguments as in $\eqref{10017}$ and $\eqref{17700}$, we have
 \begin{align}
  \widehat{\boldsymbol{\alpha}}_1\trans \widehat{\mathbf{B}}\widehat{\boldsymbol{\alpha}}_1 &=\widehat{\boldsymbol{\alpha}}_1\trans \mathbf{B}\widehat{\boldsymbol{\alpha}}_1 +2 \widehat{\boldsymbol{\alpha}}_1\trans \mathbf{S}\mathcal{B} (\mathbf{Z}\trans\boldsymbol{\varrho})\trans\widehat{\boldsymbol{\alpha}}_1+\widehat{\boldsymbol{\alpha}}_1\trans (\mathbf{Z}\trans\boldsymbol{\varrho})(\mathbf{Z}\trans\boldsymbol{\varrho})\trans\widehat{\boldsymbol{\alpha}}_1\label{10020}\\
&\le\mu_1(\boldsymbol{\Xi}) +2 \widehat{\boldsymbol{\alpha}}_1\trans \mathbf{S}\mathcal{B} (\mathbf{Z}\trans\boldsymbol{\varrho})\trans\widehat{\boldsymbol{\alpha}}_1 +\widehat{\boldsymbol{\alpha}}_1\trans (\mathbf{Z}\trans\boldsymbol{\varrho})(\mathbf{Z}\trans\boldsymbol{\varrho})\trans\widehat{\boldsymbol{\alpha}}_1 \notag\\
&\le\mu_1(\boldsymbol{\Xi}) +2 \|\widehat{\boldsymbol{\alpha}}_1\trans \mathbf{S}\mathcal{B} \|_2\|(\mathbf{Z}\trans\boldsymbol{\varrho})\trans\widehat{\boldsymbol{\alpha}}_1\|_2 +\|(\mathbf{Z}\trans\boldsymbol{\varrho})\trans\widehat{\boldsymbol{\alpha}}_1\|_2^2 \notag\\
&\le\mu_1(\boldsymbol{\Xi}) +2 \sqrt{\widehat{\boldsymbol{\alpha}}_1\trans \mathbf{B}\widehat{\boldsymbol{\alpha}}_1}C_0\varpi \|\widehat{\boldsymbol{\alpha}}_1\|_1+C_0^2\varpi^2 \|\widehat{\boldsymbol{\alpha}}_1\|_1^2 \notag\\
&\le\mu_1(\boldsymbol{\Xi})+2\sqrt{\mu_1(\boldsymbol{\Xi})}C_0\varpi \|\widehat{\boldsymbol{\alpha}}_1\|_1+C_0^2\varpi^2 \|\widehat{\boldsymbol{\alpha}}_1\|_1^2\notag\\
&=(\sqrt{\mu_1(\boldsymbol{\Xi})}+C_0\varpi \|\widehat{\boldsymbol{\alpha}}_1\|_1)^2,\notag
\end{align} 
where the inequality in the fifth line is due to $\eqref{10019}$. By $\eqref{10018}$, $\eqref{10017}$ and $\eqref{10020}$, we have 
 \begin{align}
 &\left(\mu_1(\boldsymbol{\Xi}) -2\sqrt{\mu_1(\boldsymbol{\Xi})}C_0\varpi \|\boldsymbol{\alpha}_1\|_1\right)(1+\tau^{(1)}\|\widehat{\boldsymbol{\alpha}}_1\|^2_{\lambda^{(1)}})\label{10021}\\
\le&\boldsymbol{\alpha}_1\trans \widehat{\mathbf{B}}\boldsymbol{\alpha}_1(1+\tau^{(1)}\|\widehat{\boldsymbol{\alpha}}_1\|^2_{\lambda^{(1)}})\le \widehat{\boldsymbol{\alpha}}_1\trans \widehat{\mathbf{B}}\widehat{\boldsymbol{\alpha}}_1 (1+\tau^{(1)}\|\boldsymbol{\alpha}_1\|^2_{\lambda^{(1)}})\notag\\
\le& \left(\mu_1(\boldsymbol{\Xi})+2\sqrt{\mu_1(\boldsymbol{\Xi})}C_0\varpi \|\widehat{\boldsymbol{\alpha}}_1\|_1+C_0^2\varpi^2 \|\widehat{\boldsymbol{\alpha}}_1\|_1^2\right)(1+\tau^{(1)}\|\boldsymbol{\alpha}_1\|^2_{\lambda^{(1)}}).\notag
\end{align} 
  By $\eqref{10203}$ and the inequality $ \lambda^{(1)}\|\widehat{\boldsymbol{\alpha}}_1\|^2_1\le \|\widehat{\boldsymbol{\alpha}}_1\|^2_{\lambda^{(1)}}\le \|\widehat{\boldsymbol{\alpha}}_1\|^2_1$, the left hand side of the first inequality in $\eqref{10021}$
 \begin{align}
 &\left(\mu_1(\boldsymbol{\Xi}) -2\sqrt{\mu_1(\boldsymbol{\Xi})}C_0\varpi \|\boldsymbol{\alpha}_1\|_1\right)(1+\tau^{(1)}\|\widehat{\boldsymbol{\alpha}}_1\|^2_{\lambda^{(1)}}) \label{10024}\\
= \mu_1(\boldsymbol{\Xi}) &-2\sqrt{\mu_1(\boldsymbol{\Xi})}C_0\varpi \|\boldsymbol{\alpha}_1\|_1+\left(\mu_1(\boldsymbol{\Xi}) -2\sqrt{\mu_1(\boldsymbol{\Xi})}C_0\varpi \|\boldsymbol{\alpha}_1\|_1\right)\tau^{(1)}\|\widehat{\boldsymbol{\alpha}}_1\|^2_{\lambda^{(1)}}\notag\\
\ge \mu_1(\boldsymbol{\Xi})& -2\sqrt{\mu_1(\boldsymbol{\Xi})}C_0\varpi \|\boldsymbol{\alpha}_1\|_1  +\left(\mu_1(\boldsymbol{\Xi}) -2\sqrt{\mu_1(\boldsymbol{\Xi})}\sqrt{\mu_1(\boldsymbol{\Xi})} \hbar^{-1}\right)\tau^{(1)}\|\widehat{\boldsymbol{\alpha}}_1\|^2_{\lambda^{(1)}}\notag\\
=\mu_1(\boldsymbol{\Xi}) &-2\sqrt{\mu_1(\boldsymbol{\Xi})}C_0\varpi \|\boldsymbol{\alpha}_1\|_1 +(1-2\hbar^{-1})\mu_1(\boldsymbol{\Xi})\tau^{(1)}\|\widehat{\boldsymbol{\alpha}}_1\|^2_{\lambda^{(1)}}\notag\\
\ge \mu_1(\boldsymbol{\Xi})& -2\sqrt{\mu_1(\boldsymbol{\Xi})}C_0\varpi \|\boldsymbol{\alpha}_1\|_1+\lambda^{(1)}(1-2\hbar^{-1})\mu_1(\boldsymbol{\Xi})\tau^{(1)}\|\widehat{\boldsymbol{\alpha}}_1\|^2_1 \notag\\
\ge \mu_1(\boldsymbol{\Xi})& -2\sqrt{\mu_1(\boldsymbol{\Xi})}C_0\varpi \|\boldsymbol{\alpha}_1\|_1+c^{-1}(1-2\hbar^{-1})\mu_1(\boldsymbol{\Xi})\tau^{(1)} \|\widehat{\boldsymbol{\alpha}}_1\|^2_1\notag.
\end{align} 
where we assume that $\hbar$ is large enough so that $1-2\hbar^{-1}>0$, and the last inequality is due to $\lambda^{(1)}>c^{-1}$ by the second condition in $\eqref{10216}$. The right hand side of the last inequality in $\eqref{10021}$
 \begin{align}
  &\left(\mu_1(\boldsymbol{\Xi})+2\sqrt{\mu_1(\boldsymbol{\Xi})}C_0\varpi \|\widehat{\boldsymbol{\alpha}}_1\|_1+C_0^2\varpi^2 \|\widehat{\boldsymbol{\alpha}}_1\|_1^2\right)(1+\tau^{(1)}\|\boldsymbol{\alpha}_1\|^2_{\lambda^{(1)}})\notag\\=&\mu_1(\boldsymbol{\Xi})+\mu_1(\boldsymbol{\Xi})\tau^{(1)}\|\boldsymbol{\alpha}_1\|^2_{\lambda^{(1)}} \notag\\
	&+\left( 2\sqrt{\mu_1(\boldsymbol{\Xi})}C_0\varpi \|\widehat{\boldsymbol{\alpha}}_1\|_1+C_0^2\varpi^2 \|\widehat{\boldsymbol{\alpha}}_1\|_1^2\right)(1+\tau^{(1)}\|\boldsymbol{\alpha}_1\|^2_{\lambda^{(1)}})\notag\\
	\le& \mu_1(\boldsymbol{\Xi})+\mu_1(\boldsymbol{\Xi})\tau^{(1)}\|\boldsymbol{\alpha}_1\|^2_1 \notag\\
	&+\left( 2\sqrt{\mu_1(\boldsymbol{\Xi})}C_0\varpi \|\widehat{\boldsymbol{\alpha}}_1\|_1+C_0^2\varpi^2  \|\widehat{\boldsymbol{\alpha}}_1\|_1^2 \right)(1+ A^{(1)}\hbar^{-1}),\label{10025}
\end{align} 
where the last inequality follows from the second inequality in $\eqref{10203}$. Combining $\eqref{10021}$, $\eqref{10024}$ and $\eqref{10025}$, we obtain
 \begin{align*}
 & -2\sqrt{\mu_1(\boldsymbol{\Xi})}C_0\varpi \|\boldsymbol{\alpha}_1\|_1+c^{-1}(1-2\hbar^{-1})\mu_1(\boldsymbol{\Xi})\tau^{(1)} \|\widehat{\boldsymbol{\alpha}}_1\|^2_1\\
\le&\mu_1(\boldsymbol{\Xi})\tau^{(1)}\|\boldsymbol{\alpha}_1\|^2_1 +\left( 2\sqrt{\mu_1(\boldsymbol{\Xi})}C_0\varpi \|\widehat{\boldsymbol{\alpha}}_1\|_1+C_0^2\varpi^2  \|\widehat{\boldsymbol{\alpha}}_1\|_1^2 \right)(1+ A^{(1)}\hbar^{-1})
\end{align*} 
where we have canceled a $\mu_1(\boldsymbol{\Xi})$ on each side. Multiplying $\|\boldsymbol{\alpha}_1\|_1$ on both sides of the above inequality and shifting the first term on the left to the right, by the definition of $\tau^{(1)}$,    we obtain 
 \begin{align*}
 & c^{-1}(1-2\hbar^{-1})\sqrt{\mu_1(\boldsymbol{\Xi})}A^{(1)}C_0\varpi\|\widehat{\boldsymbol{\alpha}}_1\|^2_1 \notag\\
\le& 2\sqrt{\mu_1(\boldsymbol{\Xi})}C_0\varpi \|\boldsymbol{\alpha}_1\|_1^2+\sqrt{\mu_1(\boldsymbol{\Xi})}A^{(1)}C_0 \varpi \|\boldsymbol{\alpha}_1\|_1^2 \notag\\
+&\left( 2\sqrt{\mu_1(\boldsymbol{\Xi})}C_0\varpi \|\widehat{\boldsymbol{\alpha}}_1\|_1 \|\boldsymbol{\alpha}_1\|_1+C_0^2\varpi^2 \|\boldsymbol{\alpha}_1\|_1 \|\widehat{\boldsymbol{\alpha}}_1\|_1^2\right)(1+ A^{(1)}\hbar^{-1})\notag\\
\le& \sqrt{\mu_1(\boldsymbol{\Xi})}C_0(2+A^{(1)})\varpi \|\boldsymbol{\alpha}_1\|_1^2 \notag\\
 +&\left( 2\sqrt{\mu_1(\boldsymbol{\Xi})}C_0\varpi \frac{ \|\widehat{\boldsymbol{\alpha}}_1\|_1^2+ \|\boldsymbol{\alpha}_1\|_1^2 }{2}+C_0 \varpi \sqrt{\mu_1(\boldsymbol{\Xi})} \hbar^{-1}  \|\widehat{\boldsymbol{\alpha}}_1\|_1^2\right)(1+ A^{(1)}\hbar^{-1})\notag\\
=& \sqrt{\mu_1(\boldsymbol{\Xi})}C_0[3+A^{(1)}(1+\hbar^{-1})]\varpi \|\boldsymbol{\alpha}_1\|_1^2 \notag\\
+&\sqrt{\mu_1(\boldsymbol{\Xi})}\varpi  (1+\hbar^{-1}) \|\widehat{\boldsymbol{\alpha}}_1\|_1^2 C_0(1+ A^{(1)}\hbar^{-1}),
\end{align*} 
where the third inequality follows from the first inequality in $\eqref{10203}$. The above inequality, after a simplification, gives 
 \begin{align}
 & \left\{[c^{-1}(1-2\hbar^{-1})-(1+\hbar^{-1})\hbar^{-1}]A^{(1)}-(1+\hbar^{-1})\right\}\|\widehat{\boldsymbol{\alpha}}_1\|^2_1\notag\\
\le& [3+A^{(1)}(1+\hbar^{-1})]\|\boldsymbol{\alpha}_1\|_1^2,\label{10220}
\end{align} 
where we have canceled $\sqrt{\mu_1(\boldsymbol{\Xi})}C_0\varpi$ on both sides. Because 
 \begin{align*}
 & \lim_{\substack{ A^{(1)}\to\infty\\\hbar\to\infty}}\frac{3+A^{(1)}(1+\hbar^{-1})}{[c^{-1}(1-2\hbar^{-1})-(1+\hbar^{-1})\hbar^{-1}]A^{(1)}-(1+\hbar^{-1})}= c.
\end{align*} 
 Therefore, there exist $(A_1^{L})^\prime$ and $(\hbar_0)^\prime$ only depending on $c$ such that for any $A^{(1)}\ge (A_1^{L})^\prime $ and $\hbar\ge (\hbar_0)^\prime$, we have  
 \begin{align}
 & \|\widehat{\boldsymbol{\alpha}}_1\|_1\le D_1\|\boldsymbol{\alpha}_1\|_1,\label{50023}
\end{align} 
  where $D_1=\sqrt{6c}$.  Therefore, we obtained the first inequality in the lemma. To prove the second one,  by $\eqref{10020}$ and $\eqref{50023}$,
 \begin{align*}
 &\widehat{\boldsymbol{\alpha}}_1\trans \widehat{\mathbf{B}}\widehat{\boldsymbol{\alpha}}_1 \le (\sqrt{\mu_1(\boldsymbol{\Xi})}+C_0\varpi \|\widehat{\boldsymbol{\alpha}}_1\|_1)^2\le (\sqrt{\mu_1(\boldsymbol{\Xi})}+C_0\varpi D_1\|\boldsymbol{\alpha}_1\|_1)^2\\
\le& (\sqrt{\mu_1(\boldsymbol{\Xi})}+D_1\hbar^{-1}\sqrt{\mu_1(\boldsymbol{\Xi})})^2= (1+\hbar^{-1}D_1)^2\mu_1(\boldsymbol{\Xi}),
\end{align*} 
where the first inequality in the second line is due to the first inequality in $\eqref{10203}$. 

For the last inequality in the lemma, by $\eqref{10202}$,
 \begin{align}
 &\boldsymbol{\alpha}_1\trans \widehat{\mathbf{B}}\boldsymbol{\alpha}_1-\widehat{\boldsymbol{\alpha}}_1\trans \widehat{\mathbf{B}}\widehat{\boldsymbol{\alpha}}_1\label{10011}\\
=&(\boldsymbol{\alpha}_1\trans \mathbf{B}\boldsymbol{\alpha}_1-\widehat{\boldsymbol{\alpha}}_1\trans \mathbf{B}\widehat{\boldsymbol{\alpha}}_1)+ \boldsymbol{\alpha}_1\trans (\widehat{\mathbf{B}}-\mathbf{B}) \boldsymbol{\alpha}_1- \widehat{\boldsymbol{\alpha}}_1\trans (\widehat{\mathbf{B}}-\mathbf{B})\widehat{\boldsymbol{\alpha}}_1 \notag\\
=&(\boldsymbol{\alpha}_1\trans \mathbf{B}\boldsymbol{\alpha}_1-\widehat{\boldsymbol{\alpha}}_1\trans \mathbf{B}\widehat{\boldsymbol{\alpha}}_1)+2\left(\boldsymbol{\alpha}_1\trans \mathbf{S}\mathcal{B} (\mathbf{Z}\trans\boldsymbol{\varrho})\trans\boldsymbol{\alpha}_1-\widehat{\boldsymbol{\alpha}}_1\trans \mathbf{S}\mathcal{B} (\mathbf{Z}\trans\boldsymbol{\varrho})\trans\widehat{\boldsymbol{\alpha}}_1\right) \notag\\
&+\left(\boldsymbol{\alpha}_1\trans (\mathbf{Z}\trans\boldsymbol{\varrho})(\mathbf{Z}\trans\boldsymbol{\varrho})\trans\boldsymbol{\alpha}_1-\widehat{\boldsymbol{\alpha}}_1\trans (\mathbf{Z}\trans\boldsymbol{\varrho})(\mathbf{Z}\trans\boldsymbol{\varrho})\trans\widehat{\boldsymbol{\alpha}}_1\right) \notag\\
&\ge (\boldsymbol{\alpha}_1\trans \mathbf{B}\boldsymbol{\alpha}_1-\widehat{\boldsymbol{\alpha}}_1\trans \mathbf{B}\widehat{\boldsymbol{\alpha}}_1)-2\left|\boldsymbol{\alpha}_1\trans \mathbf{S}\mathcal{B} (\mathbf{Z}\trans\boldsymbol{\varrho})\trans\boldsymbol{\alpha}_1-\widehat{\boldsymbol{\alpha}}_1\trans \mathbf{S}\mathcal{B} (\mathbf{Z}\trans\boldsymbol{\varrho})\trans\widehat{\boldsymbol{\alpha}}_1\right|\notag\\
&-\left|\boldsymbol{\alpha}_1\trans (\mathbf{Z}\trans\boldsymbol{\varrho})(\mathbf{Z}\trans\boldsymbol{\varrho})\trans\boldsymbol{\alpha}_1-\widehat{\boldsymbol{\alpha}}_1\trans (\mathbf{Z}\trans\boldsymbol{\varrho})(\mathbf{Z}\trans\boldsymbol{\varrho})\trans\widehat{\boldsymbol{\alpha}}_1\right|.\notag
\end{align} 
We will estimate the three terms on the right hand side of the last equality above. To estimate $(\boldsymbol{\alpha}_1\trans \mathbf{B}\boldsymbol{\alpha}_1-\widehat{\boldsymbol{\alpha}}_1\trans \mathbf{B}\widehat{\boldsymbol{\alpha}}_1)$, note that $\boldsymbol{\alpha}_1\trans \mathbf{B}\boldsymbol{\alpha}_1=\boldsymbol{\alpha}_1\trans\mathbf{Z}\trans \boldsymbol{\Xi}\mathbf{Z}\boldsymbol{\alpha}_1=\boldsymbol{\gamma}_1\trans\boldsymbol{\Xi}\boldsymbol{\gamma}_1$, $\widehat{\boldsymbol{\alpha}}_1\trans \mathbf{B}\widehat{\boldsymbol{\alpha}}_1=\widehat{\boldsymbol{\gamma}}_1\trans\boldsymbol{\Xi}\widehat{\boldsymbol{\gamma}}_1$, $\|\boldsymbol{\gamma}_1\|_2=\|\widehat{\boldsymbol{\gamma}}_1\|_2=1$ and the following eigen-decomposition
 \begin{align}
 \boldsymbol{\Xi}=\sum_{k=1}^K\mu_k(\boldsymbol{\Xi})\boldsymbol{\gamma}_k\boldsymbol{\gamma}_k\trans.\label{10204}
\end{align} 
We have
 \begin{align}
 &\boldsymbol{\alpha}_1\trans \mathbf{B}\boldsymbol{\alpha}_1-\widehat{\boldsymbol{\alpha}}_1\trans \mathbf{B}\widehat{\boldsymbol{\alpha}}_1 =\boldsymbol{\gamma}_1\trans\boldsymbol{\Xi}\boldsymbol{\gamma}_1-\widehat{\boldsymbol{\gamma}}_1\trans\boldsymbol{\Xi}\widehat{\boldsymbol{\gamma}}_1\label{10010}\\
&=\mu_1(\boldsymbol{\Xi})-\sum_{k=1}^K\mu_k(\boldsymbol{\Xi})(\widehat{\boldsymbol{\gamma}}_1\trans\boldsymbol{\gamma}_k)^2\notag\\
&\ge \mu_1(\boldsymbol{\Xi})-\mu_1(\boldsymbol{\Xi})(\widehat{\boldsymbol{\gamma}}_1\trans\boldsymbol{\gamma}_1)^2-\mu_2(\boldsymbol{\Xi})\sum_{k=2}^K(\widehat{\boldsymbol{\gamma}}_1\trans\boldsymbol{\gamma}_k)^2\notag\\
&=[\mu_1(\boldsymbol{\Xi})-\mu_2(\boldsymbol{\Xi})][1-(\widehat{\boldsymbol{\gamma}}_1\trans\boldsymbol{\gamma}_1)^2]+\mu_2(\boldsymbol{\Xi})[1-\sum_{k=1}^K (\widehat{\boldsymbol{\gamma}}_1\trans\boldsymbol{\gamma}_k)^2]\notag\\
&\ge [\mu_1(\boldsymbol{\Xi})-\mu_2(\boldsymbol{\Xi})][1-(\widehat{\boldsymbol{\gamma}}_1\trans\boldsymbol{\gamma}_1)^2]+\mu_2(\boldsymbol{\Xi})[1-\|\widehat{\boldsymbol{\gamma}}_1\|_2^2]\notag\\
&= [\mu_1(\boldsymbol{\Xi})-\mu_2(\boldsymbol{\Xi})][1-(\widehat{\boldsymbol{\gamma}}_1\trans\boldsymbol{\gamma}_1)^2]\ge c_2\mu_1(\boldsymbol{\Xi}) [1-(\widehat{\boldsymbol{\gamma}}_1\trans\boldsymbol{\gamma}_1)^2]\notag\\
&\ge c_2\mu_1(\boldsymbol{\Xi}) [1-(\widehat{\boldsymbol{\gamma}}_1\trans\boldsymbol{\gamma}_1)]= c_2\mu_1(\boldsymbol{\Xi}) \frac{1}{2}[2-2(\widehat{\boldsymbol{\gamma}}_1\trans\boldsymbol{\gamma}_1)]\notag\\
&=\frac{1}{2}c_2\mu_1(\boldsymbol{\Xi})[\|\widehat{\boldsymbol{\gamma}}_1\|^2_2+\|\boldsymbol{\gamma}_1\|_2^2 -2(\widehat{\boldsymbol{\gamma}}_1\trans\boldsymbol{\gamma}_1)]=\frac{1}{2}c_2\mu_1(\boldsymbol{\Xi})\|\widehat{\boldsymbol{\gamma}}_1-\boldsymbol{\gamma}_1\|_2^2,\notag
\end{align} 
where the  last inequality in the sixth line is due to Condition \ref{condition_2} (a) and the first inequality in the seventh line  is because we have assumed that $\widehat{\boldsymbol{\gamma}}_1\trans\boldsymbol{\gamma}_1\ge 0$.

For the second term on the right hand side of last inequality of $\eqref{10011}$,
 \begin{align*}
 & 2\left|\boldsymbol{\alpha}_1\trans \mathbf{S}\mathcal{B} (\mathbf{Z}\trans\boldsymbol{\varrho})\trans\boldsymbol{\alpha}_1-\widehat{\boldsymbol{\alpha}}_1\trans \mathbf{S}\mathcal{B} (\mathbf{Z}\trans\boldsymbol{\varrho})\trans\widehat{\boldsymbol{\alpha}}_1\right| \notag\\
 =&2\left|(\boldsymbol{\alpha}_1-\widehat{\boldsymbol{\alpha}}_1)\trans \mathbf{S}\mathcal{B}(\mathbf{Z}\trans\boldsymbol{\varrho})\trans\boldsymbol{\alpha}_1+ \widehat{\boldsymbol{\alpha}}_1\trans \mathbf{S}\mathcal{B} (\mathbf{Z}\trans\boldsymbol{\varrho})\trans(\boldsymbol{\alpha}_1-\widehat{\boldsymbol{\alpha}}_1) \right|\notag\\
\le &2\|(\boldsymbol{\alpha}_1-\widehat{\boldsymbol{\alpha}}_1)\trans \mathbf{S}\mathcal{B}\|_2\|(\mathbf{Z}\boldsymbol{\alpha}_1)\trans\boldsymbol{\varrho}\|_2+2\|\widehat{\boldsymbol{\alpha}}_1\trans \mathbf{S}\mathcal{B}\|_2\|(\mathbf{Z}\trans\boldsymbol{\varrho})\trans(\boldsymbol{\alpha}_1-\widehat{\boldsymbol{\alpha}}_1)\|_2, 
\end{align*} 
where 
\begin{align*}
 &\|(\boldsymbol{\alpha}_1-\widehat{\boldsymbol{\alpha}}_1)\trans \mathbf{S}\mathcal{B}\|_2=\sqrt{(\boldsymbol{\alpha}_1-\widehat{\boldsymbol{\alpha}}_1)\trans \mathbf{S}\mathcal{B}\mathcal{B}\trans \mathbf{S}(\boldsymbol{\alpha}_1-\widehat{\boldsymbol{\alpha}}_1)}\notag\\
=&\sqrt{(\boldsymbol{\alpha}_1-\widehat{\boldsymbol{\alpha}}_1)\trans \mathbf{B}(\boldsymbol{\alpha}_1-\widehat{\boldsymbol{\alpha}}_1)}=\sqrt{(\boldsymbol{\gamma}_1-\widehat{\boldsymbol{\gamma}}_1)\trans \boldsymbol{\Xi}(\boldsymbol{\gamma}_1-\widehat{\boldsymbol{\gamma}}_1)}\notag\\
\le& \sqrt{\mu_1(\boldsymbol{\Xi})(\boldsymbol{\gamma}_1-\widehat{\boldsymbol{\gamma}}_1)\trans(\boldsymbol{\gamma}_1-\widehat{\boldsymbol{\gamma}}_1)}=\sqrt{\mu_1(\boldsymbol{\Xi})(\boldsymbol{\alpha}_1-\widehat{\boldsymbol{\alpha}}_1)\trans \mathbf{S}(\boldsymbol{\alpha}_1-\widehat{\boldsymbol{\alpha}}_1)}\notag\\
\le &\sqrt{\mu_1(\boldsymbol{\Xi})\|\mathbf{S}\|_{\infty}\|\boldsymbol{\alpha}_1-\widehat{\boldsymbol{\alpha}}_1\|_1^2}=\sqrt{\mu_1(\boldsymbol{\Xi})}\|\boldsymbol{\alpha}_1-\widehat{\boldsymbol{\alpha}}_1\|_1
\end{align*} 
 and by $\eqref{10019}$, {\small $\|\widehat{\boldsymbol{\alpha}}_1\trans \mathbf{S}\mathcal{B}\|_2=\sqrt{\widehat{\boldsymbol{\alpha}}_1\trans \mathbf{S}\mathcal{B}\mathcal{B}\trans \mathbf{S}\widehat{\boldsymbol{\alpha}}_1}=\sqrt{\widehat{\boldsymbol{\alpha}}_1\trans \mathbf{B}\widehat{\boldsymbol{\alpha}}_1}\le \sqrt{\mu_1(\boldsymbol{\Xi})}$}. By the definition $\eqref{1033}$ of $\Omega$,  {\small $\|(\mathbf{Z}\trans\boldsymbol{\varrho})\trans(\boldsymbol{\alpha}_1-\widehat{\boldsymbol{\alpha}}_1)\|_2\le \max_{1\le j\le p}\|(\mathbf{Z}\trans\boldsymbol{\varrho})_j\|_2\|\boldsymbol{\alpha}_1-\widehat{\boldsymbol{\alpha}}_1\|_1\le C_0\varpi\|\boldsymbol{\alpha}_1-\widehat{\boldsymbol{\alpha}}_1\|_1$, and similarly, $\|(\mathbf{Z}\boldsymbol{\alpha}_1)\trans\boldsymbol{\varrho}\|_2\le C_0\varpi$}. Hence, the inequalities above    lead to
 \begin{align}
 & 2\left|\boldsymbol{\alpha}_1\trans \mathbf{S}\mathcal{B} (\mathbf{Z}\trans\boldsymbol{\varrho})\trans\boldsymbol{\alpha}_1-\widehat{\boldsymbol{\alpha}}_1\trans \mathbf{S}\mathcal{B} (\mathbf{Z}\trans\boldsymbol{\varrho})\trans\widehat{\boldsymbol{\alpha}}_1\right| \label{10015}\\
\le& 2\sqrt{\mu_1(\boldsymbol{\Xi})}\|\boldsymbol{\alpha}_1-\widehat{\boldsymbol{\alpha}}_1\|_1C_0  \varpi +2\sqrt{\mu_1(\boldsymbol{\Xi})}C_0\varpi\|\boldsymbol{\alpha}_1-\widehat{\boldsymbol{\alpha}}_1\|_1\notag\\
\le& 4 \sqrt{\mu_1(\boldsymbol{\Xi})}C_0\varpi\|\boldsymbol{\alpha}_1-\widehat{\boldsymbol{\alpha}}_1\|_1\notag 
\end{align} 
For the third term on the right hand side of last inequality of $\eqref{10011}$, we have 
 \begin{align}
  &\left|\boldsymbol{\alpha}_1\trans (\mathbf{Z}\trans\boldsymbol{\varrho})(\mathbf{Z}\trans\boldsymbol{\varrho})\trans\boldsymbol{\alpha}_1-\widehat{\boldsymbol{\alpha}}_1\trans (\mathbf{Z}\trans\boldsymbol{\varrho})(\mathbf{Z}\trans\boldsymbol{\varrho})\trans\widehat{\boldsymbol{\alpha}}_1\right|\notag\\
	=&\left|(\boldsymbol{\alpha}_1-\widehat{\boldsymbol{\alpha}}_1)\trans (\mathbf{Z}\trans\boldsymbol{\varrho})(\mathbf{Z}\trans\boldsymbol{\varrho})\trans(\boldsymbol{\alpha}_1+\widehat{\boldsymbol{\alpha}}_1) \right|\notag\\
	\le &\left(\max_{1\le j\le p}\|(\mathbf{Z}\trans\boldsymbol{\varrho})_j\|_2\right)^2\|\boldsymbol{\alpha}_1-\widehat{\boldsymbol{\alpha}}_1\|_1\|\boldsymbol{\alpha}_1+\widehat{\boldsymbol{\alpha}}_1\|_1\notag\\
	\le & C_0^2\varpi^2\|\boldsymbol{\alpha}_1-\widehat{\boldsymbol{\alpha}}_1\|_1(\|\boldsymbol{\alpha}_1\|_1+\|\widehat{\boldsymbol{\alpha}}_1\|_1)\notag\\
	\le &C_0^2\varpi^2\|\boldsymbol{\alpha}_1-\widehat{\boldsymbol{\alpha}}_1\|_1(1+  D_1)\|\boldsymbol{\alpha}_1\|_1\notag\\
	=&\left[C_0\varpi\|\boldsymbol{\alpha}_1\|_1\right] (1+  D_1)C_0\varpi\|\boldsymbol{\alpha}_1-\widehat{\boldsymbol{\alpha}}_1\|_1\notag\\
\le &    \hbar^{-1}\sqrt{\mu_1(\boldsymbol{\Xi})}(1+  D_1)C_0\varpi\|\boldsymbol{\alpha}_1-\widehat{\boldsymbol{\alpha}}_1\|_1\notag\\
\le &\sqrt{\mu_1(\boldsymbol{\Xi})}[\hbar^{-1} (1+D_1)]C_0\varpi\|\boldsymbol{\alpha}_1-\widehat{\boldsymbol{\alpha}}_1\|_1, \label{10027}
\end{align} 
where the inequality in the fifth line follows from $\|\widehat{\boldsymbol{\alpha}}_1\|_1\le D_1\|\boldsymbol{\alpha}_1\|_1$, and the inequality in the seventh line is due to $\eqref{10203}$. Combining $\eqref{10011}$-$\eqref{10027}$ gives
 \begin{align*}
 &\boldsymbol{\alpha}_1\trans \widehat{\mathbf{B}}\boldsymbol{\alpha}_1-\widehat{\boldsymbol{\alpha}}_1\trans \widehat{\mathbf{B}}\widehat{\boldsymbol{\alpha}}_1\notag\\
\ge& \frac{1}{2}c_2\mu_1(\boldsymbol{\Xi})\|\widehat{\boldsymbol{\gamma}}_1-\boldsymbol{\gamma}_1\|_2^2-[4+\hbar^{-1} (1+D_1)]\sqrt{\mu_1(\boldsymbol{\Xi})}C_0\varpi\|\boldsymbol{\alpha}_1-\widehat{\boldsymbol{\alpha}}_1\|_1.
\end{align*} 
We have proved the lemma.\\

 \end{proof}

\begin{proof} $\ref{lemma_1}$.\\
 Note that
 \begin{align}
  \tau^{(1)}\|\boldsymbol{\alpha}_1\|^2_{\lambda^{(1)}}-\tau^{(1)}\|\widehat{\boldsymbol{\alpha}}_1\|^2_{\lambda^{(1)}}\label{50032}\\
=\frac{A^{(1)}C_0\varpi}{\|\boldsymbol{\alpha}_1\|_1\sqrt{\mu_1(\boldsymbol{\Xi})}}\left[(1-\lambda^{(1)})(\|\boldsymbol{\alpha}_1\|^2_2-\|\widehat{\boldsymbol{\alpha}}_1\|^2_2)+\lambda^{(1)}(\|\boldsymbol{\alpha}_1\|^2_1-\|\widehat{\boldsymbol{\alpha}}_1\|^2_1)\right] .\notag
\end{align} 
Because $\|(\boldsymbol{\alpha}_1)_{J_1^c}\|_1=0$, $\|(\boldsymbol{\alpha}_1)_{J_1}\|_1=\|\boldsymbol{\alpha}_1\|_1$, $\|(\boldsymbol{\alpha}_1)_{J_1^c}\|_2=0$ and $\|(\boldsymbol{\alpha}_1)_{J_1}\|_2=\|\boldsymbol{\alpha}_1\|_2$, we have
 \begin{align}
 &\frac{A^{(1)}C_0\varpi}{\|\boldsymbol{\alpha}_1\|_1\sqrt{\mu_1(\boldsymbol{\Xi})}}(1-\lambda^{(1)})(\|\boldsymbol{\alpha}_1\|^2_2-\|\widehat{\boldsymbol{\alpha}}_1\|^2_2)\label{10031}\\
=&\frac{A^{(1)}C_0\varpi}{\|\boldsymbol{\alpha}_1\|_1\sqrt{\mu_1(\boldsymbol{\Xi})}}(1-\lambda^{(1)})(\|(\boldsymbol{\alpha}_1)_{J_1}\|^2_2-\|(\widehat{\boldsymbol{\alpha}}_1)_{J_1}\|_2^2-\|(\widehat{\boldsymbol{\alpha}}_1)_{J_1^c}\|^2_2)\notag\\
\le &\frac{A^{(1)}C_0\varpi}{\|\boldsymbol{\alpha}_1\|_1\sqrt{\mu_1(\boldsymbol{\Xi})}}(1-\lambda^{(1)})(\|(\boldsymbol{\alpha}_1)_{J_1}\|^2_2-\|(\widehat{\boldsymbol{\alpha}}_1)_{J_1}\|^2_2)\notag\\
= &\frac{A^{(1)}C_0\varpi}{\|\boldsymbol{\alpha}_1\|_1\sqrt{\mu_1(\boldsymbol{\Xi})}}(1-\lambda^{(1)})(\|( \boldsymbol{\alpha}_1)_{J_1}\|_2-\|(\widehat{\boldsymbol{\alpha}}_1 )_{J_1}\|_2)(\|(\boldsymbol{\alpha}_1)_{J_1}\|_2+\|(\widehat{\boldsymbol{\alpha}}_1)_{J_1}\|_2)\notag\\
\le &\frac{A^{(1)}C_0\varpi}{\|\boldsymbol{\alpha}_1\|_1\sqrt{\mu_1(\boldsymbol{\Xi})}}(1-\lambda^{(1)})\|(\widehat{\boldsymbol{\alpha}}_1-\boldsymbol{\alpha}_1)_{J_1}\|_2(\|\boldsymbol{\alpha}_1\|_2+\|\widehat{\boldsymbol{\alpha}}_1\|_2)\notag\\
 \le &\frac{A^{(1)}C_0\varpi}{\|\boldsymbol{\alpha}_1\|_1\sqrt{\mu_1(\boldsymbol{\Xi})}}(1-\lambda^{(1)})\|(\widehat{\boldsymbol{\alpha}}_1-\boldsymbol{\alpha}_1)_{J_1}\|_1(\|\boldsymbol{\alpha}_1\|_1+\|\widehat{\boldsymbol{\alpha}}_1\|_1)\notag\\
=& \frac{A^{(1)}C_0\varpi}{ \sqrt{\mu_1(\boldsymbol{\Xi})}}(1-\lambda^{(1)})\|(\widehat{\boldsymbol{\alpha}}_1-\boldsymbol{\alpha}_1)_{J_1}\|_1(1 +\|\widehat{\boldsymbol{\alpha}}_1\|_1/\|\boldsymbol{\alpha}_1\|_1), \notag
\end{align} 
and
 \begin{align}
 &\frac{A^{(1)}C_0\varpi}{\|\boldsymbol{\alpha}_1\|_1\sqrt{\mu_1(\boldsymbol{\Xi})}}\lambda^{(1)}(\|\boldsymbol{\alpha}_1\|^2_1-\|\widehat{\boldsymbol{\alpha}}_1\|^2_1)\label{10032}\\
=&\frac{A^{(1)}C_0\varpi}{\|\boldsymbol{\alpha}_1\|_1\sqrt{\mu_1(\boldsymbol{\Xi})}}\lambda^{(1)}(\|\boldsymbol{\alpha}_1\|_1-\|\widehat{\boldsymbol{\alpha}}_1\|_1)(\|\boldsymbol{\alpha}_1\|_1+\|\widehat{\boldsymbol{\alpha}}_1\|_1) \notag\\
=&\frac{A^{(1)}C_0\varpi}{ \sqrt{\mu_1(\boldsymbol{\Xi})}}\lambda^{(1)}(\|(\boldsymbol{\alpha}_1)_{J_1}\|_1-\|(\widehat{\boldsymbol{\alpha}}_1)_{J_1}\|_1-\|(\widehat{\boldsymbol{\alpha}}_1)_{J_1^c}\|_1)(1 +\|\widehat{\boldsymbol{\alpha}}_1\|_1/\|\boldsymbol{\alpha}_1\|_1) \notag\\
\le &\frac{A^{(1)}C_0\varpi}{ \sqrt{\mu_1(\boldsymbol{\Xi})}}(\lambda^{(1)}\|(\boldsymbol{\alpha}_1-\widehat{\boldsymbol{\alpha}}_1)_{J_1}\|_1-\lambda^{(1)}\|(\widehat{\boldsymbol{\alpha}}_1)_{J_1^c}\|_1)(1 +\|\widehat{\boldsymbol{\alpha}}_1\|_1/\|\boldsymbol{\alpha}_1\|_1) .\notag
\end{align} 
 By $\eqref{50032}$-$\eqref{10032}$, we have
 \begin{align}
 & \qquad \tau^{(1)}\|\boldsymbol{\alpha}_1\|^2_{\lambda^{(1)}}-\tau^{(1)}\|\widehat{\boldsymbol{\alpha}}_1\|^2_{\lambda^{(1)}} \label{356}\\
\le &\frac{A^{(1)}C_0\varpi}{ \sqrt{\mu_1(\boldsymbol{\Xi})}}( \|(\boldsymbol{\alpha}_1-\widehat{\boldsymbol{\alpha}}_1)_{J_1}\|_1-\lambda^{(1)}\|(\widehat{\boldsymbol{\alpha}}_1)_{J_1^c}\|_1)(1 +\|\widehat{\boldsymbol{\alpha}}_1\|_1/\|\boldsymbol{\alpha}_1\|_1) .\notag
\end{align} 

In the following, we will consider Cases (a) and (b), separately.\\

{\bf Case (a)}: $\|(\boldsymbol{\alpha}_1-\widehat{\boldsymbol{\alpha}}_1)_{J_1}\|_1<\lambda^{(1)} \|(\widehat{\boldsymbol{\alpha}}_1)_{J_1^c}\|_1$.\\

 In this case, by $\eqref{356}$, we have  $\tau^{(1)}\|\boldsymbol{\alpha}_1\|^2_{\lambda^{(1)}}-\tau^{(1)}\|\widehat{\boldsymbol{\alpha}}_1\|^2_{\lambda^{(1)}}<0$. By the following inequality (which is the first inequality of $\eqref{10029}$),
 \begin{align*}
 \widehat{\boldsymbol{\alpha}}_1\trans \widehat{\mathbf{B}}\widehat{\boldsymbol{\alpha}}_1(\tau^{(1)}\|\boldsymbol{\alpha}_1\|^2_{\lambda^{(1)}}-\tau^{(1)}\|\widehat{\boldsymbol{\alpha}}_1\|^2_{\lambda^{(1)}})\ge (\boldsymbol{\alpha}_1\trans \widehat{\mathbf{B}}\boldsymbol{\alpha}_1-\widehat{\boldsymbol{\alpha}}_1\trans \widehat{\mathbf{B}}\widehat{\boldsymbol{\alpha}}_1)(1+ \tau^{(1)}\|\widehat{\boldsymbol{\alpha}}_1\|^2_{\lambda^{(1)}}),
\end{align*} 
we have $\boldsymbol{\alpha}_1\trans \widehat{\mathbf{B}}\boldsymbol{\alpha}_1<\widehat{\boldsymbol{\alpha}}_1\trans \widehat{\mathbf{B}}\widehat{\boldsymbol{\alpha}}_1$. Now by $\eqref{10017}$ and $\eqref{10203}$,
 \begin{align}
 &\widehat{\boldsymbol{\alpha}}_1\trans \widehat{\mathbf{B}}\widehat{\boldsymbol{\alpha}}_1>\boldsymbol{\alpha}_1\trans \widehat{\mathbf{B}}\boldsymbol{\alpha}_1 \ge\mu_1(\boldsymbol{\Xi}) -2\sqrt{\mu_1(\boldsymbol{\Xi})}C_0\varpi \|\boldsymbol{\alpha}_1\|_1\ge (1-2\hbar^{-1})\mu_1(\boldsymbol{\Xi})\notag
\end{align} 
which together with $\eqref{356}$ give 
 \begin{align}
 &\widehat{\boldsymbol{\alpha}}_1\trans \widehat{\mathbf{B}}\widehat{\boldsymbol{\alpha}}_1(\tau^{(1)}\|\boldsymbol{\alpha}_1\|^2_{\lambda^{(1)}}-\tau^{(1)}\|\widehat{\boldsymbol{\alpha}}_1\|^2_{\lambda^{(1)}})\notag\\
\le&   (1-2\hbar^{-1})\mu_1(\boldsymbol{\Xi})\frac{A^{(1)}C_0\varpi}{ \sqrt{\mu_1(\boldsymbol{\Xi})}}\left[\|(\boldsymbol{\alpha}_1-\widehat{\boldsymbol{\alpha}}_1)_{J_1}\|_1-\lambda^{(1)} \|(\widehat{\boldsymbol{\alpha}}_1)_{J_1^c}\|_1\right](1 +\|\widehat{\boldsymbol{\alpha}}_1\|_1/\|\boldsymbol{\alpha}_1\|_1) \notag\\
\le &  (1-2\hbar^{-1}) A^{(1)}C_0\mu_1(\boldsymbol{\Xi})^{1/2}\varpi(\|(\boldsymbol{\alpha}_1-\widehat{\boldsymbol{\alpha}}_1)_{J_1}\|_1-\lambda^{(1)}\|(\widehat{\boldsymbol{\alpha}}_1)_{J_1^c}\|_1)\le 0,\notag
\end{align} 
where we use $1 +\|\widehat{\boldsymbol{\alpha}}_1\|_1/\|\boldsymbol{\alpha}_1\|_1\ge 1$.
\vspace{10mm}

{\bf Case (b)}: $\|(\boldsymbol{\alpha}_1-\widehat{\boldsymbol{\alpha}}_1)_{J_1}\|_1\ge \lambda^{(1)} \|(\widehat{\boldsymbol{\alpha}}_1)_{J_1^c}\|_1$.\\

 By Lemma \ref{lemma_3}, we have $\|\widehat{\boldsymbol{\alpha}}_1\|_1\le D_1\|\boldsymbol{\alpha}_1\|_1$ and $\widehat{\boldsymbol{\alpha}}_1\trans \widehat{\mathbf{B}}\widehat{\boldsymbol{\alpha}}_1 \le(1+\hbar^{-1}D_1)^2\mu_1(\boldsymbol{\Xi})$. Therefore,     by $\eqref{356}$, we have
 \begin{align*}
 &\widehat{\boldsymbol{\alpha}}_1\trans \widehat{\mathbf{B}}\widehat{\boldsymbol{\alpha}}_1(\tau^{(1)}\|\boldsymbol{\alpha}_1\|^2_{\lambda^{(1)}}-\tau^{(1)}\|\widehat{\boldsymbol{\alpha}}_1\|^2_{\lambda^{(1)}})\notag\\
\le &  \widehat{\boldsymbol{\alpha}}_1\trans \widehat{\mathbf{B}}\widehat{\boldsymbol{\alpha}}_1\frac{A^{(1)}C_0\varpi}{ \sqrt{\mu_1(\boldsymbol{\Xi})}}(\|(\boldsymbol{\alpha}_1-\widehat{\boldsymbol{\alpha}}_1)_{J_1}\|_1-\lambda^{(1)}\|(\widehat{\boldsymbol{\alpha}}_1)_{J_1^c}\|_1)(1 +\|\widehat{\boldsymbol{\alpha}}_1\|_1/\|\boldsymbol{\alpha}_1\|_1)\notag\\
\le &  (1+\hbar^{-1}D_1)^2 A^{(1)}C_0\mu_1(\boldsymbol{\Xi})^{1/2}\varpi(\|(\boldsymbol{\alpha}_1-\widehat{\boldsymbol{\alpha}}_1)_{J_1}\|_1-\lambda^{(1)}\|(\widehat{\boldsymbol{\alpha}}_1)_{J_1^c}\|_1)(1 +D_1)\notag\\
\le &  (1+\hbar^{-1}D_1)^2(1 +D_1) A^{(1)}C_0\mu_1(\boldsymbol{\Xi})^{1/2}\varpi\|(\boldsymbol{\alpha}_1-\widehat{\boldsymbol{\alpha}}_1)_{J_1}\|_1.
\end{align*} 
The proof is completed.\\

\end{proof}

\begin{proof} $\ref{lemma_20}$.\\

Because in this case,
 \begin{align}
 &\|(\widehat{\boldsymbol{\alpha}}_1-\boldsymbol{\alpha}_1)_{J_1^c}\|_1=\|(\widehat{\boldsymbol{\alpha}}_1)_{J_1^c}\|_1\notag\\
\le& (\lambda^{(1)})^{-1}\|(\widehat{\boldsymbol{\alpha}}_1-\boldsymbol{\alpha}_1)_{J_1}\|_1< c\|(\widehat{\boldsymbol{\alpha}}_1-\boldsymbol{\alpha}_1)_{J_1}\|_1.\label{10735}
\end{align} 
by Condition \ref{condition_3},  
{\small\begin{align}
 \kappa\|(\widehat{\boldsymbol{\alpha}}_1-\boldsymbol{\alpha}_1)_{J_1}\|_2\le \|\frac{\mathbf{X}}{\sqrt{n}}(\widehat{\boldsymbol{\alpha}}_1-\boldsymbol{\alpha}_1)\|_2=\|\mathbf{Z}(\widehat{\boldsymbol{\alpha}}_1-\boldsymbol{\alpha}_1)\|_2=\|\widehat{\boldsymbol{\gamma}}_1-\boldsymbol{\gamma}_1\|_2.\label{10036}
\end{align}}
Moreover, because $\|\widehat{\boldsymbol{\alpha}}_1-\boldsymbol{\alpha}_1\|_1=\|(\widehat{\boldsymbol{\alpha}}_1-\boldsymbol{\alpha}_1)_{J_1^c}\|_1+\|(\widehat{\boldsymbol{\alpha}}_1-\boldsymbol{\alpha}_1)_{J_1}\|_1< (1+c)\|(\widehat{\boldsymbol{\alpha}}_1-\boldsymbol{\alpha}_1)_{J_1}\|_1$,
by $\eqref{10029}$, Lemma \ref{lemma_1} and  $\eqref{10036}$, we have 
 \begin{align}
 &  \frac{1}{2}c_2\mu_1(\boldsymbol{\Xi})\|\widehat{\boldsymbol{\gamma}}_1-\boldsymbol{\gamma}_1\|_2^2\notag\\
\le& \left[(1+\hbar^{-1}D_1)^2(1 +D_1) A^{(1)}+(1+c)[4+\hbar^{-1} (1+D_1)] (1+ D_1^2A^{(1)}\hbar^{-1}) \right]\notag\\
&\qquad \times \sqrt{\mu_1(\boldsymbol{\Xi})}C_0\varpi\|(\widehat{\boldsymbol{\alpha}}_1-\boldsymbol{\alpha}_1)_{J_1}\|_1\notag\\
\le& \left[(1+\hbar^{-1}D_1)^2(1 +D_1) A^{(1)}+(1+c)[4+\hbar^{-1} (1+D_1)] (1+ D_1^2A^{(1)}\hbar^{-1}) \right]\notag\\
&\qquad \times \sqrt{\mu_1(\boldsymbol{\Xi})}C_0\varpi\sqrt{s}\|(\widehat{\boldsymbol{\alpha}}_1-\boldsymbol{\alpha}_1)_{J_1}\|_2\notag\\
 \le &\left[(1+\hbar^{-1}D_1)^2(1 +D_1) A^{(1)}+(1+c)[4+\hbar^{-1} (1+D_1)] (1+ D_1^2A^{(1)}\hbar^{-1}) \right]\notag\\
&\qquad \times\sqrt{\mu_1(\boldsymbol{\Xi})}C_0\varpi\sqrt{s}\|\widehat{\boldsymbol{\gamma}}_1-\boldsymbol{\gamma}_1\|_2/\kappa.\notag
\end{align} 
 Similar to Case (a), we obtain
 \begin{align*}
 & \|\widehat{\boldsymbol{\gamma}}_1-\boldsymbol{\gamma}_1\|_2\le D_3C_0\kappa^{-1}\mu_1(\boldsymbol{\Xi})^{-1/2}\varpi\sqrt{s},\notag\\
& \|\mathbf{Z}(\widehat{\boldsymbol{\alpha}}_1-\boldsymbol{\alpha}_1)\|_2^2=\|\widehat{\boldsymbol{\gamma}}_1-\boldsymbol{\gamma}_1\|_2^2\le D_3^2C_0^2\kappa^{-2}\mu_1(\boldsymbol{\Xi})^{-1}\varpi^2s,\notag\\
 &\|\mathbf{X}(\widehat{\boldsymbol{\alpha}}_1-\boldsymbol{\alpha}_1)\|_2^2=n\|\widehat{\boldsymbol{\gamma}}_1-\boldsymbol{\gamma}_1\|_2^2\le nD_3^2C_0^2\kappa^{-2}\mu_1(\boldsymbol{\Xi})^{-1}\varpi^2s,\notag\\
&\|\widehat{\boldsymbol{\alpha}}_1-\boldsymbol{\alpha}_1\|_1 \le  (1+c)\|(\widehat{\boldsymbol{\alpha}}_1-\boldsymbol{\alpha}_1)_{J_1}\|_1\le (1+c)\sqrt{s}\|(\widehat{\boldsymbol{\alpha}}_1-\boldsymbol{\alpha}_1)_{J_1}\|_2\notag\\
&\le (1+c)\sqrt{s}\kappa^{-1}\|\widehat{\boldsymbol{\gamma}}_1-\boldsymbol{\gamma}_1\|_2 \le (1+c)D_3C_0\kappa^{-2} \mu_1(\boldsymbol{\Xi})^{-1/2}\varpi s  
\end{align*} 
where   
\begin{align*}
 D_3=&2c_2^{-1}\left[(1+\hbar^{-1}D_1)^2(1 +D_1) A^{(1)} \right\notag\\
&\left +(1+c)[4+\hbar^{-1} (1+D_1)] (1+ D_1^2A^{(1)}\hbar^{-1}) \right]. 
\end{align*} 
Note that
\begin{align}
 & \lim_{\substack{ A^{(1)}\to\infty\\\hbar\to\infty}}\frac{D_3}{A^{(1)}}\le 2c_2^{-1}(1+D_1)=2c_2^{-1}(1+\sqrt{6c}).\label{50034}
\end{align}
 Hence, there exist $(A_1^{L})^{\prime\prime\prime\prime}$ and $(\hbar_0)^{\prime\prime\prime\prime}$ only depending on $\delta_0$, $c$ and $c_2$ such that for any $A^{(1)}\ge (A_1^{L})^{\prime\prime\prime\prime}$ and $\hbar\ge (\hbar_0)^{\prime\prime\prime\prime}$, we have  
 \begin{align*}
 & D_3\le 4c_2^{-1}(1+\sqrt{6c})A^{(1)},
\end{align*} 
and the inequalities in Theorem \ref{theorem_3} (a) follows in this case. \\

\end{proof}

\begin{proof}  $\ref{lemma_18}$.\\

Let $\mathbf{D}_i$ denote the $i$-th column of $\mathbf{D}$, where $1\le i\le q$. By the definition of the Frobenius norm,
\begin{align*}
&\|\mathbf{M}\mathbf{D}\|_F^2=\sum_{i=1}^q\|\mathbf{M}\mathbf{D}_i\|_2^2\le \sum_{i=1}^q\|\mathbf{M}\|^2\|\mathbf{D}_i\|_2^2=\|\mathbf{M}\|^2\|\mathbf{D}\|_F^2.
\end{align*} 
Then the lemma follows.\\

\end{proof}

\begin{proof} $\ref{lemma_8}$.\vspace{3mm}

By $\eqref{10098}$,
 \begin{align}
& \|\widehat{\mathbf{C}}-\mathbf{C}\|=\|\mathbf{S}^{-1/2}(\widehat{\mathbf{B}}-\mathbf{B})\mathbf{S}^{-1/2}\|\notag\\
\le &\|\mathbf{S}^{-1/2}\left[(\mathbf{S}\mathcal{B})(\mathbf{Z}\trans\boldsymbol{\varrho})\trans +(\mathbf{Z}\trans\boldsymbol{\varrho})(\mathbf{S}\mathcal{B})\trans +(\mathbf{Z}\trans\boldsymbol{\varrho})(\mathbf{Z}\trans\boldsymbol{\varrho})\trans\right]\mathbf{S}^{-1/2}\|. \label{10099}
\end{align} 
To estimate the right hand side of $\eqref{10099}$, for any vector $\mathbf{a}\in\mathbb{R}^p$ with $\|\mathbf{a}\|_2=1$, we will estimate $\left|\mathbf{a}\trans\mathbf{S}^{-1/2}\left[[(\mathbf{S}\mathcal{B})(\mathbf{Z}\trans\boldsymbol{\varrho})\trans +(\mathbf{Z}\trans\boldsymbol{\varrho})(\mathbf{S}\mathcal{B})\trans +(\mathbf{Z}\trans\boldsymbol{\varrho})(\mathbf{Z}\trans\boldsymbol{\varrho})\trans\right]\mathbf{S}^{-1/2}\mathbf{a} \right|$. By the definition $\eqref{10008}$ of $\mathbf{B}$, the definition $\eqref{50012}$ of $\mathbf{C}$, and the Cauchy-Schwarz inequality, we have
 \begin{align}
& \left|\mathbf{a}\trans\mathbf{S}^{-1/2}\left[[(\mathbf{S}\mathcal{B})(\mathbf{Z}\trans\boldsymbol{\varrho})\trans +(\mathbf{Z}\trans\boldsymbol{\varrho})(\mathbf{S}\mathcal{B})\trans +(\mathbf{Z}\trans\boldsymbol{\varrho})(\mathbf{Z}\trans\boldsymbol{\varrho})\trans\right]\mathbf{S}^{-1/2}\mathbf{a} \right|\label{10100}\\
\le & 2\left| \mathbf{a}\trans\mathbf{S}^{-1/2}\mathbf{S}\mathcal{B} (\mathbf{Z}\trans\boldsymbol{\varrho})\trans\mathbf{S}^{-1/2}\mathbf{a} \right|+  \mathbf{a}\trans\mathbf{S}^{-1/2}(\mathbf{Z}\trans\boldsymbol{\varrho})(\mathbf{Z}\trans\boldsymbol{\varrho})\trans\mathbf{S}^{-1/2}\mathbf{a}\notag\\
\le& 2\sqrt{ \mathbf{a}\trans\mathbf{S}^{-1/2}\mathbf{S}\mathcal{B} \mathcal{B}\trans\mathbf{S}\mathbf{S}^{-1/2}\mathbf{a} }\sqrt{ \mathbf{a}\trans\mathbf{S}^{-1/2}(\mathbf{Z}\trans\boldsymbol{\varrho})(\mathbf{Z}\trans\boldsymbol{\varrho})\trans\mathbf{S}^{-1/2}\mathbf{a}}\notag\\
&+ \mathbf{a}\trans\mathbf{S}^{-1/2}(\mathbf{Z}\trans\boldsymbol{\varrho})(\mathbf{Z}\trans\boldsymbol{\varrho})\trans\mathbf{S}^{-1/2}\mathbf{a}\notag\\
=&2\sqrt{\mathbf{a}\trans\mathbf{S}^{-1/2}\mathbf{B}\mathbf{S}^{-1/2}\mathbf{a}}\|(\mathbf{Z}\trans\boldsymbol{\varrho})\trans\mathbf{S}^{-1/2}\mathbf{a}\|_2+\|(\mathbf{Z}\trans\boldsymbol{\varrho})\trans\mathbf{S}^{-1/2}\mathbf{a}\|_2^2\notag\\
=&2\sqrt{\mathbf{a}\trans \mathbf{C} \mathbf{a}}\|(\mathbf{Z}\trans\boldsymbol{\varrho})\trans\mathbf{S}^{-1/2}\mathbf{a}\|_2+\|(\mathbf{Z}\trans\boldsymbol{\varrho})\trans\mathbf{S}^{-1/2}\mathbf{a}\|_2^2\notag\\
\le &2\sqrt{\|\mathbf{C}\|}\|\mathbf{a}\|_2\|(\mathbf{Z}\trans\boldsymbol{\varrho})\trans\mathbf{S}^{-1/2}\mathbf{a}\|_2+\|(\mathbf{Z}\trans\boldsymbol{\varrho})\trans\mathbf{S}^{-1/2}\mathbf{a}\|_2^2\notag\\
=&2\sqrt{\lambda_{max}(\mathbf{C})} \|(\mathbf{Z}\trans\boldsymbol{\varrho})\trans\mathbf{S}^{-1/2}\mathbf{a}\|_2+\|(\mathbf{Z}\trans\boldsymbol{\varrho})\trans\mathbf{S}^{-1/2}\mathbf{a}\|_2^2,\notag\\
=&2\sqrt{\mu_1(\boldsymbol{\Xi})} \|(\mathbf{Z}\trans\boldsymbol{\varrho})\trans\mathbf{S}^{-1/2}\mathbf{a}\|_2+\|(\mathbf{Z}\trans\boldsymbol{\varrho})\trans\mathbf{S}^{-1/2}\mathbf{a}\|_2^2,\notag
\end{align} 
where the last inequality is due to $\eqref{10101}$. Let $(\mathbf{Z}\trans\boldsymbol{\varrho})_j$ denote the $j$-th row of the $p\times q$ matrix $ \mathbf{Z}\trans\boldsymbol{\varrho} $ and $(\mathbf{S}^{-1/2}\mathbf{a})_j$ denote the $j$-th coordinates of the $p$-dimensional vector   $\mathbf{S}^{-1/2}\mathbf{a}$, respectively. Then by $\eqref{50011}$,
  \begin{align}
&\|(\mathbf{Z}\trans\boldsymbol{\varrho})\trans\mathbf{S}^{-1/2}\mathbf{a}\|_2\label{10103}\\
\le& \|\sum_{j=1}^p(\mathbf{Z}\trans\boldsymbol{\varrho})_j (\mathbf{S}^{-1/2}\mathbf{a})_j\|_2\le \sum_{j=1}^p\|(\mathbf{Z}\trans\boldsymbol{\varrho})_j\|_2|(\mathbf{S}^{-1/2}\mathbf{a})_j|\notag\\
\le&  \max_{1\le j\le p}\|(\mathbf{Z}\trans\boldsymbol{\varrho})_j\|_2\sum_{j=1}^p|(\mathbf{S}^{-1/2}\mathbf{a})_j|\le \max_{1\le j\le p}\|(\mathbf{Z}\trans\boldsymbol{\varrho})_j\|_2\sqrt{p}\|\mathbf{S}^{-1/2}\mathbf{a}\|_2\notag\\
\le &\max_{1\le j\le p}\|(\mathbf{Z}\trans\boldsymbol{\varrho})_j\|_2\sqrt{p}\|\mathbf{S}^{-1/2}\|\|\mathbf{a}\|_2\le c_0^{-1/2}\sqrt{p}\max_{1\le j\le p}\|(\mathbf{Z}\trans\boldsymbol{\varrho})_j\|_2,\notag
\end{align} 
where the second inequality in the third line is because of the Cauchy–Schwarz inequality. For any $1\le j\le p$,  because
 \begin{align*}
(\mathbf{Z}\trans\boldsymbol{\varrho})_j=\frac{1}{\sqrt{n}}(\mathbf{X}\trans\boldsymbol{\varrho})_j=\frac{1}{n}\sum_{i=1}^n\mathbf{X}_{ij}(\boldsymbol{\varepsilon}_i-\bar{\boldsymbol{\varepsilon}}) \\
=\frac{1}{n}\sum_{i=1}^n\mathbf{X}_{ij}\boldsymbol{\varepsilon}_i-\bar{\mathbf{X}}_{\cdot j}\bar{\boldsymbol{\varepsilon}}=\frac{1}{n}\sum_{i=1}^n\mathbf{X}_{ij}\boldsymbol{\varepsilon}_i
\end{align*} 
where $\mathbf{X}_{ij}$ is the $(i,j)$-th entry of $\mathbf{X}$ and $\bar{\mathbf{X}}_{\cdot j}$ is the mean of the $j$-th column of $\mathbf{X}$ which is equal to zero,  $(\mathbf{Z}\trans\boldsymbol{\varrho})_j$ has the same distribution as $\sqrt{\sum_{i=1}^n\mathbf{X}_{ij}^2/n^2}\boldsymbol{\varepsilon}_1=\sqrt{\mathbf{S}_{jj}/n}\boldsymbol{\varepsilon}_1=\boldsymbol{\varepsilon}_1/\sqrt{n}$, where  $\mathbf{S}_{jj}$ is the $j$-th diagonal element of $\mathbf{S}$ which is equal to 1 by Condition \ref{condition_1}. By the inequality in Lemma 3.1 of Section 3.1 in \citet{ledoux2011probability} for the tail probability of Gaussian variables, we have for any $t>M_{\epsilon}$,
 \begin{align}
  &P\left(\|(\mathbf{Z}\trans\boldsymbol{\varrho})_j\|_2>\frac{t}{\sqrt{n}}\right)\notag\\
	&=P\left(\|\frac{\boldsymbol{\varepsilon}_1}{\sqrt{n}}\|_2> \frac{t}{\sqrt{n}}\right)=P\left(\| \boldsymbol{\varepsilon}_1 \|_2> t\right)\le e^{-\frac{(t-M_{\epsilon})^2}{2\sigma^2}}.\label{10200}
\end{align} 
which leads to
 \begin{align}
  &P\left(\max_{1\le j\le p}\|(\mathbf{Z}\trans\boldsymbol{\varrho})_j\|_2>\frac{t}{\sqrt{n}}\right)\le pe^{-\frac{(t-M_{\epsilon})^2}{2\sigma^2}}\notag\\
	&\text{and hence } \max_{1\le j\le p}\|(\mathbf{Z}\trans\boldsymbol{\varrho})_j\|_2=O_p(1/\sqrt{n}).\label{10104}
\end{align} 
It follows from $\eqref{10099}$-$\eqref{10104}$ that 
 \begin{align*}
& \|\widehat{\mathbf{C}}-\mathbf{C}\|=\sqrt{\mu_1(\boldsymbol{\Xi})}O_p(1/\sqrt{n}) .
\end{align*} 
The proof is completed.\\

\end{proof}

\begin{proof}$\ref{lemma_10}$.\\

 By $\eqref{362}$ and the similar arguments as in  $\eqref{10017}$, we can obtain
 \begin{align}
&\boldsymbol{\beta}_k\trans  \widehat{\mathbf{B}}\boldsymbol{\beta}_k  \ge\boldsymbol{\beta}_k\trans \mathbf{B}\boldsymbol{\beta}_k -2\sqrt{\mu_1(\boldsymbol{\Xi})}C_0\varpi \|\boldsymbol{\beta}_k\|_1\label{10246}\\
= &\boldsymbol{\beta}_k\trans\mathbf{Z}\trans\boldsymbol{\Xi}\mathbf{Z}\boldsymbol{\beta}_k -2\sqrt{\mu_1(\boldsymbol{\Xi})}C_0\varpi \|\boldsymbol{\beta}_k\|_1 \notag\\
=& \boldsymbol{\gamma}_k\trans(\mathbf{I}-\widehat{\mathbf{P}}_{k-1})\boldsymbol{\Xi}(\mathbf{I}-\widehat{\mathbf{P}}_{k-1})\boldsymbol{\gamma}_k-2\sqrt{\mu_1(\boldsymbol{\Xi})}C_0\varpi \|\boldsymbol{\beta}_k\|_1\notag\\
=& \boldsymbol{\gamma}_k\trans \boldsymbol{\Xi} \boldsymbol{\gamma}_k-2\boldsymbol{\gamma}_k\trans \widehat{\mathbf{P}}_{k-1}\boldsymbol{\Xi}\boldsymbol{\gamma}_k+\boldsymbol{\gamma}_k\trans\widehat{\mathbf{P}}_{k-1}\boldsymbol{\Xi}\widehat{\mathbf{P}}_{k-1}\boldsymbol{\gamma}_k-2\sqrt{\mu_1(\boldsymbol{\Xi})}C_0\varpi \|\boldsymbol{\beta}_k\|_1\notag\\
\ge&\boldsymbol{\gamma}_k\trans \boldsymbol{\Xi} \boldsymbol{\gamma}_k-2\boldsymbol{\gamma}_k\trans \widehat{\mathbf{P}}_{k-1}\boldsymbol{\Xi}\boldsymbol{\gamma}_k -2\sqrt{\mu_1(\boldsymbol{\Xi})}C_0\varpi \|\boldsymbol{\beta}_k\|_1\notag\\
=&\mu_k(\boldsymbol{\Xi})-2\mu_k(\boldsymbol{\Xi})\boldsymbol{\gamma}_k\trans \widehat{\mathbf{P}}_{k-1} \boldsymbol{\gamma}_k -2\sqrt{\mu_1(\boldsymbol{\Xi})}C_0\varpi \|\boldsymbol{\beta}_k\|_1\notag\\
=&\mu_k(\boldsymbol{\Xi})-2\mu_k(\boldsymbol{\Xi}) \| \widehat{\mathbf{P}}_{k-1} \boldsymbol{\gamma}_k\|_2^2 -2\sqrt{\mu_1(\boldsymbol{\Xi})}C_0\varpi \|\boldsymbol{\beta}_k\|_1\notag\\
\ge &\mu_k(\boldsymbol{\Xi})-2\mu_1(\boldsymbol{\Xi}) \| \widehat{\mathbf{P}}_{k-1} \boldsymbol{\gamma}_k\|_2^2 -2 \sqrt{\mu_1(\boldsymbol{\Xi})}M_{k,1}\kappa^{-1}C_0\varpi \sqrt{s}, \notag
\end{align} 
 where the last equality is due to $\eqref{10245}$. Because $\mathbf{P}_{k-1}\boldsymbol{\gamma}_k=\mathbf{0}$,  
 \begin{align}
&\| \widehat{\mathbf{P}}_{k-1} \boldsymbol{\gamma}_k\|_2^2=\| (\widehat{\mathbf{P}}_{k-1}-\mathbf{P}_{k-1}) \boldsymbol{\gamma}_k\|_2^2\le \| \widehat{\mathbf{P}}_{k-1}-\mathbf{P}_{k-1} \|^2\label{10250}\\
\le& b_{k-1,3}C_0^2\kappa^{-2}\mu_1(\boldsymbol{\Xi})^{-1}\varpi^2 s=  b_{k-1,3}C_0\kappa^{-2}\mu_1(\boldsymbol{\Xi})^{-1/2}\varpi \sqrt{s}\frac{C_0 \varpi \sqrt{s}}{\mu_1(\boldsymbol{\Xi})^{1/2}} \notag\\
\le& b_{k-1,3}C_0\kappa^{-2}\mu_1(\boldsymbol{\Xi})^{-1/2}\varpi \sqrt{s}\hbar^{-1}\kappa=b_{k-1,3}\hbar^{-1}C_0\kappa^{-1}\mu_1(\boldsymbol{\Xi})^{-1/2}\varpi \sqrt{s},\notag
\end{align} 
 where the first inequality in the second line follows from $\eqref{1053}$ and the first inequality in the last line is due to $\eqref{10247}$. Combining $\eqref{10250}$ and  $\eqref{10246}$ gives the second inequality in the lemma.
 \begin{align}
&\boldsymbol{\beta}_k\trans  \widehat{\mathbf{B}}\boldsymbol{\beta}_k  \ge  \mu_k(\boldsymbol{\Xi})-  N_{k,3}C_0\kappa^{-1}\mu_1(\boldsymbol{\Xi})^{1/2}\varpi \sqrt{s}, \label{10248}
\end{align} 
where $N_{k,3}=2(b_{k-1,3}\hbar^{-1}+ M_{k,1})$. On the other hand, 
 \begin{align}
&\boldsymbol{\beta}_k\trans  \mathbf{S}\boldsymbol{\beta}_k  = \boldsymbol{\beta}_k\trans\mathbf{Z}\trans \mathbf{Z}\boldsymbol{\beta}_k= \boldsymbol{\gamma}_k\trans(\mathbf{I}-\widehat{\mathbf{P}}_{k-1}) (\mathbf{I}-\widehat{\mathbf{P}}_{k-1})\boldsymbol{\gamma}_k\notag\\
 =& \boldsymbol{\gamma}_k\trans (\mathbf{I}-\widehat{\mathbf{P}}_{k-1})\boldsymbol{\gamma}_k= \|\boldsymbol{\gamma}_k\|_2^2- \boldsymbol{\gamma}_k\trans\widehat{\mathbf{P}}_{k-1} \boldsymbol{\gamma}_k \notag\\
=&\|\boldsymbol{\gamma}_k\|_2^2- \|\widehat{\mathbf{P}}_{k-1} \boldsymbol{\gamma}_k\|_2^2 \le \|\boldsymbol{\gamma}_k\|_2^2=1, \label{10249}
\end{align} 
which is the first inequality in the lemma. Now we prove the third inequality in the lemma. By  the similar arguments as in  $\eqref{10020}$, we have
 \begin{align}
 &\widehat{\boldsymbol{\alpha}}_k\trans \widehat{\mathbf{B}}\widehat{\boldsymbol{\alpha}}_k\label{364}\\
 \le& \widehat{\boldsymbol{\alpha}}_k\trans \mathbf{B}\widehat{\boldsymbol{\alpha}}_k+2\sqrt{\mu_1(\boldsymbol{\Xi})}C_0\varpi \|\widehat{\boldsymbol{\alpha}}_k\|_1+C_0^2\varpi^2 \|\widehat{\boldsymbol{\alpha}}_k\|_1^2\notag\\
= &\widehat{\boldsymbol{\alpha}}_k\trans \mathbf{B}\widehat{\boldsymbol{\alpha}}_k+2\sqrt{\mu_1(\boldsymbol{\Xi})}\frac{C_0\varpi}{\sqrt{s}}\sqrt{s} \|\widehat{\boldsymbol{\alpha}}_k\|_1+(C_0\varpi)C_0\varpi \|\widehat{\boldsymbol{\alpha}}_k\|_1^2\notag\\
\le & \widehat{\boldsymbol{\alpha}}_k\trans \mathbf{B}\widehat{\boldsymbol{\alpha}}_k+\sqrt{\mu_1(\boldsymbol{\Xi})}\frac{C_0\varpi}{\sqrt{s}}[\kappa^{-1}s+ \kappa\|\widehat{\boldsymbol{\alpha}}_k\|_1^2]+(C_0\varpi)C_0\varpi \|\widehat{\boldsymbol{\alpha}}_k\|_1^2\notag\\
\le & \widehat{\boldsymbol{\alpha}}_k\trans \mathbf{B}\widehat{\boldsymbol{\alpha}}_k+\sqrt{\mu_1(\boldsymbol{\Xi})}\frac{C_0\varpi}{\sqrt{s}}[\kappa^{-1}s+ \kappa\|\widehat{\boldsymbol{\alpha}}_k\|_1^2]+\frac{\kappa\mu_1(\boldsymbol{\Xi})^{1/2}\hbar^{-1}}{\sqrt{s}}C_0\varpi \|\widehat{\boldsymbol{\alpha}}_k\|_1^2\notag\\
=&\widehat{\boldsymbol{\alpha}}_k\trans \mathbf{B}\widehat{\boldsymbol{\alpha}}_k+\sqrt{\mu_1(\boldsymbol{\Xi})}C_0\kappa^{-1}\varpi\sqrt{s}+\frac{(1+\hbar^{-1})}{\sqrt{s}}\kappa\sqrt{\mu_1(\boldsymbol{\Xi})}C_0\varpi \|\widehat{\boldsymbol{\alpha}}_k\|_1^2\notag,
\end{align} 
where the inequality in the fourth line is due to the inequality: $a^2+b^2\ge 2ab$,  and the inequality in the fourth line follows from $\eqref{10247}$. We estimate $\widehat{\boldsymbol{\alpha}}_k\trans \mathbf{B}\widehat{\boldsymbol{\alpha}}_k$. By the eigen-decomposition of $\boldsymbol{\Xi}$,
 \begin{align}
  &\widehat{\boldsymbol{\alpha}}_k\trans \mathbf{B}\widehat{\boldsymbol{\alpha}}_k= \widehat{\boldsymbol{\gamma}}_k\trans \boldsymbol{\Xi}\widehat{\boldsymbol{\gamma}}_k=\widehat{\boldsymbol{\gamma}}_k\trans (\sum_{i=1}^K\mu_i(\boldsymbol{\Xi})\boldsymbol{\gamma}_i \boldsymbol{\gamma}_i\trans)\widehat{\boldsymbol{\gamma}}_k=\sum_{i=1}^K\mu_i(\boldsymbol{\Xi})(\boldsymbol{\gamma}_i\trans\widehat{\boldsymbol{\gamma}}_k)^2\label{10251}\\
	\le& \mu_1(\boldsymbol{\Xi})\sum_{i=1}^{k-1}(\boldsymbol{\gamma}_i\trans\widehat{\boldsymbol{\gamma}}_k)^2+\mu_k(\boldsymbol{\Xi})\sum_{i=k}^{K}(\boldsymbol{\gamma}_i\trans\widehat{\boldsymbol{\gamma}}_k)^2\notag\\
	=&\mu_1(\boldsymbol{\Xi})\|\mathbf{P}_{k-1}\widehat{\boldsymbol{\gamma}}_k\|_2^2+\mu_k(\boldsymbol{\Xi})\sum_{i=k}^{K}(\boldsymbol{\gamma}_i\trans\widehat{\boldsymbol{\gamma}}_k)^2 	\notag\\
	\le& \mu_1(\boldsymbol{\Xi})\|\mathbf{P}_{k-1}\widehat{\boldsymbol{\gamma}}_k\|_2^2+\mu_k(\boldsymbol{\Xi})\| \widehat{\boldsymbol{\gamma}}_k\|_2^2\notag\\
	\le& b_{k-1,3}\hbar^{-1}C_0\kappa^{-1}\mu_1(\boldsymbol{\Xi})^{1/2}\varpi \sqrt{s}+\mu_k(\boldsymbol{\Xi}),\notag
\end{align} 
where the last inequality follows from $\eqref{10250}$. Combining $\eqref{10251}$ and $\eqref{364}$ gives 
 \begin{align*}
 &\widehat{\boldsymbol{\alpha}}_k\trans \widehat{\mathbf{B}}\widehat{\boldsymbol{\alpha}}_k \le \mu_k(\boldsymbol{\Xi})+b_{k-1,3}\hbar^{-1}C_0\kappa^{-1}\mu_1(\boldsymbol{\Xi})^{1/2}\varpi \sqrt{s}+\sqrt{\mu_1(\boldsymbol{\Xi})}C_0\kappa^{-1}\varpi\sqrt{s}\notag\\
&+\frac{(1+\hbar^{-1})}{\sqrt{s}}\kappa\sqrt{\mu_1(\boldsymbol{\Xi})}C_0\varpi \|\widehat{\boldsymbol{\alpha}}_k\|_1^2\notag\\
=&\mu_k(\boldsymbol{\Xi})+(b_{k-1,3}\hbar^{-1}+1)\sqrt{\mu_1(\boldsymbol{\Xi})}\kappa^{-1}C_0\varpi\sqrt{s}\notag\\
&+\frac{(1+\hbar^{-1})}{\sqrt{s}}\kappa\sqrt{\mu_1(\boldsymbol{\Xi})}C_0\varpi \|\widehat{\boldsymbol{\alpha}}_k\|_1^2 
\end{align*} 
The proof is completed.\\

\end{proof}

\begin{proof} $\ref{lemma_12}$.\\

Consider the left hand side of the first inequality in $\eqref{10252}$. We first note that by the definition of $ \tau^{(k)}$,
\begin{align}
& \tau^{(k)}\|\widehat{\boldsymbol{\alpha}}_k\|^2_{\lambda^{(k)}}= \frac{A^{(k)}C_0\varpi}{\|\boldsymbol{\alpha}_1\|_1\sqrt{\mu_1(\boldsymbol{\Xi})}}\|\widehat{\boldsymbol{\alpha}}_k\|^2_{\lambda^{(k)}}\notag\\
\ge&\frac{A^{(k)}C_0\varpi}{\|\boldsymbol{\alpha}_1\|_1\sqrt{\mu_1(\boldsymbol{\Xi})}}\lambda^{(k)}\|\widehat{\boldsymbol{\alpha}}_k\|^2_1\ge \kappa\frac{A^{(k)}C_0\varpi}{\sqrt{s\mu_1(\boldsymbol{\Xi})}}c^{-1}\|\widehat{\boldsymbol{\alpha}}_k\|^2_1, \label{50039}
\end{align}  
where the last inequality is due to $\lambda^{(k)}>c^{-1}$ and and $\|\boldsymbol{\alpha}_1\|_1\le \kappa^{-1}\sqrt{s}$ by Lemma \ref{lemma_4}.  By the second inequality in Lemma \ref{lemma_10} and $\eqref{10247}$, we have
 \begin{align}
&(\boldsymbol{\beta}_k\trans \widehat{\mathbf{B}}\boldsymbol{\beta}_k)\left(1+\tau^{(k)}\|\widehat{\boldsymbol{\alpha}}_k\|^2_{\lambda^{(k)}}\right)\label{10254}\\
\ge& \left(\mu_k(\boldsymbol{\Xi})-  N_{k,3}C_0\kappa^{-1}\sqrt{\mu_1(\boldsymbol{\Xi})}\varpi \sqrt{s}\right)\left(1+\kappa\frac{A^{(k)}C_0\varpi}{\sqrt{s\mu_1(\boldsymbol{\Xi})}}c^{-1}\|\widehat{\boldsymbol{\alpha}}_k\|^2_1\right)\notag\\
=&\mu_k(\boldsymbol{\Xi})-  N_{k,3}C_0\kappa^{-1}\sqrt{\mu_1(\boldsymbol{\Xi})}\varpi \sqrt{s}\notag\\
&+ \left(\mu_k(\boldsymbol{\Xi})-  N_{k,3}\kappa^{-1}\sqrt{\mu_1(\boldsymbol{\Xi})}C_0\varpi \sqrt{s}\right) \kappa\frac{A^{(k)}C_0\varpi}{\sqrt{s\mu_1(\boldsymbol{\Xi})}}c^{-1}\|\widehat{\boldsymbol{\alpha}}_k\|^2_1\notag\\
\ge&  \mu_k(\boldsymbol{\Xi}) -  N_{k,3}C_0\kappa^{-1}\sqrt{\mu_1(\boldsymbol{\Xi})}\varpi \sqrt{s}\notag\\
&+ \left(c_3^{-1}\mu_1(\boldsymbol{\Xi})-  N_{k,3}\kappa^{-1}\sqrt{\mu_1(\boldsymbol{\Xi})}\hbar^{-1} \kappa\mu_1(\boldsymbol{\Xi})^{1/2} \right) \kappa\frac{A^{(k)}C_0\varpi}{\sqrt{s\mu_1(\boldsymbol{\Xi})}}c^{-1}\|\widehat{\boldsymbol{\alpha}}_k\|^2_1\notag\\
  \ge & \mu_k(\boldsymbol{\Xi})-N_{k,3}C_0\kappa^{-1}\sqrt{\mu_1(\boldsymbol{\Xi})}\varpi \sqrt{s}\notag\\
&+ \left(c_3^{-1}-  N_{k,3}\hbar^{-1} \right)\sqrt{\mu_1(\boldsymbol{\Xi})}\kappa\frac{A^{(k)}C_0\varpi}{\sqrt{s}}c^{-1}\|\widehat{\boldsymbol{\alpha}}_k\|^2_1,\notag
\end{align}  
where the inequality in the fifth line is due to $\mu_k(\boldsymbol{\Xi})\ge c_3^{-1}\mu_1(\boldsymbol{\Xi})$ by Condition \ref{condition_2}(b).   Now we estimate the right hand side of the second inequality in $\eqref{10252}$. By the third inequality in Lemma  \ref{lemma_10} and the first inequality in the second line of $\eqref{10257}$,  
 \begin{align*}
 &(\widehat{\boldsymbol{\alpha}}_k\trans \widehat{\mathbf{B}}\widehat{\boldsymbol{\alpha}}_k)(1+\tau^{(k)}\|\boldsymbol{\beta}_k\|^2_{\lambda^{(k)}})\le (\widehat{\boldsymbol{\alpha}}_k\trans \widehat{\mathbf{B}}\widehat{\boldsymbol{\alpha}}_k)(1+\tau^{(k)}\|\boldsymbol{\beta}_k\|^2_1) \\
\le &\left(\mu_k(\boldsymbol{\Xi})+(b_{k-1,3}\hbar^{-1}+1)\sqrt{\mu_1(\boldsymbol{\Xi})}\kappa^{-1}C_0\varpi\sqrt{s}+\frac{(1+\hbar^{-1})}{\sqrt{s}}\kappa\sqrt{\mu_1(\boldsymbol{\Xi})}C_0\varpi \|\widehat{\boldsymbol{\alpha}}_k\|_1^2\right)\notag\\
&  \times \left(1+A^{(k)}M_{k,1}^2\kappa^{-1}\frac{C_0 \varpi\sqrt{s}}{\sqrt{\mu_1(\boldsymbol{\Xi})}}\right)=\mu_k(\boldsymbol{\Xi})+\mu_k(\boldsymbol{\Xi})A^{(k)}M_{k,1}^2\kappa^{-1}\frac{C_0 \varpi\sqrt{s}}{\sqrt{\mu_1(\boldsymbol{\Xi})}}\notag\\
&+\left(1+A^{(k)}M_{k,1}^2\kappa^{-1}\frac{ C_0\varpi\sqrt{s}}{\sqrt{\mu_1(\boldsymbol{\Xi})}}\right)\notag\\
&\times \left[(b_{k-1,3}\hbar^{-1}+1)\sqrt{\mu_1(\boldsymbol{\Xi})}\kappa^{-1}C_0\varpi\sqrt{s}+\frac{(1+\hbar^{-1})}{\sqrt{s}}\kappa\sqrt{\mu_1(\boldsymbol{\Xi})}C_0\varpi \|\widehat{\boldsymbol{\alpha}}_k\|_1^2\right] \notag\\
\le &\mu_k(\boldsymbol{\Xi})+\mu_1(\boldsymbol{\Xi})A^{(k)}M_{k,1}^2\kappa^{-1}\frac{C_0 \varpi\sqrt{s}}{\sqrt{\mu_1(\boldsymbol{\Xi})}}\notag\\
&+ \left(1+A^{(k)}M_{k,1}^2\hbar^{-1} \right)(b_{k-1,3}\hbar^{-1}+1)C_0\sqrt{\mu_1(\boldsymbol{\Xi})}\kappa^{-1}\varpi\sqrt{s}\notag\\
&+ \left(1+A^{(k)}M_{k,1}^2\hbar^{-1} \right)\frac{(1+\hbar^{-1})}{\sqrt{s}}\kappa\sqrt{\mu_1(\boldsymbol{\Xi})}C_0\varpi \|\widehat{\boldsymbol{\alpha}}_k\|_1^2\notag\\
=&\mu_k(\boldsymbol{\Xi})+\left[A^{(k)}M_{k,1}^2 + \left(1+A^{(k)}M_{k,1}^2\hbar^{-1} \right)(b_{k-1,3}\hbar^{-1}+1)\right]C_0\sqrt{\mu_1(\boldsymbol{\Xi})}\kappa^{-1}\varpi\sqrt{s}\notag\\
&+\left(1+A^{(k)}M_{k,1}^2\hbar^{-1} \right)\frac{(1+\hbar^{-1})}{\sqrt{s}}\kappa\sqrt{\mu_1(\boldsymbol{\Xi})}C_0\varpi \|\widehat{\boldsymbol{\alpha}}_k\|_1^2,\notag
\end{align*} 
where the second inequality is due to the last inequality in the second line of $\eqref{10257}$. The above inequality, together with $\eqref{10252}$ and $\eqref{10254}$, leads to
 \begin{align}
&  \left[\left(c_3^{-1}-  N_{k,3}\hbar^{-1} \right)A^{(k)}c^{-1}-\left(1+A^{(k)}M_{k,1}^2\hbar^{-1} \right)(1+\hbar^{-1})\right] \notag\\
&\times\sqrt{\mu_1(\boldsymbol{\Xi})}\kappa\frac{C_0\varpi}{\sqrt{s}} \|\widehat{\boldsymbol{\alpha}}_k\|^2_1\notag\\
\le &\left[N_{k,3}+A^{(k)}M_{k,1}^2 + \left(1+A^{(k)}M_{k,1}^2\hbar^{-1} \right)(b_{k-1,3}\hbar^{-1}+1) \right]\notag\\
&\times\sqrt{\mu_1(\boldsymbol{\Xi})}\kappa^{-1}C_0\varpi\sqrt{s},\label{10256}
\end{align} 
where we have canceled $\mu_k(\boldsymbol{\Xi})$ on both sides. Because $N_{k,3}=2(b_{k-1,3}\hbar^{-1}+ M_{k,1})$ and $M_{k,1}$ does not depend on $\hbar$, we have
 \begin{align*}
&\lim_{\substack{ A^{(k)}\to\infty\\\hbar\to\infty}}\sqrt{\frac{N_{k,3}+A^{(k)}M_{k,1}^2 + \left(1+A^{(k)}M_{k,1}^2\hbar^{-1} \right)(b_{k-1,3}\hbar^{-1}+1)}{\left(c_3^{-1}-  N_{k,3}\hbar^{-1} \right)A^{(k)}c^{-1}-\left(1+A^{(k)}M_{k,1}^2\hbar^{-1} \right)(1+\hbar^{-1})}}\notag\\
=&M_{k,1}\sqrt{c_3c}.
\end{align*} 
Therefore, there exist $(A_k^{L})^\prime$ and $(\hbar_0)^\prime$ only depending on $\delta_0$, $c$ and $c_2\sim c_5$ such that for any $A^{(k)}\ge (A_k^{L})^\prime $ and $\hbar\ge (\hbar_0)^\prime$,  we have 
 \begin{align*}
\|\widehat{\boldsymbol{\alpha}}_k\|_1\le D_{k,1}\kappa^{-1}\sqrt{s}. 
\end{align*} 
where $D_{k,1}=2M_{k,1}\sqrt{c_3c}$.\\
\end{proof}

\begin{proof} $\ref{lemma_11}$.\\

We first estimate the left hand side of $\eqref{20252}$. Note that
 \begin{align}
&\boldsymbol{\beta}_k\trans \widehat{\mathbf{B}}\boldsymbol{\beta}_k-\widehat{\boldsymbol{\alpha}}_k\trans \widehat{\mathbf{B}}\widehat{\boldsymbol{\alpha}}_k\notag\\
\ge& \boldsymbol{\beta}_k\trans \mathbf{B}\boldsymbol{\beta}_k-\widehat{\boldsymbol{\alpha}}_k\trans \mathbf{B}\widehat{\boldsymbol{\alpha}}_k-|\boldsymbol{\beta}_k\trans (\widehat{\mathbf{B}}-\mathbf{B})\boldsymbol{\beta}_k-\widehat{\boldsymbol{\alpha}}_k\trans (\widehat{\mathbf{B}}-\mathbf{B})\widehat{\boldsymbol{\alpha}}_k|\label{10265}
\end{align} 
By the first line  of $\eqref{10251}$ and $\widehat{\mathbf{P}}_{k-1}\widehat{\boldsymbol{\gamma}}_k=\mathbf{0}$,
 \begin{align}
  &\widehat{\boldsymbol{\alpha}}_k\trans \mathbf{B}\widehat{\boldsymbol{\alpha}}_k =\sum_{i=1}^K\mu_i(\boldsymbol{\Xi})(\boldsymbol{\gamma}_i\trans\widehat{\boldsymbol{\gamma}}_k)^2\le \mu_1(\boldsymbol{\Xi})\sum_{i=1}^{k-1}(\boldsymbol{\gamma}_i\trans\widehat{\boldsymbol{\gamma}}_k)^2+\sum_{i=k}^{K}\mu_i(\boldsymbol{\Xi})(\boldsymbol{\gamma}_i\trans\widehat{\boldsymbol{\gamma}}_k)^2 \label{10259}\\
	\le& \mu_1(\boldsymbol{\Xi})\|\mathbf{P}_{k-1}\widehat{\boldsymbol{\gamma}}_k\|_2^2+\mu_k(\boldsymbol{\Xi})(\boldsymbol{\gamma}_k\trans\widehat{\boldsymbol{\gamma}}_k)^2+\mu_{k+1}(\boldsymbol{\Xi})\sum_{i=k+1}^{K}(\boldsymbol{\gamma}_i\trans\widehat{\boldsymbol{\gamma}}_k)^2\notag\\
	=&\mu_1(\boldsymbol{\Xi})\|\mathbf{P}_{k-1}\widehat{\boldsymbol{\gamma}}_k-\widehat{\mathbf{P}}_{k-1}\widehat{\boldsymbol{\gamma}}_k\|_2^2+\mu_k(\boldsymbol{\Xi})(\boldsymbol{\gamma}_k\trans\widehat{\boldsymbol{\gamma}}_k)^2 +\mu_{k+1}(\boldsymbol{\Xi})\sum_{i=k+1}^{K}(\boldsymbol{\gamma}_i\trans\widehat{\boldsymbol{\gamma}}_k)^2\notag\\
	\le &\mu_1(\boldsymbol{\Xi})\|\mathbf{P}_{k-1} -\widehat{\mathbf{P}}_{k-1} \|^2+\mu_k(\boldsymbol{\Xi})(\boldsymbol{\gamma}_k\trans\widehat{\boldsymbol{\gamma}}_k)^2+\mu_{k+1}(\boldsymbol{\Xi})\left[\|\widehat{\boldsymbol{\gamma}}_k\|_2^2-(\boldsymbol{\gamma}_k\trans\widehat{\boldsymbol{\gamma}}_k)^2 \right]\notag\\
	= &\mu_1(\boldsymbol{\Xi})\|\mathbf{P}_{k-1} -\widehat{\mathbf{P}}_{k-1} \|^2+\mu_k(\boldsymbol{\Xi})(\boldsymbol{\gamma}_k\trans\widehat{\boldsymbol{\gamma}}_k)^2+\mu_{k+1}(\boldsymbol{\Xi})\left[1-(\boldsymbol{\gamma}_k\trans\widehat{\boldsymbol{\gamma}}_k)^2 \right]\notag\\
	=&\mu_1(\boldsymbol{\Xi})\|\mathbf{P}_{k-1} -\widehat{\mathbf{P}}_{k-1} \|^2+\mu_k(\boldsymbol{\Xi})-(\mu_k(\boldsymbol{\Xi})-\mu_{k+1}(\boldsymbol{\Xi}))\left[1-(\boldsymbol{\gamma}_k\trans\widehat{\boldsymbol{\gamma}}_k)^2 \right].\notag
\end{align} 
On the other hand, by $\mathbf{P}_{k-1}\boldsymbol{\gamma}_k=\mathbf{0}$ and the similar arguments as in $\eqref{10246}$,
 \begin{align}
& \boldsymbol{\beta}_k\trans \mathbf{B}\boldsymbol{\beta}_k  = \boldsymbol{\beta}_k\trans\mathbf{Z}\trans\boldsymbol{\Xi}\mathbf{Z}\boldsymbol{\beta}_k \ge \mu_k(\boldsymbol{\Xi})-2\mu_1(\boldsymbol{\Xi}) \| \widehat{\mathbf{P}}_{k-1} \boldsymbol{\gamma}_k\|_2^2\notag\\
=&\mu_k(\boldsymbol{\Xi})-2\mu_k(\boldsymbol{\Xi}) \| \widehat{\mathbf{P}}_{k-1}\boldsymbol{\gamma}_k-\mathbf{P}_{k-1}\boldsymbol{\gamma}_k\|_2^2\notag\\
=&\mu_k(\boldsymbol{\Xi})-2\mu_k(\boldsymbol{\Xi}) \| \widehat{\mathbf{P}}_{k-1} -\mathbf{P}_{k-1} \|^2. \label{10260}
\end{align} 
Combining  $\eqref{10259}$ and $\eqref{10260}$ gives 
 \begin{align}
&(\boldsymbol{\beta}_k\trans \mathbf{B}\boldsymbol{\beta}_k-\widehat{\boldsymbol{\alpha}}_k\trans \mathbf{B}\widehat{\boldsymbol{\alpha}}_k)\ge (\mu_k(\boldsymbol{\Xi})-\mu_{k+1}(\boldsymbol{\Xi}))\left[1-(\boldsymbol{\gamma}_k\trans\widehat{\boldsymbol{\gamma}}_k)^2 \right]\label{50040}\\
&\qquad -(2\mu_k(\boldsymbol{\Xi})+\mu_1(\boldsymbol{\Xi})) \| \widehat{\mathbf{P}}_{k-1} -\mathbf{P}_{k-1} \|^2\notag\\
\ge& (\mu_k(\boldsymbol{\Xi})-\mu_{k+1}(\boldsymbol{\Xi}))\left[1+(\boldsymbol{\gamma}_k\trans\widehat{\boldsymbol{\gamma}}_k) \right]\left[1-(\boldsymbol{\gamma}_k\trans\widehat{\boldsymbol{\gamma}}_k) \right]-3\mu_1(\boldsymbol{\Xi}) \| \widehat{\mathbf{P}}_{k-1} -\mathbf{P}_{k-1} \|^2 \notag\\
\ge& c_2\mu_k(\boldsymbol{\Xi})\left[1-(\boldsymbol{\gamma}_k\trans\widehat{\boldsymbol{\gamma}}_k) \right]-3\mu_1(\boldsymbol{\Xi}) b_{k-1,3}C_0^2\kappa^{-2}\mu_1(\boldsymbol{\Xi})^{-1}\varpi^2 s \notag\\
 =&\frac{1}{2}c_2\mu_k(\boldsymbol{\Xi})\|\boldsymbol{\gamma}_k-\widehat{\boldsymbol{\gamma}}_k\|_2^2-3 b_{k-1,3}C_0^2\kappa^{-2} \varpi^2 s,\notag
\end{align} 
where the inequality in the third line is due to Condition \ref{condition_2}(a), $\eqref{1053}$ and $\boldsymbol{\gamma}_k\trans\widehat{\boldsymbol{\gamma}}_k\ge 0$.  For the last term on the right hand side of $\eqref{10265}$, by the similar arguments as in $\eqref{10011}$, $\eqref{10015}$ and $\eqref{10027}$ in the proof of Lemma \ref{lemma_3}, we can obtain
 \begin{align}
 &|\boldsymbol{\beta}_k\trans (\widehat{\mathbf{B}}-\mathbf{B})\boldsymbol{\beta}_k-\widehat{\boldsymbol{\alpha}}_k\trans (\widehat{\mathbf{B}}-\mathbf{B})\widehat{\boldsymbol{\alpha}}_k|\label{10261}\\
\le& 4 \sqrt{\mu_1(\boldsymbol{\Xi})}C_0\varpi\|\boldsymbol{\beta}_k-\widehat{\boldsymbol{\alpha}}_k\|_1+ C_0^2\varpi^2\|\boldsymbol{\beta}_k-\widehat{\boldsymbol{\alpha}}_k\|_1(\|\boldsymbol{\beta}_k\|_1+\|\widehat{\boldsymbol{\alpha}}_k\|_1)\notag\\
\le& 4 \sqrt{\mu_1(\boldsymbol{\Xi})}C_0\varpi\|\boldsymbol{\beta}_k-\widehat{\boldsymbol{\alpha}}_k\|_1+ C_0^2\varpi^2\|\boldsymbol{\beta}_k-\widehat{\boldsymbol{\alpha}}_k\|_1(M_{k,1}+D_{k,1})\kappa^{-1}\sqrt{s}\notag\\
=& 4 \sqrt{\mu_1(\boldsymbol{\Xi})}C_0\varpi\|\boldsymbol{\beta}_k-\widehat{\boldsymbol{\alpha}}_k\|_1+ (M_{k,1}+D_{k,1})C_0\varpi\|\boldsymbol{\beta}_k-\widehat{\boldsymbol{\alpha}}_k\|_1\kappa^{-1}(C_0\varpi\sqrt{s})\notag\\
\le& 4 \sqrt{\mu_1(\boldsymbol{\Xi})}C_0\varpi\|\boldsymbol{\beta}_k-\widehat{\boldsymbol{\alpha}}_k\|_1+ (M_{k,1}+D_{k,1})C_0\varpi\|\boldsymbol{\beta}_k-\widehat{\boldsymbol{\alpha}}_k\|_1 \hbar^{-1} \mu_1(\boldsymbol{\Xi})^{1/2} \notag\\
=&[4+ (M_{k,1}+D_{k,1})\hbar^{-1}]\sqrt{\mu_1(\boldsymbol{\Xi})}C_0\varpi\|\boldsymbol{\beta}_k-\widehat{\boldsymbol{\alpha}}_k\|_1,\notag
\end{align} 
where the third inequality follows from $\eqref{10245}$ and $\eqref{10258}$, and the fifth one is due to $\eqref{10247}$. By $\eqref{10252}$,
 \begin{align}
 & 1+\tau^{(k)}\|\widehat{\boldsymbol{\alpha}}_k\|^2_{\lambda^{(k)}}\le \frac{\widehat{\boldsymbol{\alpha}}_k\trans \widehat{\mathbf{B}}\widehat{\boldsymbol{\alpha}}_k}{\boldsymbol{\beta}_k\trans \widehat{\mathbf{B}}\boldsymbol{\beta}_k}(1+\tau^{(k)}\|\boldsymbol{\beta}_k\|^2_{\lambda^{(k)}}), \label{10262}
\end{align} 
By the second   inequality in Lemma \ref{lemma_10},   we have 
\begin{align}
 \boldsymbol{\beta}_k\trans \widehat{\mathbf{B}}\boldsymbol{\beta}_k&\ge  \mu_k(\boldsymbol{\Xi})-  N_{k,3}\mu_1(\boldsymbol{\Xi})^{1/2}C_0\kappa^{-1}\varpi \sqrt{s}\notag\\
&\ge \mu_k(\boldsymbol{\Xi})-  N_{k,3}\mu_1(\boldsymbol{\Xi})^{1/2}\hbar^{-1}\mu_1(\boldsymbol{\Xi})^{1/2}\notag\\
&\ge c_3\mu_1(\boldsymbol{\Xi})-  N_{k,3} \hbar^{-1}\mu_1(\boldsymbol{\Xi}) =N_{k,7}\mu_1(\boldsymbol{\Xi}), \label{370}
\end{align} 
where the second  inequality is due to $\eqref{10247}$ and $N_{k,7}= c_3-N_{k,3}\hbar^{-1} =c_3-2(b_{k-1,3}\hbar^{-1}+ M_{k,1})\hbar^{-1}$ (see the definition of $N_{k,3}$ in Lemma \ref{lemma_10}), and by the third  inequality in Lemma \ref{lemma_10},
 \begin{align}
   \widehat{\boldsymbol{\alpha}}_k\trans \widehat{\mathbf{B}}\widehat{\boldsymbol{\alpha}}_k&\le \mu_k(\boldsymbol{\Xi})+(b_{k-1,3}\hbar^{-1}+1)\sqrt{\mu_1(\boldsymbol{\Xi})}\kappa^{-1}C_0\varpi\sqrt{s}\notag\\
&+\frac{(1+\hbar^{-1})}{\sqrt{s}}\kappa\sqrt{\mu_1(\boldsymbol{\Xi})}C_0\varpi \|\widehat{\boldsymbol{\alpha}}_k\|_1^2\notag\\
&\le \mu_1(\boldsymbol{\Xi})+(b_{k-1,3}\hbar^{-1}+1)\sqrt{\mu_1(\boldsymbol{\Xi})}\kappa^{-1}C_0\varpi\sqrt{s}\notag\\
&+\frac{(1+\hbar^{-1})}{\sqrt{s}}\kappa\sqrt{\mu_1(\boldsymbol{\Xi})}C_0\varpi D_{k,1}^2\kappa^{-2} s\notag\\
&=\mu_1(\boldsymbol{\Xi})+[(b_{k-1,3}\hbar^{-1}+1)+(1+\hbar^{-1}) D_{k,1}^2]\sqrt{\mu_1(\boldsymbol{\Xi})}\kappa^{-1}C_0\varpi\sqrt{s}\notag\\
&\le \mu_1(\boldsymbol{\Xi})+[(b_{k-1,3}\hbar^{-1}+1)+(1+\hbar^{-1}) D_{k,1}^2]\sqrt{\mu_1(\boldsymbol{\Xi})}\hbar^{-1}\sqrt{\mu_1(\boldsymbol{\Xi})}\notag\\
& =N_{k,6}\mu_1(\boldsymbol{\Xi}),\label{369}
\end{align} 
where the second  inequality is due to $\eqref{10258}$ and $N_{k,6}=1+(b_{k-1,3}\hbar^{-1}+1)\hbar^{-1}+(1+\hbar^{-1}) D_{k,1}^2\hbar^{-1}$. Now $\eqref{10262}-\eqref{369}$ and $\eqref{10257}$ give
 \begin{align}
   1+\tau^{(k)}\|\widehat{\boldsymbol{\alpha}}_k\|^2_{\lambda^{(k)}}&\le \frac{\widehat{\boldsymbol{\alpha}}_k\trans \widehat{\mathbf{B}}\widehat{\boldsymbol{\alpha}}_k}{\boldsymbol{\beta}_k\trans \widehat{\mathbf{B}}\boldsymbol{\beta}_k}(1+\tau^{(k)}\|\boldsymbol{\beta}_k\|^2_{\lambda^{(k)}})\notag\\
&\le N_{k,6}N_{k,7}^{-1}(1+A^{(k)}M_{k,1}^2\hbar^{-1}). \label{371}
\end{align} 
Therefore, by $\eqref{10265}$-$\eqref{10261}$ and $\eqref{371}$,  
 \begin{align}
&(\boldsymbol{\beta}_k\trans \widehat{\mathbf{B}}\boldsymbol{\beta}_k-\widehat{\boldsymbol{\alpha}}_k\trans \widehat{\mathbf{B}}\widehat{\boldsymbol{\alpha}}_k)(1+\tau^{(k)}\|\widehat{\boldsymbol{\alpha}}_k\|^2_{\lambda^{(k)}})\label{10263}\\
 \ge& \left(\frac{1}{2}c_2\mu_k(\boldsymbol{\Xi})\|\boldsymbol{\gamma}_k-\widehat{\boldsymbol{\gamma}}_k\|_2^2-3 b_{k-1,3}C_0^2\kappa^{-2} \varpi^2 s\right\notag\\
  &\left -[4+ (M_{k,1}+D_{k,1})\hbar^{-1}]\sqrt{\mu_1(\boldsymbol{\Xi})}C_0\varpi\|\boldsymbol{\beta}_k-\widehat{\boldsymbol{\alpha}}_k\|_1\right)(1+\tau^{(k)}\|\widehat{\boldsymbol{\alpha}}_k\|^2_{\lambda^{(k)}})\notag\\
\ge& \frac{1}{2}c_2\mu_k(\boldsymbol{\Xi})\|\boldsymbol{\gamma}_k-\widehat{\boldsymbol{\gamma}}_k\|_2^2-N_{k,4}C_0^2\kappa^{-2} \varpi^2 s-N_{k,5}C_0\sqrt{\mu_1(\boldsymbol{\Xi})}\varpi\|\boldsymbol{\beta}_k-\widehat{\boldsymbol{\alpha}}_k\|_1,\notag
\end{align} 
where 
 \begin{align*}
 &N_{k,4}=3b_{k-1,3}N_{k,6}N_{k,7}^{-1}(1+A^{(k)}M_{k,1}^2\hbar^{-1}),\\
& N_{k,5}=[4+ (M_{k,1}+D_{k,1})\hbar^{-1}]N_{k,6}N_{k,7}^{-1}(1+A^{(k)}M_{k,1}^2\hbar^{-1}).
\end{align*} 
Therefore, by $\eqref{20252}$ and $\eqref{10263}$, we obtain
 \begin{align}
& \frac{1}{2}c_2\mu_k(\boldsymbol{\Xi})\|\boldsymbol{\gamma}_k-\widehat{\boldsymbol{\gamma}}_k\|_2^2\notag\\
\le& N_{k,4}C_0^2\kappa^{-2} \varpi^2 s+N_{k,5}C_0\sqrt{\mu_1(\boldsymbol{\Xi})}\varpi\|\boldsymbol{\beta}_k-\widehat{\boldsymbol{\alpha}}_k\|_1\notag\\
&+ (\widehat{\boldsymbol{\alpha}}_k\trans \widehat{\mathbf{B}}\widehat{\boldsymbol{\alpha}}_k )[\tau^{(k)}\|\boldsymbol{\beta}_k\|^2_{\lambda^{(k)}}-\tau^{(k)}\|\widehat{\boldsymbol{\alpha}}_k\|^2_{\lambda^{(k)}}].\label{10267}
\end{align} 
 To estimate the term $(\widehat{\boldsymbol{\alpha}}_k\trans \widehat{\mathbf{B}}\widehat{\boldsymbol{\alpha}}_k )[\tau^{(k)}\|\boldsymbol{\beta}_k\|^2_{\lambda^{(k)}}-\tau^{(k)}\|\widehat{\boldsymbol{\alpha}}_k\|^2_{\lambda^{(k)}}]$ in $\eqref{10267}$, we consider the following three possible situations, separately:
\begin{itemize}
\item {\bf Situation 1:} $\boldsymbol{\beta}_k\trans \widehat{\mathbf{B}}\boldsymbol{\beta}_k\ge \widehat{\boldsymbol{\alpha}}_k\trans \widehat{\mathbf{B}}\widehat{\boldsymbol{\alpha}}_k$. \\
In this case, by $\eqref{20252}$ and $\eqref{369}$, we have 
 \begin{align*}
0&\le (\boldsymbol{\beta}_k\trans \widehat{\mathbf{B}}\boldsymbol{\beta}_k-\widehat{\boldsymbol{\alpha}}_k\trans \widehat{\mathbf{B}}\widehat{\boldsymbol{\alpha}}_k)(1+\tau^{(k)}\|\widehat{\boldsymbol{\alpha}}_k\|^2_{\lambda^{(k)}})\notag\\
&\le  (\widehat{\boldsymbol{\alpha}}_k\trans \widehat{\mathbf{B}}\widehat{\boldsymbol{\alpha}}_k )[\tau^{(k)}\|\boldsymbol{\beta}_k\|^2_{\lambda^{(k)}}-\tau^{(k)}\|\widehat{\boldsymbol{\alpha}}_k\|^2_{\lambda^{(k)}}]\notag\\
&\le N_{k,6}\mu_1(\boldsymbol{\Xi})[\tau^{(k)}\|\boldsymbol{\beta}_k\|^2_{\lambda^{(k)}}-\tau^{(k)}\|\widehat{\boldsymbol{\alpha}}_k\|^2_{\lambda^{(k)}}]. 
\end{align*} 
\item {\bf Situation 2:} $\boldsymbol{\beta}_k\trans \widehat{\mathbf{B}}\boldsymbol{\beta}_k< \widehat{\boldsymbol{\alpha}}_k\trans \widehat{\mathbf{B}}\widehat{\boldsymbol{\alpha}}_k$ and $\|\boldsymbol{\beta}_k\|^2_{\lambda^{(k)}}<\|\widehat{\boldsymbol{\alpha}}_k\|^2_{\lambda^{(k)}}$. \\
In this case, by   $\eqref{370}$, we have 
 \begin{align*}
&  (\widehat{\boldsymbol{\alpha}}_k\trans \widehat{\mathbf{B}}\widehat{\boldsymbol{\alpha}}_k )[\tau^{(k)}\|\boldsymbol{\beta}_k\|^2_{\lambda^{(k)}}-\tau^{(k)}\|\widehat{\boldsymbol{\alpha}}_k\|^2_{\lambda^{(k)}}]\notag\\
<& (\boldsymbol{\beta}_k\trans \widehat{\mathbf{B}}\boldsymbol{\beta}_k)[\tau^{(k)}\|\boldsymbol{\beta}_k\|^2_{\lambda^{(k)}}-\tau^{(k)}\|\widehat{\boldsymbol{\alpha}}_k\|^2_{\lambda^{(k)}}]\notag\\
\le& N_{k,7}\mu_1(\boldsymbol{\Xi})[\tau^{(k)}\|\boldsymbol{\beta}_k\|^2_{\lambda^{(k)}}-\tau^{(k)}\|\widehat{\boldsymbol{\alpha}}_k\|^2_{\lambda^{(k)}}]. 
\end{align*} 
\item {\bf Situation 3:} $\boldsymbol{\beta}_k\trans \widehat{\mathbf{B}}\boldsymbol{\beta}_k< \widehat{\boldsymbol{\alpha}}_k\trans \widehat{\mathbf{B}}\widehat{\boldsymbol{\alpha}}_k$ and $\|\boldsymbol{\beta}_k\|^2_{\lambda^{(k)}}\ge\|\widehat{\boldsymbol{\alpha}}_k\|^2_{\lambda^{(k)}}$. \\
In this case, by   $\eqref{369}$, we have 
 \begin{align*}
&  (\widehat{\boldsymbol{\alpha}}_k\trans \widehat{\mathbf{B}}\widehat{\boldsymbol{\alpha}}_k )[\tau^{(k)}\|\boldsymbol{\beta}_k\|^2_{\lambda^{(k)}}-\tau^{(k)}\|\widehat{\boldsymbol{\alpha}}_k\|^2_{\lambda^{(k)}}] \notag\\
\le& N_{k,6}\mu_1(\boldsymbol{\Xi})[\tau^{(k)}\|\boldsymbol{\beta}_k\|^2_{\lambda^{(k)}}-\tau^{(k)}\|\widehat{\boldsymbol{\alpha}}_k\|^2_{\lambda^{(k)}}]. 
\end{align*} 
\end{itemize} 
Hence, in all the three situations, we have either 
 \begin{align*}
&  (\widehat{\boldsymbol{\alpha}}_k\trans \widehat{\mathbf{B}}\widehat{\boldsymbol{\alpha}}_k )[\tau^{(k)}\|\boldsymbol{\beta}_k\|^2_{\lambda^{(k)}}-\tau^{(k)}\|\widehat{\boldsymbol{\alpha}}_k\|^2_{\lambda^{(k)}}] \notag\\
\le&N_{k,6}\mu_1(\boldsymbol{\Xi})[\tau^{(k)}\|\boldsymbol{\beta}_k\|^2_{\lambda^{(k)}}-\tau^{(k)}\|\widehat{\boldsymbol{\alpha}}_k\|^2_{\lambda^{(k)}}]. 
\end{align*} 
or
 \begin{align*}
&  (\widehat{\boldsymbol{\alpha}}_k\trans \widehat{\mathbf{B}}\widehat{\boldsymbol{\alpha}}_k )[\tau^{(k)}\|\boldsymbol{\beta}_k\|^2_{\lambda^{(k)}}-\tau^{(k)}\|\widehat{\boldsymbol{\alpha}}_k\|^2_{\lambda^{(k)}}] \notag\\
\le&N_{k,7}\mu_1(\boldsymbol{\Xi})[\tau^{(k)}\|\boldsymbol{\beta}_k\|^2_{\lambda^{(k)}}-\tau^{(k)}\|\widehat{\boldsymbol{\alpha}}_k\|^2_{\lambda^{(k)}}],
\end{align*} 
which together with $\eqref{10267}$ lead to the conclusion.\\

 \end{proof}

\begin{proof}  $\ref{lemma_13}$.\\

We first consider the second term on the right hand side of $\eqref{383}$. Recall that $\boldsymbol{\beta}_k=\boldsymbol{\alpha}_k-\boldsymbol{\delta}_k$. By $\eqref{50043}$, 
 \begin{align}
  &N_{k,5}C_0\sqrt{\mu_1(\boldsymbol{\Xi})}\varpi\|\boldsymbol{\beta}_k-\widehat{\boldsymbol{\alpha}}_k\|_1=N_{k,5}C_0\sqrt{\mu_1(\boldsymbol{\Xi})}\varpi\|\boldsymbol{\alpha}_k-\boldsymbol{\delta}_k-\widehat{\boldsymbol{\alpha}}_k\|_1\label{10270}\\
	\le& N_{k,5}C_0\sqrt{\mu_1(\boldsymbol{\Xi})}\varpi\|\boldsymbol{\alpha}_k-\widehat{\boldsymbol{\alpha}}_k\|_1+N_{k,5}C_0\sqrt{\mu_1(\boldsymbol{\Xi})}\varpi\|\boldsymbol{\delta}_k\|_1\notag\\
	\le& N_{k,5}C_0\sqrt{\mu_1(\boldsymbol{\Xi})}\varpi\|\boldsymbol{\alpha}_k-\widehat{\boldsymbol{\alpha}}_k\|_1+N_{k,5}C_0\sqrt{\mu_1(\boldsymbol{\Xi})}\varpi M_{k,0}C_0\kappa^{-2}\mu_1(\boldsymbol{\Xi})^{-1/2}\varpi s\notag\\
	=&	N_{k,5}C_0\sqrt{\mu_1(\boldsymbol{\Xi})}\varpi\|\boldsymbol{\alpha}_k-\widehat{\boldsymbol{\alpha}}_k\|_1+N_{k,5} M_{k,0}C_0^2\kappa^{-2} \varpi^2 s.\notag
\end{align} 
Now we consider the last term on the right hand side of $\eqref{383}$. By $\eqref{10245}$,
 \begin{align}
  & \|\boldsymbol{\beta}_k\|_2^2\le (\|\boldsymbol{\alpha}_k\|_2+\|\boldsymbol{\delta}_k\|_2)^2\le  (\|\boldsymbol{\alpha}_k\|_2+\|\boldsymbol{\delta}_k\|_1)^2\notag\\
	\le& \|\boldsymbol{\alpha}_k\|_2^2+2\|\boldsymbol{\alpha}_k\|_1\|\boldsymbol{\delta}_k\|_1+\|\boldsymbol{\delta}_k\|_1^2\notag\\
	\le &\|\boldsymbol{\alpha}_k\|_2^2+2(\|\boldsymbol{\alpha}_k\|_1+\|\boldsymbol{\delta}_k\|_1)\|\boldsymbol{\delta}_k\|_1\notag\\
	\le &\|\boldsymbol{\alpha}_k\|_2^2+2M_{k,1}\|\boldsymbol{\alpha}_1\|_1M_{k,0}\kappa^{-2}\mu_1(\boldsymbol{\Xi})^{-1/2}C_0\varpi s \label{10271}
\end{align} 
Similarly,
 \begin{align}
  & \|\boldsymbol{\beta}_k\|_1^2\le  (\|\boldsymbol{\alpha}_k\|_1+\|\boldsymbol{\delta}_k\|_1)^2= \|\boldsymbol{\alpha}_k\|_1^2+2\|\boldsymbol{\alpha}_k\|_1\|\boldsymbol{\delta}_k\|_1+\|\boldsymbol{\delta}_k\|_1^2\notag\\
	\le & \|\boldsymbol{\alpha}_k\|_1^2+2M_{k,1}\|\boldsymbol{\alpha}_1\|_1M_{k,0}\kappa^{-2}\mu_1(\boldsymbol{\Xi})^{-1/2}C_0\varpi s .\label{10272}
\end{align} 
Combining $\eqref{10271}$ and $\eqref{10272}$ gives
 \begin{align*}
& \tau^{(k)}\|\boldsymbol{\beta}_k\|^2_{\lambda^{(k)}}=\tau^{(k)}(1-\lambda^{(k)})\|\boldsymbol{\beta}_k\|^2_2+\tau^{(k)}\lambda^{(k)}\|\boldsymbol{\beta}_k\|^2_1\notag\\
\le& \tau^{(k)}(1-\lambda^{(k)})\left[\|\boldsymbol{\alpha}_k\|_2^2+2M_{k,1}\|\boldsymbol{\alpha}_1\|_1M_{k,0}\kappa^{-2}\mu_1(\boldsymbol{\Xi})^{-1/2}C_0\varpi s \right]\notag\\
&+\tau^{(k)}\lambda^{(k)}\left[\|\boldsymbol{\alpha}_k\|_1^2+2M_{k,1}\|\boldsymbol{\alpha}_1\|_1M_{k,0}\kappa^{-2}\mu_1(\boldsymbol{\Xi})^{-1/2}C_0\varpi s \right]\notag\\
=&\tau^{(k)} \|\boldsymbol{\alpha}_k\|_{\lambda^{(k)}}^2+\tau^{(k)}2M_{k,1}\|\boldsymbol{\alpha}_1\|_1M_{k,0}\kappa^{-2}\mu_1(\boldsymbol{\Xi})^{-1/2}C_0\varpi s \notag\\
=&\tau^{(k)} \|\boldsymbol{\alpha}_k\|_{\lambda^{(k)}}^2+\frac{A^{(k)}C_0\varpi}{\|\boldsymbol{\alpha}_1\|_1\sqrt{\mu_1(\boldsymbol{\Xi})}}2M_{k,1}\|\boldsymbol{\alpha}_1\|_1M_{k,0}\kappa^{-2}\mu_1(\boldsymbol{\Xi})^{-1/2}C_0\varpi s \notag\\
=&\tau^{(k)} \|\boldsymbol{\alpha}_k\|_{\lambda^{(k)}}^2+ 2A^{(k)}M_{k,1} M_{k,0}\kappa^{-2}\mu_1(\boldsymbol{\Xi})^{-1}C_0^2\varpi^2 s 
\end{align*} 
which together with $\eqref{369}$ lead to
 \begin{align}
&  N_{k,6}\mu_1(\boldsymbol{\Xi})\tau^{(k)}\|\boldsymbol{\beta}_k\|^2_{\lambda^{(k)}}\label{10273}\\
\le& N_{k,6}\mu_1(\boldsymbol{\Xi})\tau^{(k)}\|\boldsymbol{\alpha}_k\|^2_{\lambda^{(k)}}+ N_{k,6}\mu_1(\boldsymbol{\Xi})2A^{(k)}M_{k,1} M_{k,0}\kappa^{-2}\mu_1(\boldsymbol{\Xi})^{-1}C_0^2\varpi^2 s \notag\\
\le & N_{k,6}\mu_1(\boldsymbol{\Xi})\tau^{(k)}\|\boldsymbol{\alpha}_k\|^2_{\lambda^{(k)}}+ 2A^{(k)}N_{k,6}M_{k,1} M_{k,0}\kappa^{-2} C_0^2\varpi^2 s\notag
\end{align} 
By $\eqref{383}$ and $\eqref{10270}$-$\eqref{10273}$,
 \begin{align}
& \frac{1}{2}c_2\mu_k(\boldsymbol{\Xi})\|\boldsymbol{\gamma}_k-\widehat{\boldsymbol{\gamma}}_k\|_2^2\label{10274}\\
\le& N_{k,4}C_0^2\kappa^{-2} \varpi^2 s+N_{k,5}C_0\sqrt{\mu_1(\boldsymbol{\Xi})}\varpi\|\boldsymbol{\beta}_k-\widehat{\boldsymbol{\alpha}}_k\|_1\notag\\
&+ N_{k,6}\mu_1(\boldsymbol{\Xi})[\tau^{(k)}\|\boldsymbol{\beta}_k\|^2_{\lambda^{(k)}}-\tau^{(k)}\|\widehat{\boldsymbol{\alpha}}_k\|^2_{\lambda^{(k)}}]\notag\\
\le & N_{k,4}C_0^2\kappa^{-2} \varpi^2 s+N_{k,5}C_0\sqrt{\mu_1(\boldsymbol{\Xi})}\varpi\|\boldsymbol{\alpha}_k-\widehat{\boldsymbol{\alpha}}_k\|_1+N_{k,5} M_{k,0}C_0^2\kappa^{-2} \varpi^2 s\notag\\
&+2A^{(k)}N_{k,6} M_{k,1} M_{k,0}\kappa^{-2} C_0^2\varpi^2 s+ N_{k,6}\mu_1(\boldsymbol{\Xi})[\tau^{(k)}\|\boldsymbol{\alpha}_k\|^2_{\lambda^{(k)}}-\tau^{(k)}\|\widehat{\boldsymbol{\alpha}}_k\|^2_{\lambda^{(k)}}]\notag\\
=&N_{k,1}C_0^2\kappa^{-2} \varpi^2 s+N_{k,5}C_0\sqrt{\mu_1(\boldsymbol{\Xi})}\varpi\|\boldsymbol{\alpha}_k-\widehat{\boldsymbol{\alpha}}_k\|_1\notag\\
& + N_{k,6}\mu_1(\boldsymbol{\Xi})[\tau^{(k)}\|\boldsymbol{\alpha}_k\|^2_{\lambda^{(k)}}-\tau^{(k)}\|\widehat{\boldsymbol{\alpha}}_k\|^2_{\lambda^{(k)}}],\notag
\end{align} 
where $N_{k,1}=N_{k,4}+N_{k,5} M_{k,0} +2A^{(k)}N_{k,6}M_{k,1} M_{k,0}$. Next, we calculate
 \begin{align*}
 &\tau^{(k)}\|\boldsymbol{\alpha}_k\|^2_{\lambda^{(k)}}-\tau^{(k)}\|\widehat{\boldsymbol{\alpha}}_k\|^2_{\lambda^{(k)}}\\
 =&\frac{A^{(k)}C_0\varpi}{\|\boldsymbol{\alpha}_1\|_1\mu_1(\boldsymbol{\Xi})^{1/2}}\left[(1-\lambda^{(k)})(\|\boldsymbol{\alpha}_k\|^2_2-\|\widehat{\boldsymbol{\alpha}}_k\|^2_2)+\lambda^{(k)}(\|\boldsymbol{\alpha}_k\|^2_1-\|\widehat{\boldsymbol{\alpha}}_k\|^2_1)\right].\notag
\end{align*} 
We have
 \begin{align}
 &\frac{A^{(k)}C_0\varpi}{\|\boldsymbol{\alpha}_1\|_1\mu_1(\boldsymbol{\Xi})^{1/2}}(1-\lambda^{(k)})(\|\boldsymbol{\alpha}_k\|^2_2-\|\widehat{\boldsymbol{\alpha}}_k\|^2_2)\label{10031}\\
=&\frac{A^{(k)}C_0\varpi}{\|\boldsymbol{\alpha}_1\|_1\mu_1(\boldsymbol{\Xi})^{1/2}}(1-\lambda^{(k)})(\|(\boldsymbol{\alpha}_k)_{J_k}\|^2_2-\|(\widehat{\boldsymbol{\alpha}}_k)_{J_k}\|_2^2-\|(\widehat{\boldsymbol{\alpha}}_k)_{J_k^c}\|^2_2)\notag\\
\le &\frac{A^{(k)}C_0\varpi}{\|\boldsymbol{\alpha}_1\|_1\mu_1(\boldsymbol{\Xi})^{1/2}}(1-\lambda^{(k)})(\|(\boldsymbol{\alpha}_k)_{J_k}\|^2_2-\|(\widehat{\boldsymbol{\alpha}}_k)_{J_k}\|^2_2)\notag\\
= &\frac{A^{(k)}C_0\varpi}{\|\boldsymbol{\alpha}_1\|_1\mu_1(\boldsymbol{\Xi})^{1/2}}(1-\lambda^{(k)})(\|( \boldsymbol{\alpha}_k)_{J_k}\|_2-\|(\widehat{\boldsymbol{\alpha}}_k )_{J_k}\|_2)(\|(\boldsymbol{\alpha}_k)_{J_k}\|_2+\|(\widehat{\boldsymbol{\alpha}}_k)_{J_k}\|_2)\notag\\
= &\frac{A^{(k)}C_0\varpi}{\|\boldsymbol{\alpha}_1\|_1\mu_1(\boldsymbol{\Xi})^{1/2}}(1-\lambda^{(k)}) [(\|( \boldsymbol{\alpha}_k)_{J_k}\|_2-\|(\widehat{\boldsymbol{\alpha}}_k )_{J_k}\|_2)(2\|(\boldsymbol{\alpha}_k)_{J_k}\|_2)\notag\\
&\qquad\qquad \qquad\qquad\qquad\qquad-(\|( \boldsymbol{\alpha}_k)_{J_k}\|_2-\|(\widehat{\boldsymbol{\alpha}}_k )_{J_k}\|_2)^2 ]\notag\\
\le &\frac{A^{(k)}C_0\varpi}{\|\boldsymbol{\alpha}_1\|_1\mu_1(\boldsymbol{\Xi})^{1/2}}(1-\lambda^{(k)})\left[(\|( \boldsymbol{\alpha}_k)_{J_k}\|_2-\|(\widehat{\boldsymbol{\alpha}}_k )_{J_k}\|_2)(2\|(\boldsymbol{\alpha}_k)_{J_k}\|_2) \right]\notag\\
\le &A^{(k)}C_0\mu_1(\boldsymbol{\Xi})^{-1/2}\varpi(1-\lambda^{(k)})\|(\widehat{\boldsymbol{\alpha}}_k-\boldsymbol{\alpha}_k)_{J_k}\|_2(2\|(\boldsymbol{\alpha}_k)_{J_k}\|_2/\|\boldsymbol{\alpha}_1\|_1)\notag\\
 \le &A^{(k)}C_0\mu_1(\boldsymbol{\Xi})^{-1/2}\varpi(1-\lambda^{(k)})\|(\widehat{\boldsymbol{\alpha}}_k-\boldsymbol{\alpha}_k)_{J_k}\|_1(2\|\boldsymbol{\alpha}_k\|_1/\|\boldsymbol{\alpha}_1\|_1)\notag\\
=& A^{(k)}C_0\mu_1(\boldsymbol{\Xi})^{-1/2}\varpi(1-\lambda^{(k)})\|(\widehat{\boldsymbol{\alpha}}_k-\boldsymbol{\alpha}_k)_{J_k}\|_1(2\|\boldsymbol{\alpha}_k\|_1/\|\boldsymbol{\alpha}_1\|_1) , \notag
\end{align} 
and
 \begin{align}
 &\frac{A^{(k)}C_0\varpi}{\|\boldsymbol{\alpha}_1\|_1\mu_1(\boldsymbol{\Xi})^{1/2}}\lambda^{(k)}(\|\boldsymbol{\alpha}_k\|^2_1-\|\widehat{\boldsymbol{\alpha}}_k\|^2_1)\label{170032}\\
=&\frac{A^{(k)}C_0\varpi}{\|\boldsymbol{\alpha}_1\|_1\mu_1(\boldsymbol{\Xi})^{1/2}}\lambda^{(k)}(\|\boldsymbol{\alpha}_k\|_1-\|\widehat{\boldsymbol{\alpha}}_k\|_1)(\|\boldsymbol{\alpha}_k\|_1+\|\widehat{\boldsymbol{\alpha}}_k\|_1) \notag\\
=&\frac{A^{(k)}C_0\varpi}{\|\boldsymbol{\alpha}_1\|_1\mu_1(\boldsymbol{\Xi})^{1/2}}\lambda^{(k)}\left[(\|\boldsymbol{\alpha}_k\|_1-\|\widehat{\boldsymbol{\alpha}}_k\|_1)(2\|\boldsymbol{\alpha}_k\|_1)-(\|\boldsymbol{\alpha}_k\|_1-\|\widehat{\boldsymbol{\alpha}}_k\|_1)^2\right] \notag\\
\le &\frac{A^{(k)}C_0\varpi}{\|\boldsymbol{\alpha}_1\|_1\mu_1(\boldsymbol{\Xi})^{1/2}}\lambda^{(k)}\left[(\|\boldsymbol{\alpha}_k\|_1-\|\widehat{\boldsymbol{\alpha}}_k\|_1)(2\|\boldsymbol{\alpha}_k\|_1) \right] \notag\\
=&A^{(k)}C_0\mu_1(\boldsymbol{\Xi})^{-1/2}\varpi\lambda^{(k)}(\|(\boldsymbol{\alpha}_k)_{J_k}\|_1-\|(\widehat{\boldsymbol{\alpha}}_k)_{J_k}\|_1-\|(\widehat{\boldsymbol{\alpha}}_k)_{J_k^c}\|_1)(2\|\boldsymbol{\alpha}_k\|_1/\|\boldsymbol{\alpha}_1\|_1)  \notag\\
\le &A^{(k)}C_0\mu_1(\boldsymbol{\Xi})^{-1/2}\varpi\lambda^{(k)} (\|(\boldsymbol{\alpha}_k-\widehat{\boldsymbol{\alpha}}_k)_{J_k}\|_1-\|(\widehat{\boldsymbol{\alpha}}_k)_{J_k^c}\|_1)(2\|\boldsymbol{\alpha}_k\|_1/\|\boldsymbol{\alpha}_1\|_1) .\notag
\end{align} 
 Therefore, combining $\eqref{10031}$ and $\eqref{170032}$, we obtain
 \begin{align*}
 & \tau^{(k)}\|\boldsymbol{\alpha}_k\|^2_{\lambda^{(k)}}-\tau^{(k)}\|\widehat{\boldsymbol{\alpha}}_k\|^2_{\lambda^{(k)}} \\
\le &A^{(k)}C_0\mu_1(\boldsymbol{\Xi})^{-1/2}\varpi\left[\|(\boldsymbol{\alpha}_k-\widehat{\boldsymbol{\alpha}}_k)_{J_k}\|_1-\lambda^{(k)} \|(\widehat{\boldsymbol{\alpha}}_k)_{J_k^c}\|_1\right](2\|\boldsymbol{\alpha}_k\|_1/\|\boldsymbol{\alpha}_1\|_1),\notag
\end{align*} 
 which leads to
 \begin{align}
 & N_{k,6}\mu_1(\boldsymbol{\Xi})[\tau^{(k)}\|\boldsymbol{\alpha}_k\|^2_{\lambda^{(k)}}-\tau^{(k)}\|\widehat{\boldsymbol{\alpha}}_k\|^2_{\lambda^{(k)}}]\notag\\
\le &N_{k,2}\mu_1(\boldsymbol{\Xi})^{1/2}C_0\varpi\left[\|(\boldsymbol{\alpha}_k-\widehat{\boldsymbol{\alpha}}_k)_{J_k}\|_1-\lambda^{(k)} \|(\widehat{\boldsymbol{\alpha}}_k)_{J_k^c}\|_1\right],\label{35600}
\end{align} 
 where $N_{k,2}=A^{(k)}N_{k,6}(2\|\boldsymbol{\alpha}_k\|_1/\|\boldsymbol{\alpha}_1\|_1)$. By $\eqref{10274}$, $\eqref{35600}$ and noting that $\|\boldsymbol{\alpha}_k-\widehat{\boldsymbol{\alpha}}_k\|_1=\|(\boldsymbol{\alpha}_k-\widehat{\boldsymbol{\alpha}}_k)_{J_k}\|_1+\|(\widehat{\boldsymbol{\alpha}}_k)_{J_k^c}\|_1$, we have
 \begin{align*}
& \frac{1}{2}c_2\mu_k(\boldsymbol{\Xi})\|\boldsymbol{\gamma}_k-\widehat{\boldsymbol{\gamma}}_k\|_2^2\notag\\
\le& N_{k,1}C_0^2\kappa^{-2} \varpi^2 s+ (N_{k,2}+N_{k,5})\mu_1(\boldsymbol{\Xi})^{1/2}C_0\varpi \|(\boldsymbol{\alpha}_k-\widehat{\boldsymbol{\alpha}}_k)_{J_k}\|_1\notag\\
&-(\lambda^{(k)}N_{k,2}-N_{k,5})\mu_1(\boldsymbol{\Xi})^{1/2}C_0\varpi \|(\widehat{\boldsymbol{\alpha}}_k)_{J_k^c}\|_1 , 
\end{align*} 

\end{proof}